\documentclass[useAMS,usenatbib]{mn2e}
\usepackage{graphicx,rotating,subfigure,amsmath,wasysym,amssymb}
\usepackage[retainorgcmds]{IEEEtrantools}
\usepackage{ctable}
\usepackage{multirow}

\newcommand{\rem}[1]{}  

\newcommand{\markChange}[1]{{#1}} 

\newcommand  \dist[1]   {\ifmmode d_{#1}\else $d_{#1}$\fi} 

\newcommand{\ts}{\textsuperscript}
\def \second    {2\ts{\scriptsize{nd}} }  

\def \log       {\ifmmode \mbox{log}\else log\fi}
\def \dex       {\ifmmode \mbox{dex}\else dex\fi}

\def  \sun	{\ifmmode \odot \else \mbox{$\odot$} \fi}
\def  \earth	{\ifmmode \oplus \else \mbox{$\oplus$} \fi}

\def  \Hubble   {\ifmmode H_{0}\else $H_{0}$\fi} 
\def  \h        {\ifmmode h\else $h$\fi} 

\def  \deg	{\ifmmode ^{\circ}\else        $^{\circ}$\fi}
\def  \arcmin	{\ifmmode ^\prime\else         $^\prime$\fi}
\def  \arcsec	{\ifmmode ^{\prime\prime}\else $^{\prime\prime}$\fi}
\def  \RA       {\ifmmode \alpha\else          $\alpha$\fi}  
\def  \Dec      {\ifmmode \delta\else          $\delta$\fi} 

\def  \sr       {\ifmmode sr\else              $sr$\fi}                     
\def  \sqDeg    {\ifmmode \mbox{deg}^{2}\else  $\mbox{deg}^{2}$\fi}         

\def  \yr       {\ifmmode \mbox{yr} \else $\mbox{yr}$ \fi} 
\def  \Myr      {\ifmmode \mbox{Myr}\else $\mbox{Myr}$\fi} 
\def  \Gyr      {\ifmmode \mbox{Gyr}\else $\mbox{Gyr}$\fi} 

\def  \GHz      {\ifmmode \mbox{GHz} \else GHz \fi} 

\def  \um      {\ifmmode \mu\!\mbox{m}\else $\mu\!\mbox{m}$\fi} 
\def  \nm      {\ifmmode \mbox{nm}\else nm\fi} 
\def  \Angst   {\ifmmode \mbox{\AA}\else  $\mbox{\AA}$\fi} 
\def  \pc      {\ifmmode \mbox{pc}  \else $\mbox{pc}$ \fi} 
\def  \kpc     {\ifmmode \mbox{kpc}\else $\mbox{kpc}$\fi} 
\def  \Mpc     {\ifmmode \mbox{Mpc}\else $\mbox{Mpc}$\fi} 

\def  \Mpch    {\ifmmode \h^{-1}\,\Mpc\else $\h^{-1}\,\Mpc$\fi} 
\def  \VolMpch {\ifmmode \h^{-3}\,\Mpc^{3}\else $\h^{-3}\,\Mpc^{3}$\fi} 
\def  \NumCntMpch {\ifmmode \h^{3}\,\Mpc^{-3}\else $\h^{3}\,\Mpc^{-3}$\fi} 

\def  \kms	{\ifmmode {\rm km\,s}^{-1}\else {km\,s$^{-1}$}\fi}

\def  \z       {\ifmmode \mbox{z}\else  z\fi}  
\def  \cz      {\ifmmode \mbox{cz}\else cz\fi} 

\def  \Mass	 {\ifmmode M                \else $M$                 \fi}
\def  \Mearth	 {\ifmmode \Mass_\earth     \else $\Mass_\earth$      \fi}
\def  \Msun      {\ifmmode \Mass_\sun\else        $\Mass_\sun$\fi}
\def  \Msunh     {\ifmmode \h^{-1}\,\Msun\else    $\h^{-1}\,\Msun$\fi}      
\def  \MHI       {\ifmmode \Mass_{HI}\else        $\Mass_{HI}$\fi}
\def  \Mstar     {\ifmmode \Mass_{*}\else         $\Mass_{*}$\fi}  
\def  \Mhalo     {\ifmmode \Mass_{halo}\else      $\Mass_{halo}$\fi} 

\def  \density         {\ifmmode \rho        \else $\rho$        \fi} 
\def  \densityMsunMpch {\ifmmode {\h^{2}\,\Msun\,\Mpc^{-3}}\else 
                                 $\h^{2}\,\Msun\,\Mpc^{-3}$ \fi} 

\def  \Lunimosity {\ifmmode L\else $L$\fi}
\def  \SpecLum    {\ifmmode \Lunimosity_\nu \else $\Lunimosity_\nu$ \fi} 
\def  \Lsun       {\ifmmode \Lunimosity_\sun \else $\Lunimosity_\sun$ \fi}
\def  \Lsunh      {\ifmmode \h^{-2}\,\Lsun\else    $\h^{-2}\,\Lsun$\fi}      

\def \U           {\ifmmode U\else $U$\fi}   
\def \B           {\ifmmode B\else $B$\fi}   
\def \V           {\ifmmode V\else $V$\fi}   
\def \R           {\ifmmode R\else $R$\fi}   
\def \I           {\ifmmode I\else $I$\fi}   
\def \FUV         {\ifmmode FUV\else $FUV$\fi} 
\def \NUV         {\ifmmode NUV\else $NUV$\fi} 
\def \J           {\ifmmode J\else $J$\fi}     
\def \H           {\ifmmode H\else $H$\fi}     
\def \Ks          {\ifmmode K_{s}\else $K_{s}$\fi} 
\def \WThree      {\ifmmode W3\else $W3$\fi}  
\def \WFour       {\ifmmode W4\else $W4$\fi}  

\def \m           {\mbox{m}} 
\def \magu        {\U} 
\def \magb        {\B} 
\def \magv        {\V} 
\def \magr        {\R} 
\def \magi        {\I} 
\def \SDSSu       {\ifmmode u\else $u$\fi} 
\def \SDSSg       {\ifmmode g\else $g$\fi} 
\def \SDSSr       {\ifmmode r\else $r$\fi} 
\def \SDSSi       {\ifmmode i\else $i$\fi} 
\def \SDSSz       {\ifmmode z\else $z$\fi} 
\def \magFUV      {\ifmmode \m_{\mbox{\scriptsize FUV}}\else $\mbox{m}_{\mbox{\scriptsize FUV}}$\fi} 
\def \magNUV      {\ifmmode \m_{\mbox{\scriptsize NUV}}\else $\mbox{m}_{\mbox{\scriptsize NUV}}$\fi} 
\def \magJ        {\J}   
\def \magH        {\H}   
\def \magKs       {\Ks} 
\def \magWThree   {\ifmmode \m_{\WThree}\else $\mbox{m}_{\WThree}$\fi} 
\def \magWFour    {\ifmmode \m_{\WFour}\else $\mbox{m}_{\WFour}$\fi} 

\def \AbsM       {\mbox{M}} 
\def \AbsMFUV    {\ifmmode \mbox{M}_{\mbox{\scriptsize FUV}}\else $\mbox{M}_{\mbox{\scriptsize FUV}}$\fi} 
\def \AbsMNUV    {\ifmmode \mbox{M}_{\mbox{\scriptsize NUV}}\else $\mbox{M}_{\mbox{\scriptsize NUV}}$\fi} 
\def \AbsMJ      {\ifmmode \mbox{M}_{\mbox{\scriptsize J}}\else $\mbox{M}_{\mbox{\scriptsize J}}$\fi}  
\def \AbsMH      {\ifmmode \mbox{M}_{\mbox{\scriptsize H}}\else $\mbox{M}_{\mbox{\scriptsize H}}$\fi}  
\def \AbsMKs     {\ifmmode \mbox{M}_{\mbox{\scriptsize K_s}}\else $\mbox{M}_{\mbox{\scriptsize K_s}}$\fi} 
\def \AbsMWThree {\ifmmode \mbox{M}_{\WThree}\else $\mbox{M}_{\WThree}$\fi} 
\def \AbsMWFour  {\ifmmode \mbox{M}_{\WFour}\else $\mbox{M}_{\WFour}$\fi} 
\def \AbsMg       {\ifmmode \mbox{M}_{g}\else $\mbox{M}_{g}$\fi} 

\def \AbsMgMax    {\ifmmode \mbox{M}_{g,max}\else $\mbox{M}_{g,max}$\fi} 
\def \AbsMgSun    {\ifmmode \mbox{M}_{g,\sun}\else $\mbox{M}_{g,\sun}$\fi} 

\def \magnitude {\mbox{mag}} 
\def \magAsecSq {\ifmmode \mbox{mag\,arcsec}^{-2}\else  $\mbox{mag\,arcsec}^{-2}$\fi} 

\def \BMinR       {\ifmmode \left({\B}-{\R}\right)\else $\left({\B}-{\R}\right)$\fi} 
\def \BMinV       {\ifmmode \left({\B}-{\V}\right)\else $\left({\B}-{\V}\right)$\fi} 
\def \gMinr       {\ifmmode \left({\SDSSg}-{\SDSSr}\right)\else $\left({\SDSSg}-{\SDSSr}\right)$\fi} 
\def \gMini       {\ifmmode \left({\SDSSg}-{\SDSSi}\right)\else $\left({\SDSSg}-{\SDSSi}\right)$\fi} 
\def \uMinr       {\ifmmode \left({\SDSSu}-{\SDSSr}\right)\else $\left({\SDSSu}-{\SDSSr}\right)$\fi} 
\def \NUVMinu     {\ifmmode \left({\NUV}-{\SDSSu}\right)\else $\left({\NUV}-{\SDSSu}\right)$\fi} 
\def \NUVMinr     {\ifmmode \left({\NUV}-{\SDSSr}\right)\else $\left({\NUV}-{\SDSSr}\right)$\fi} 

\def \EBV         {\ifmmode \mbox{E}\BMinV\else $\mbox{E}\BMinV$\fi} 

\def \F          {\mbox{F}} 
\def \L          {\mbox{L}} 

\def \Whalf      {\ifmmode W_{50}\else $W_{50}$\fi}          
\def \FHI        {\ifmmode \mbox{F}_{HI} \else $\F_{HI}$\fi} 
\def \SNR        {\mbox{SNR}}                                
\def \logMHI     {\ifmmode \log\left(\MHI/\Msun\right)\else $\log\left(\MHI/\Msun\right)$\fi } 

\def  \ergPerS     {\ifmmode {\rm ergs\,s}^{-1} \else ergs s$^{-1}$ \fi}
\def  \ergcms      {\ifmmode {\rm ergs\,cm}^{-2}\,{\rm s}^{-1} 
                       \else ergs\,cm$^{-2}$\,s$^{-1}$ \fi}
\def  \ergcmsA     {\ifmmode{\rm ergs\,cm}^{-2}\,{\rm s}^{-1}\,{\rm\AA}^{-1}
                       \else ergs\,cm$^{-2}$\,s$^{-1}$\,\AA$^{-1}$ \fi}
\def  \ergcmsHz    {\ifmmode{\rm ergs\,cm}^{-2}\,{\rm s}^{-1}\,{\rm Hz}^{-1}
                       \else ergs\,cm$^{-2}$\,s$^{-1}$\,Hz$^{-1}$ \fi}
\def  \Jy          {\mbox{Jy}} 

\def  \Metallicity {\ifmmode Z\else $Z$\fi}
\def  \Zsun        {\ifmmode \Metallicity_\sun\else $\Metallicity_\sun$\fi}

\def \SFR       {\ifmmode \mbox{SFR}\else $\mbox{SFR}$\fi} 
\def \SSFR      {\ifmmode \mbox{SSFR}\else $\mbox{SSFR}$\fi} 
\def \SFE       {\ifmmode \mbox{SFE}\else $\mbox{SFE}$\fi} 
\def \MsunPerYr {\ifmmode \Msun\,\yr^{-1}\else $\Msun\,yr^{-1}$\fi} 

\def  \Halpha   {\ifmmode {{\rm H}_\alpha}\else {H$_\alpha$}\fi}  
\def  \nHa      {\ifmmode {{\rm nH}_\alpha}\else {nH$_\alpha$}\fi}  
\def  \Hbeta    {\ifmmode {\rm H}_\beta \else H$_\beta$ \fi}
\def  \Hgamma   {\ifmmode {\rm H}_\gamma \else H$_\gamma$ \fi}
\def  \Hdelta   {\ifmmode {\rm H}_\delta \else H$_\delta$ \fi}
\def  \Lya      {\ifmmode {\rm Ly}_\alpha \else Ly$_\alpha$ \fi}
\def  \Lyb      {\ifmmode {\rm Ly}_\beta \else Ly$_\beta$ \fi}
\def  \Pa       {\ifmmode {\rm P}_\alpha \else P$_\alpha$ \fi}
\def  \Pb       {\ifmmode {\rm P}_\beta \else P$_\beta$ \fi}
\def  \Bra      {\ifmmode {\rm Br}_\alpha \else Br$_\alpha$ \fi}
\def  \Brg      {\ifmmode {\rm Br}_\gamma \else Br$_\gamma$ \fi}

\title[EIG - II.]
      {EIG - II. Intriguing characteristics of the most extremely isolated galaxies}
\author[O. Spector and N. Brosch]
{O. Spector$^{}$\thanks{E-mail: odedspec@wise.tau.ac.il} and N. Brosch \\
$^{}$Wise Observatory and the Beverly and Raymond Sackler School of Physics and Astronomy, \\
       Tel Aviv University, Tel Aviv 69978, Israel\\
}

\begin{document}

\date{Accepted 2017 March 17. Received 2017 March 17; in original form 2016 March 10}

\pagerange{\pageref{firstpage}--\pageref{lastpage}} \pubyear{2017}

\maketitle

\label{firstpage}

\begin{abstract}

We have selected a sample of 41 extremely isolated galaxies (EIGs) from the local universe using both optical and HI ALFALFA redshifts \citep{2016MNRAS.456..885S}.
Narrow band {\Halpha} and wide band imaging along with public data were used to derive star formation rates (SFRs), star formation histories (SFHs), and morphological classifications for the EIGs.
We have found that the extreme isolation of the EIGs does not affect considerably their star-formation compared to field galaxies. EIGs are typically `blue cloud' galaxies that fit the `main sequence of star forming galaxies' and may show asymmetric star formation and strong compact star-forming regions.
We discovered surprising environmental dependencies of the HI content, \MHI, and of the morphological type of EIGs; The most isolated galaxies (of subsample EIG-1) have lower {\MHI} on average (with $2.5\,\sigma$ confidence) and a higher tendency to be early-types (with 0.94 confidence) compared to the less isolated galaxies of subsample EIG-2.
To the best of our knowledge this is the first study that finds an effect in which an isolated sample shows a higher fraction of early-types compared to a less isolated sample.
Both early-type and late-type EIGs follow the same colour-to-{\Mstar}, {\SFR}-to-{\Mstar} (`main sequence') and {\MHI}-to-{\Mstar} relations.
This indicates that the mechanisms and factors governing star formation, colour and the {\MHI}-to-{\Mstar} relation are similar in early-type and late-type EIGs, and that the morphological type of EIGs is not governed by their {\MHI} content, colour or {\SFR}.

\end{abstract}

\begin{keywords}
galaxies: star formation -- galaxies: evolution -- galaxies: structure
\end{keywords}

\section{Introduction}
\label{sec:Introduction}

The research described here is part of an extensive study of star formation properties and evolution of galaxies in different environments and of various morphological types, conducted in the past few decades \citep[e.g.,][]{1983PhDT.........1B, 1995PhDT........86A, 1998MNRAS.298..920A, 1998ApJ...504..720B, 2001PhDT..Ana_Heller, 2006MNRAS.368..864B, 2008arXiv0806.2722B, 2008MNRAS.390..408Z}. Specifically, we studied galaxies in the most extremely underdense regions of the local Universe. These galaxies are particularly interesting since they evolved with little or no environmental interference in the last few {\Gyr}, and are therefore useful for validating and calibrating galaxy evolution models. Furthermore, when compared to galaxies in denser regions, they illuminate the overall effects of the environment on the evolution of galaxies.

It is well-known that extremely dense environments can greatly influence the star formation (SF) in galaxies. Tidal interactions and mergers of galaxies can trigger extreme starbursts with SFR up to $10^{3}\,\MsunPerYr$, while isolated galaxies hardly ever exhibit {SFR $>$20\,\MsunPerYr} \citep{1998ARA&A..36..189K}.
Although the effect on SFR may be extreme during mergers in clusters as well as in pairs and loose groups, the effect on SFR averaged over the whole history of a galaxy may be small \citep{2003A&A...405...31B, 2009ApJ...692..556R}. 
In cluster environments, apart from the higher rate of interactions, ram pressure by the intracluster medium strips the galaxies of their gas and, therefore, reduces SF. It has also been suggested that in some cases the ram pressure might increase SF \citep{1985ApJ...294L..89G}.

Galaxies in isolated environments are generally considered to be gas-rich, fainter, bluer, of later type, and exhibit higher specific star formation rates (SSFRs; SFRs per unit stellar mass) than galaxies in average density environments \citep{1980ApJ...236..351D, 1999AJ....118.2561G, 2002A&A...389..405P, 2004ApJ...617...50R, 2005ApJ...624..571R, 2012ApJ...753..166D, 2012AJ....144...16K, 2014MNRAS.438..548M, 2016arXiv160104092M}. Some claim that this is not just an effect of the higher abundance of late-type galaxies, and that the late-type galaxies themselves are fainter in under-dense regions than in average density regions \citep{2004A&A...420..873V, 2005MNRAS.356.1155C, IAU:949432}.

\markChange{Numerous other studies also indicate that the properties of galaxies are influenced by their neighbourhood.}
\cite{1982ApJ...253..526B} found that the inner regions of isolated galaxies are bluer, compared to `field' galaxies. This was later suggested to be a consequence of intensive formation of massive stars in the nuclei \citep{1982A&A...113..231B}. 
\cite{2004A&A...420..873V} found that bars are less frequent in isolated galaxies than in perturbed galaxies.
\cite{2014ApJ...788L..39F} found that bluer pseudo-bulges tend to reside in neighbourhoods with a higher probability of tidal perturbation. They suggest that the environment could be playing a role in rejuvenating pseudo-bulges.
\cite{2012MNRAS.424.2574W} found that satellite galaxies around isolated bright primary galaxies are systematically redder than field galaxies of the same stellar mass, except around primaries with $log\left(\Mstar/\Msun \right) < 10.8$, where the satellites' colours were similar or even bluer.

\vspace{12pt}

This work attempts, among other things, to help resolve the question of `Nature vs. Nurture'; does the evolution of galaxies depend only on their content or do their large-scale environments have a significant evolutionary influence.
Some argue that galaxy formation is driven predominantly by the mass of the host DM halo, and is nearly independent of the larger-scale halo environment (e.g., \citealt{2008MNRAS.386.2285C, 2009ApJ...691..633T}). This is supported by their simulation models that produce void galaxies conforming to some observed statistical properties (e.g., colour distribution, luminosity function and nearest neighbour statistics). However, since there are many galaxy properties that most simulations cannot predict (e.g., HI content), and since the halo mass of galaxies cannot be directly measured, this hypothesis is hard to prove or disprove.

\vspace{12pt}

We have chosen a sample of extremely isolated galaxies (EIGs) from the local universe based on a simple isolation criterion. The neighbourhood properties of this sample were analysed using both observational data and cosmological simulations. The cosmological simulations were further used to estimate the properties and histories of the dark matter (DM) haloes in which the sample EIGs reside.
The sample and its analysis are described in detail in the first paper of this series, \cite{2016MNRAS.456..885S} (SB16), and are summarized here in Section \ref{sec:Sample}.

Extensive optical observations of the sample EIGs in broad-band and rest-frame {\Halpha} were performed using the one meter telescope of the Florence and George Wise Observatory\footnote{IAU code 097 - http://wise-obs.tau.ac.il/} (WO). 
Section \ref{ch:ObsNProcess} describes these observations and their processing.
The results of these observations, along with public observational data, were used to measure the current SFRs and to estimate SFHs. These observational results are described in section \ref{ch:Results}. Analysis of these results is presented in section \ref{ch:Analysis}, and the findings are discussed in section \ref{s:DisConc}.

\vspace{12pt}

Throughout this work, unless indicated otherwise, $\Lambda$ cold dark matter ($\Lambda$CDM) cosmology with the seven-year Wilkinson Microwave Anisotropy Probe data (WMAP7, \citealt{2011ApJS..192...17B}) parameters are used, including the dimensionless Hubble parameter $\h = 0.704$.
We adopt here the solar {\SDSSg}-band absolute magnitude of $\AbsMgSun = +5.12$ (according to the Sloan Digital Sky Survey, SDSS, DR7 web site\footnote{www.sdss.org/dr7/algorithms/sdssUBVRITransform.html\\\#vega\_sun\_colors}).

\section{The Sample}
\label{sec:Sample}

We have chosen the sample of EIGs using a simple isolation criterion: a galaxy is considered an EIG and is included in the sample if it has no known neighbours closer than \markChange{a certain neighbour distance limit in 3D redshift space ({200\,\kms} or {300\,\kms} as explained below)} and if its redshift is in the range $2000<\cz<7000\,\kms$.
No magnitude, HI mass or size limit was used in the selection of candidate neighbours. The use of such limits would have somewhat reduced the level of isolation of the sample (especially for the closer EIGs) and therefore was not preferred. Not using such limits, however, complicates somewhat the analysis of the sample's isolation level (described in section 3 of SB16 \markChange{and summarized below}). \markChange{It also causes the sensitivity limits (listed below) and the isolation level to depend on redshift. Higher redshift EIGs are less isolated on average than lower redshift EIGs. For this reason, the redshift of EIGs was limited to {7000\,\kms}.}

One of the unique advantages of the EIG sample we study here is that, apart from the optical redshift data commonly used to estimate environment density, it also utilized HI redshifts from the Arecibo Legacy Fast ALFA survey (ALFALFA) survey.
The ALFALFA survey is a second-generation untargeted extragalactic HI survey initiated in 2005 \citep{2005AJ....130.2598G, 2007AJ....133.2569G, 2007AJ....133.2087S}. This survey utilizes the superior sensitivity and angular resolution of the Arecibo 305\,m radio telescope to conduct the deepest ever census of the local HI Universe.
ALFALFA was particularly useful in verifying the isolation of the target galaxies, since by being an HI survey it easily measures redshifts of low surface brightness galaxies (LSBs) and other low-luminosity late-type neighbours that are often difficult to detect optically but abound with HI\rem{\citep{2013JApA...34...19D}}.
\markChange{
The ALFALFA dataset we used was the ``$\alpha$.40 HI source catalogue'' \citep[$\alpha$.40;][]{2011AJ....142..170H}. This catalogue covers 40 percent of the final ALFALFA survey area ($\sim$2800\,\sqDeg) and contains 15855 sources.
The sensitivity limit of the ALFALFA dataset is given by eq. (6) and (7) of \cite{2011AJ....142..170H} as function of the velocity width of the HI line profile, {\Whalf}. For a typical value $\Whalf = 100\,\kms$ the sensitivity limit of the ALFALFA dataset is {$\sim$0.6\,\Jy\,\kms}. For the redshift range of the EIG sample this translates to HI mass, $\log \left( \MHI / \Msun \right)$, sensitivity limit of {$\sim$8.0} (for $\cz = 2000\,\kms$) to {$\sim$9.1} (for $\cz = 7000\,\kms$).
}

The search criterion was applied to two sky regions, one in the spring sky (Spring) and the other in the autumn sky (Autumn) as described in Table \ref{T:Sample-Regions}. These particular regions were selected since they are covered by the $\alpha$.40 ALFALFA catalogue \citep{2011AJ....142..170H}. Both regions include mainly high Galactic latitudes. The Spring region is almost fully covered by spectroscopic data in SDSS DR10 \citep{2014ApJS..211...17A}.

\begin{ctable} 
[
  caption = Sample search regions,
  doinside = \small,
  label   = T:Sample-Regions
]
{@{}cccc@{}}
{}
{
  \FL
          & {\RA} (J2000)      & {\Dec} (J2000)         & \cz~$\left[ \kms \right]$
  \ML
  Spring  & 7h30m--16h30m   & $4\deg$--$16\deg$   & 2000--7000
  \NN
  Autumn  & 22h00m--03h00m  & $24\deg$--$28\deg$  & 2000--7000
  \LL
}

\end{ctable}

In addition to ALFALFA, the NASA/IPAC Extragalactic Database\footnote{http://ned.ipac.caltech.edu/} (NED) was also used as a source for coordinates and redshifts in and around the search regions. 
\markChange{The NED dataset we used includes data downloaded from NED on November 13, 2012 for object types: galaxies, galaxy clusters, galaxy pairs, galaxy triples, galaxy groups, and QSO.
The completeness functions derived in section 3.2 of SB16 indicate that the sensitivity limit of the NED dataset in terms of $\SDSSg$ magnitude is $\sim$18.5 for the Spring sky region and $\sim$17 for the Autumn sky region. For the redshift range of the EIG sample this translates to absolute $\SDSSg$ magnitude sensitivity limit of {$\sim$-13.8} (for $\cz = 2000\,\kms$) to {$\sim$-16.5} (for $\cz = 7000\,\kms$) for the Spring sky region, and {$\sim$-15.3} to {$\sim$-18.0} for the Autumn sky region. A rough conversion to stellar mass, {\Mstar}, assuming $\SDSSg$ luminosity to mass ratio as that of the sun, gives a $\log \left( \Mstar / \Msun \right)$ sensitivity limit of {$\sim$7.6} (for $\cz = 2000\,\kms$) to {$\sim$8.6} (for $\cz = 7000\,\kms$) for the Spring sky region, and {$\sim$8.2} to {$\sim$9.2} for the Autumn sky region.
}

The EIGs, studied here, were divided to three subsamples:
\begin{enumerate}
  \item[1.] Galaxies that passed the criterion using both NED and ALFALFA data \markChange{with a neighbour distance limit of {300\,\kms}. This translates to not having any known neighbour within a distance of $3\,\Mpch \cong 4.26\,\Mpc$.}
  \item[2.] Galaxies that passed the criterion using NED data \markChange{with a neighbour distance limit of {3\,\Mpch}}, but did not pass using ALFALFA data (had neighbours closer than {3\,\Mpch} in the ALFALFA database).
  \item[3.] Galaxies for which the distance to the closest neighbour in NED's data is {2 -- 3\,\Mpch} (regardless of the distance to the closest neighbour in ALFALFA's data).
  \item[]
\end{enumerate}

\markChange{Subsamples 1 and 2 contain all catalogued galaxies that passed their criteria in the studied sky regions. Subsample 3 contains only those galaxies that seemed to be isolated in the various searches performed over the years, but were later found to have neighbours in the range {2 -- 3\,\Mpch} (with the 2012 NED dataset described above)}.
It also contains a galaxy, EIG 3s-06, which was found by searching the ALFALFA data alone, but had neighbours in the range {2 -- 3\,\Mpch} in the NED dataset.

The galaxies were named according to their subsample and sky region, using the following format:
\begin{equation*}
\mbox{EIG BR-XX}
\end{equation*}
where:
\begin{description}
  \item[B] is the galaxy's subsample (1, 2 or 3, as described above);
  \item[R] is the sky region: `s' - Spring, `a' - Autumn;
  \item[XX] is the serial number of the galaxy in the subsample.
  \item[]
\end{description} 

So, for example, object EIG 3s-06 is the sixth galaxy in subsample 3 of the spring sky region.

The galaxies of the different subsamples are listed in Tables 2 through 7 of SB16. Subsample EIG-1 contains 21 galaxies (14 Spring and 7 Autumn galaxies). Subsample EIG-2 contains 11 galaxies (7 Spring and 4 Autumn galaxies). Subsample EIG-3 contains 9 galaxies (7 Spring and 2 Autumn galaxies). \markChange{In total, the sample contains 41 EIGs.}
Notes regarding specific EIGs are listed in Appendix \ref{App:EIGdata}.

\vspace{12pt}

The use of the ALFALFA unbiased HI data significantly improved the quality of the sample. Out of 32 galaxies that passed the \markChange{3\,\Mpch} criterion using NED data alone, 11 galaxies did not pass the criterion when tested with ALFALFA data.

Neighbourhood properties of the sample EIGs were analysed using both observational data and cosmological simulations. The analysis based on observational data is described in detail in section 2.4 of SB16. Tables 8 and 9 of SB16 list properties such as the distance to the closest neighbour and neighbour counts for each sample EIG. A comparison to random galaxies show that on average the neighbourhood density of EIGs is about one order of magnitude lower than that of field galaxies.
Observational neighbourhood data further indicates that EIGs tend to reside close to walls and filaments rather than in centres of voids.

Using cosmological simulations, we confirmed that the EIG-1 and EIG-2 subsamples are a subset of galaxies significantly more isolated than the general galaxy population. Apart from the low density regions in which they reside, EIGs  are characterized by normal mass haloes, which have evolved gradually with little or no major mergers or major mass-loss events. As a result of their low-density environments, the tidal acceleration exerted on EIGs is typically about one order of magnitude lower than the average tidal acceleration exerted on the general population of galaxies.
The level of contamination in the sample, i.e. the fraction of EIGs which are not in extremely isolated environments or which experienced strong interactions in the last {3\,\Gyr}, was found to be {5\%--10\%}. The Spring EIGs seem to be more isolated than the Autumn EIGs.
For further details about the analysis using cosmological simulations and its results see section 3 of SB16.

\vspace{12pt}

For similar purposes, other samples of isolated galaxies were defined and studied 
in `the Analysis of the interstellar Medium of Isolated GAlaxies' (AMIGA) international project \citep{2007A&A...472..121V, 2013MNRAS.434..325F}\rem{http://amiga.iaa.es}, 
in the `Two Micron Isolated Galaxy' catalogue (2MIG; \citealt{2010AstBu..65....1K}), 
in the `Local Orphan Galaxies' catalogue (LOG; \citealt{2011AstBu..66....1K, 2013AstBu..68..243K}), 
and in the 'Void Galaxy Survey' (VGS;  \citealt{2012AJ....144...16K}).
In section 2.5 of SB16 these were discussed and compared to the EIG sample studied here. The comparison showed that the EIG sample galaxies are significantly more isolated than the AMIGA, 2MIG and LOG galaxies (in terms of the distance to the closest neighbour) and that the \mbox{EIG-1} galaxies are more isolated than the VGS galaxies.
\markChange{Other notable isolated galaxy samples, not analysed in SB16, are the UNAM-KIAS catalogue of isolated galaxies \citep{2010AJ....139.2525H} and the catalogues of isolated galaxies, isolated pairs, and isolated triplets in the local Universe of \cite{2015A&A...578A.110A}.} 

\section{Observations and Data Processing}
\label{ch:ObsNProcess}

\subsection{Instrumentation}

Optical observations were performed using the 1\,meter (40\,inch) telescope of the WO. The telescope was equipped with a $1300\times1340$ back-illuminated Princeton Instruments CCD with pixel size of $0.57 \pm 0.01$ \,arcsec\,pixel$^{-1}$ and an overall field of view of $\sim12.5\,\arcmin$.
EIGs were imaged using wide band Bessell U, B, V, R and I filters and a set of narrow-band rest-frame {\Halpha} filters for various redshifts. A thorough description of the {\Halpha} filter set is provided in appendix A of \cite{2015PhDT}.

\subsection{Observations}

\markChange{We observed 34 of the EIGs} in the {\R} and in one or two appropriate {\Halpha} narrow bands. \markChange{Of these EIGs}, those not covered by SDSS (and a few that are) were imaged also in the U, B, V and I bands. At least six dithered exposures were obtained in each filter. Exposures were 20 minutes long for the {\Halpha} and U bands, and 10 minutes long for the B, V, and I bands. Exposures in the R band were 5 minutes long for EIGs that were observed only in R and {\Halpha}, and 10 minutes long for EIGs observed in all bands.
Whenever possible, the exposures of the R and {\Halpha} bands were taken in time proximity so that their atmospheric conditions and air-masses would remain similar. This is important for the accurate measurement of the {\Halpha} equivalent width (EW), as described in \cite{2012MNRAS.419.2156S} (S12).

Photometric calibrations of the wide bands were performed for EIGs not covered by SDSS, and for a few that are, using \cite{1992AJ....104..340L}\rem{Landolt} standards.
Spectrophotometric calibrations of the {\Halpha} band were performed using \cite{1990AJ.....99.1621O} standard stars that have well known spectra, are stable, and have as few features around {\Halpha} as possible.

Images were processed using the Image Reduction and Analysis Facility ({\small IRAF}) software\footnote{http://iraf.noao.edu/}.
The reduction pipeline included standard bias subtraction, flat-fielding and image alignment. For images taken in the I band, a fringe removal step was added.

\subsection{{Net-\Halpha} images}
\label{s:ObsNPrc_NetHa}

{Net-\Halpha} ({\nHa}) data were derived from the measurements using the recipes described in S12.
EW values were derived using eq.~12 and 16 of S12. The {\nHa} fluxes were derived using eq.~7 and 12 of S12 after applying the photometric calibrations described in section 3 of S12.
Eq. 12, 16 and 7 of SB12 are shown here for reference:

\begin{equation}
\begin{IEEEeqnarraybox*}{lCl}
\mbox{cps}_{N,line} 
    & \cong &
        \left(\mbox{cps}_{N} - \frac { \mbox{cps}_{W} }{ \mbox{WNCR} } \right) \\
    &&  \times \left[1 -  \frac{1}{\mbox{WNCR}} \cdot 
                   \frac{ \mbox{T}_{atm,W}(\lambda_{line}) \: \mbox{T}_{W}(\lambda_{line}) } 
                        { \mbox{T}_{atm,N}(\lambda_{line}) \: \mbox{T}_{N}(\lambda_{line}) } 
        \right]^{-1}
\end{IEEEeqnarraybox*}
\label{e:cps_line_N_solved}
\end{equation}

\begin{equation}
\mbox{EW} \cong  \frac { \mbox{cps}_{N,line} } 
                {\mbox{cps}_{N} - \mbox{cps}_{N,line}}
          \cdot
          \frac {\int_0^\infty \mbox{T}_{N}(\lambda) \: d\lambda}
                { \mbox{T}_{N}(\lambda_{line}) } 
\label{e:EW_solved}
\end{equation}

\begin{equation}
\mbox{F}_{line}  \cong  \frac { \mbox{cps}_{N,line} } 
                              { \mbox{T}_{atm,N}(\lambda_{line}) \: \mbox{T}_{N}(\lambda_{line}) \:
                                \mbox{R}_{\lambda}(\lambda_{line}) }
\label{e:F_line}
\end{equation}
where:
\begin{description}
  \item[$\mbox{cps}_{N}$, $\mbox{cps}_{W}$] are the measured count rates of the narrow-band (N) and wide-band (W) filters (respectively) in instrumental units (typically analogue to digital units per second, ADU s$^{-1}$);
  \item[$\mbox{cps}_{N,line}$] is the line contribution to $\mbox{cps}_{N}$ (see also eq. 3 of SB12); 
  \item[WNCR] (wide to narrow continuum ratio) is the ratio between the count rate contributed by the continuum in the W band and the count rate contributed by the continuum in the N band (see also eq. 10 of SB12);
  \item[$\mbox{T}_{N}(\lambda)$, $\mbox{T}_{W}(\lambda)$] are the transmittance functions of the N and W bands, respectively;
  \item[$\mbox{T}_{atm,N}(\lambda)$, $\mbox{T}_{atm,W}(\lambda)$] are the atmospheric transmittance as function of wavelength, including effects of weather, elevation and airmass of observation, when the N and W bands (respectively) were imaged;
  \item[$\mbox{R}_{\lambda}$] is the responsivity as function of wavelength of the rest of the electro-optical system (i.e. the telescope and sensors, excluding the transmittance effect of the filters) typically in ADU~erg$^{-1}$~cm$^{2}$;
  \item[$\lambda_{line}$] is the central wavelength of the emission line;
  \item[$\mbox{F}_{line}$] is the emission-line's flux.
  \item[]
\end{description}

The WNCR required for \eqref{e:cps_line_N_solved} was estimated using the method of WNCR-to-colour linear fit suggested in section 6 of S12 (sixth paragraph).
The process included selecting a reference wavelength band for each EIG, the first band with a good quality image from the following list: {\V}, {\I}, {\B}, SDSS {\SDSSg} and SDSS {\SDSSi}. In the combined images of these {\R} and reference bands foreground stars were identified (using their intensity profiles), and their instrumental colours (reference minus {\R}) were measured along with that of the EIG.

All {\nHa} measurements were performed on the individual {\Halpha} images, each paired with an {\R} image taken at the closest time and \markChange{airmass} available. For each such pair, the foreground stars were measured in both the {\R} and {\Halpha} images, and their WNCR values were calculated (the {\R} to {\Halpha} cps ratio). A linear relation between WNCR and the uncalibrated colour was fitted to the results of these foreground stars. The WNCR of the pair of {\Halpha} and {\R} images was then calculated using the fit and the EIG's measured uncalibrated colour.

Next, an {\nHa} image was created for each {\Halpha} and {\R} image pair. The sky level, measured around the EIG, was first subtracted from the {\Halpha} and {\R} images. Then, the images were scaled by their exposure time. Finally, the pixel values of the {\nHa} images were calculated using \eqref{e:cps_line_N_solved}.

\subsection{Photometry}
\label{sec:Wise_Photometry}

Apertures for the photometric measurements of the EIGs were defined as polygons or as elliptical isophotes fitted to the combined R-band images.
The polygonal apertures approximately trace the $\R = 26\,\magAsecSq$ isophote of the EIGs, but exclude foreground Galactic stars and galaxies, projected close to the EIG. This resulted in some reduction in the measured flux from the EIGs, which was significant only for EIG 2s-06 that has a foreground star of magnitude $\SDSSr = 15.6$ projected close to its centre.
Polygonal apertures were also defined for some resolved HII regions and other regions of interest within the EIGs (see figures \ref{f:RHacolorImgEIG-1}, \ref{f:RHacolorImgEIG-2} and \ref{f:RHacolorImgEIG-3} below).

Wherever possible, photometric measurements were made using SDSS calibrated images using the same apertures defined for the WO images.
The SDSS calibrations were tested by comparing results of seven EIGs that had photometric calibrations performed at the WO as well as SDSS data.
The {\magu \magb \magv \magr \magi} magnitudes were converted to {\SDSSu \SDSSg \SDSSr \SDSSi \SDSSz} SDSS magnitudes using the transformation recommended in Table 1 of \cite{2005AJ....130..873J} for all stars with $\R-\I < 1.15$. This is the transformation recommended by SDSS for galaxies.\footnote{http://www.sdss3.org/dr9/algorithms/sdssUBVRITransform.php}
On average, the results were similar to the SDSS calibrated magnitudes. The standard deviation of the difference between the WO calibration (converted to {\SDSSu \SDSSg \SDSSr \SDSSi \SDSSz}) and the SDSS calibration was {0.05\,\magnitude} for {\SDSSr} and {\SDSSi}, {0.07\,\magnitude} for {\SDSSg}, and {0.11\,\magnitude} for {\SDSSu} and {\SDSSz}.

Where available, SDSS measurements were used to calibrate the {\nHa} flux using the method described in section 3 of S12, in which {\SDSSg}, {\SDSSr} and {\SDSSi} are used to estimate the continuum flux at the rest-frame {\Halpha} wavelength. This continuum flux estimate is then multiplied by the equivalent width to obtain the {\nHa} flux.

Thirteen EIGs had both spectrophotometric and SDSS {\nHa} calibrations. We found random deviations between the results of the two calibrations, which the original uncertainty propagation estimates did not predict. These may be attributed to the inaccuracy introduced by estimating the continuum at {\Halpha} using an interpolation of two or three SDSS magnitude measurements (see section 3 of S12).
This was compensated for by adding 0.2 relative uncertainty to the SDSS calibrations.

\subsection{Absolute magnitudes and luminosities}
\label{s:ObsNPrc_AbsMagLum}

To calculate the absolute magnitudes and luminosities, the calibrated apparent magnitudes and fluxes were first corrected for foreground Galactic extinction. The Wise Observatory (\U\V\B\R\I) and SDSS magnitudes were corrected using Galactic extinction NED data based on \cite{2011ApJ...737..103S}. 

The Galactic extinctions of the {\Halpha} fluxes, $A_{\lambda_\Halpha}$, were estimated using an interpolation between the {\SDSSr} and {\SDSSi} extinctions. The interpolation was linear in $ln \left( A_{\lambda} \right)$ vs.~{$\lambda$}, since this fits well $A_{\lambda}$ of {\SDSSu\SDSSg\SDSSr\SDSSi\SDSSz} and {\U\B\V\R\I}.

This work utilizes data from the Galaxy Evolution Explorer mission (GALEX; \citealt{2005ApJ...619L...1M}), the Two Micron All Sky Survey (2MASS; \citealt{2006AJ....131.1163S}) and the Wide-field Infrared Survey Explorer (WISE; \citealt{2010AJ....140.1868W}).
The GALEX ({\NUV} and {\FUV}) and 2MASS ({\J}, {\H} and {\Ks}) Galactic extinctions were calculated using the $A_{\mbox{\scriptsize \B}}$ and $A_{\mbox{\scriptsize \V}}$ of \cite{2011ApJ...737..103S} and the \second column of Table 2 of \cite{2013MNRAS.430.2188Y}, which gives $R_{band} = A_{band} / \left( A_{\mbox{\scriptsize \B}}-A_{\mbox{\scriptsize \V}} \right)$ for each band.
The Galactic extinction of the WISE {\WThree} and {\WFour} bands were estimated using the calculated $A \left( \Ks \right)$ and the values for $A_{\mbox{\scriptsize 12\,\um}} / A_{\mbox{\scriptsize K}}$ and $A_{\mbox{\scriptsize 22\,\um}} / A_{\mbox{\scriptsize K}}$ quoted in column 2 of Table 2 of \cite{2009ApJ...693L..81M}.

\vspace{12pt}

Distance estimates, required for calculating absolute magnitudes and fluxes, were based on the local velocity field model of \cite{2000ApJ...529..786M} that includes terms for the influence of the Virgo Cluster, the Great Attractor, and the Shapley Supercluster.
As customary in this field \citep[e.g., ][]{2012A&A...540A..47F, 2011AJ....142..170H, 2012AJ....144...16K} uncertainties were not estimated for these distances.
At the low redshifts of the EIGs ($\z < 0.024$) K-corrections are not significant compared to the uncertainty that they introduce. Therefore, K-corrections were not applied to the measured magnitudes and fluxes.
The apparent magnitudes, {\m}, and fluxes, {\F} (after correcting for Galactic extinction) were converted to absolute magnitudes, {\AbsM}, and luminosities, {\L}, using: 
$ \AbsM = \m - 5 \cdot \log \frac{\left( 1 + \z \right) D_m}{10\,\pc} $ and 
$ \L = 4 \pi D_m^2  \left( 1 + \z \right)^{2} \cdot \F $, where $D_m$ is the distance estimate (comoving transverse distance).

\vspace{24pt}

\rem{
To conclude, we observed 36 EIGs, produced deep maps of their {\Halpha} content and measured their total {\Halpha} luminosity and equivalent width. These data were not available in the literature or in public databases.
We have also produced {\U\B\V\R\I} images and absolute magnitude measurements of some of these EIGs, deeper than available before.
}

Further details about the observations and data processing can be found in section 5 of \cite{2015PhDT}.

\section{Results}
\label{ch:Results}

\subsection{Apparent magnitudes and fluxes}

Table \ref{T:EIG_ugriz} lists the SDSS apparent magnitudes of the \markChange{39} EIGs measured as described in section \ref{sec:Wise_Photometry}.\footnote{\markChange{Two of the EIGs, 1a-05 and 1a-06, are not in the SDSS footprint and were not imaged in the WO.}}
Photometrically calibrated {\U\B\V\R\I} (Bessell) measurements were made for eight of the EIGs. Their apparent magnitudes are listed in Table \ref{T:EIG_ubvri}.

\begin{ctable} 
[
  caption = {Apparent SDSS magnitudes},
  cap     = {Apparent SDSS magnitudes},
  doinside = \small,
  star,
  label   = {T:EIG_ugriz}
]
{ @{}cr@{$\,\pm\,$}lr@{$\,\pm\:$}lr@{$\,\pm\,$}lr@{$\,\pm\,$}lr@{$\,\pm\:$}l@{}
}
{
  \tnote[a]
  {
    No optical counterpart was found for EIG 1s-05 (an ALFALFA object).
  }
  \tnote[b]
  {
    EIG 1s-14 is projected close to a foreground bright star. Uncertainties were estimated to be 0.1\,$mag$.
  }
  \tnote[c]
  { 
    Converted from Landolt ({\magu}{\magb}{\magv}{\magr}{\magi}) magnitudes.
  }
  \tnote[d]
  {
    EIG 2s-06 has a significant foreground star in front of it. Uncertainties were estimated to be 0.2\,$mag$.
  }
}
{
 \FL
 EIG  & \multicolumn {2}{c}{\SDSSu} & \multicolumn {2}{c}{\SDSSg} & \multicolumn {2}{c}{\SDSSr} 
      & \multicolumn {2}{c}{\SDSSi} & \multicolumn {2}{c}{\SDSSz}
 \ML
 1s-01 & 17.37 & 0.03 & 16.633 & 0.007 & 16.315 & 0.007 & 16.090 & 0.008 & 15.99 & 0.03
 \NN
 1s-02 & 16.766 & 0.008 & 15.911 & 0.003 & 15.552 & 0.003 & 15.378 & 0.003 & 15.243 & 0.006
 \NN
 1s-03 & 17.29 & 0.03 & 16.200 & 0.007 & 15.61 & 0.01 & 15.231 & 0.008 & 14.99 & 0.02
 \NN
 1s-04 & 18.43 & 0.05 & 17.47 & 0.01 & 17.09 & 0.01 & 16.87 & 0.01 & 16.90 & 0.05
 \NN
 \;\;1s-05 \tmark[a] &  \multicolumn {2}{c}{---}  &  \multicolumn {2}{c}{---}  &  \multicolumn {2}{c}{---}  &  \multicolumn {2}{c}{---}  &  \multicolumn {2}{c}{---} 
 \NN
 1s-06 &  \multicolumn {2}{c}{---}  & 16.719 & 0.006 & 16.360 & 0.007 & 16.14 & 0.01 & 15.87 & 0.03
 \NN
 1s-07 &  \multicolumn {2}{c}{---}  & 17.482 & 0.006 & 17.005 & 0.005 & 16.698 & 0.006 & 16.50 & 0.02
 \NN
 1s-08 & 18.86 & 0.03 & 17.839 & 0.007 & 17.416 & 0.007 & 17.247 & 0.008 & 16.81 & 0.02
 \NN
 1s-09 &  \multicolumn {2}{c}{---}  & 16.900 & 0.005 & 16.679 & 0.006 & 16.518 & 0.007 & 16.50 & 0.02
 \NN
 1s-10 & 18.20 & 0.02 & 17.480 & 0.005 & 17.273 & 0.007 & 17.17 & 0.01 & 17.06 & 0.02
 \NN
 1s-11 &  \multicolumn {2}{c}{---}  & 16.37 & 0.09 & 15.81 & 0.01 & 15.6 & 0.1 & 15.35 & 0.08
 \NN
 1s-12 & 18.77 & 0.02 & 17.904 & 0.006 & 17.689 & 0.007 & 17.51 & 0.01 & 17.53 & 0.03
 \NN
 1s-13 & 18.99 & 0.09 & 17.92 & 0.02 & 17.69 & 0.02 & 17.59 & 0.03 & 17.74 & 0.07
 \NN
 \;\;1s-14 \tmark[b] & 17.0 & 0.1 & 15.7 & 0.1 & 15.1 & 0.1 & 14.8 & 0.1 & 14.6 & 0.1
 \ML
 1a-01 & 16.79 & 0.02 & 15.686 & 0.004 & 15.194 & 0.004 & 14.931 & 0.004 & 14.72 & 0.01
 \NN
 1a-02 & 18.05 & 0.03 & 16.988 & 0.006 & 16.464 & 0.005 & 16.212 & 0.006 & 16.08 & 0.02
 \NN
 1a-03 &  \multicolumn {2}{c}{---}  & 17.50 & 0.02 & 17.0 & 0.2 & 16.91 & 0.06 & 16.7 & 0.1
 \NN
 \;\;1a-04 \tmark[c] & 15.40 & 0.06 & 13.49 & 0.02 & 12.62 & 0.03 & 12.13 & 0.03 & 11.72 & 0.03
 \NN
 1a-07 & 17.56 & 0.02 & 16.649 & 0.004 & 16.327 & 0.005 & 16.138 & 0.005 & 15.99 & 0.02
 \ML
 2s-01 & 18.45 & 0.04 & 17.64 & 0.01 & 17.35 & 0.01 & 17.13 & 0.02 & 16.96 & 0.05
 \NN
 2s-02 & 18.23 & 0.05 & 17.084 & 0.008 & 16.591 & 0.009 & 16.34 & 0.01 & 16.34 & 0.03
 \NN
 2s-04 &  \multicolumn {2}{c}{---}  & 18.20 & 0.02 & 17.86 & 0.02 & 17.52 & 0.02 & 17.5 & 0.1
 \NN
 2s-05 & 16.53 & 0.01 & 15.562 & 0.003 & 15.118 & 0.003 & 14.908 & 0.003 & 14.738 & 0.008
 \NN
 \;\;2s-06 \tmark[d] & 17.5 & 0.2 & 16.5 & 0.2 & 16.1 & 0.2 & 15.8 & 0.2 & 15.7 & 0.2
 \NN
 2s-07 & 18.13 & 0.02 & 16.926 & 0.005 & 16.476 & 0.004 & 16.268 & 0.005 & 16.10 & 0.01
 \NN
 2s-08 & 18.04 & 0.02 & 17.450 & 0.005 & 17.435 & 0.006 & 17.606 & 0.008 & 17.48 & 0.02
 \ML
 2a-01 & 17.05 & 0.03 & 15.642 & 0.004 & 14.794 & 0.003 & 14.319 & 0.004 & 13.94 & 0.01
 \NN
 2a-02 & 17.90 & 0.08 & 16.73 & 0.01 & 16.09 & 0.01 & 15.74 & 0.01 & 15.46 & 0.03
 \NN
 2a-03 & 16.99 & 0.02 & 15.906 & 0.003 & 15.441 & 0.003 & 15.179 & 0.004 & 15.00 & 0.01
 \NN
 2a-04 & 19.3 & 0.2 & 17.85 & 0.03 & 17.43 & 0.02 & 17.20 & 0.04 & 16.94 & 0.09
 \ML
 3s-01 & 18.47 & 0.06 & 17.59 & 0.02 & 17.15 & 0.02 & 16.89 & 0.02 & 16.82 & 0.08
 \NN
 3s-02 &  \multicolumn {2}{c}{---}  & 17.134 & 0.008 & 16.84 & 0.01 & 16.67 & 0.01 & 16.55 & 0.04
 \NN
 3s-03 & 16.97 & 0.01 & 16.136 & 0.004 & 15.832 & 0.004 & 15.713 & 0.005 & 15.58 & 0.01
 \NN
 3s-04 &  \multicolumn {2}{c}{---}  & 19.03 & 0.02 & 18.77 & 0.03 & 18.82 & 0.04 & 18.9 & 0.1
 \NN
 3s-05 &  \multicolumn {2}{c}{---}  & 16.305 & 0.006 & 15.805 & 0.006 & 15.573 & 0.007 & 15.37 & 0.02
 \NN
 3s-06 &  \multicolumn {2}{c}{---}  & 17.422 & 0.009 & 17.03 & 0.01 & 16.88 & 0.01 & 16.65 & 0.03
 \NN
 3s-07 &  \multicolumn {2}{c}{---}  & 17.318 & 0.007 & 16.891 & 0.007 & 16.618 & 0.009 & 16.51 & 0.03
 \ML
 3a-01 & 16.41 & 0.02 & 15.466 & 0.004 & 15.138 & 0.004 & 14.944 & 0.004 & 14.62 & 0.02
 \NN
 3a-02 & 17.36 & 0.03 & 16.126 & 0.004 & 15.468 & 0.004 & 15.075 & 0.004 & 14.772 & 0.009
 \LL
}
\end{ctable}
 \rem{label = {T:EIG_ugriz}}

\begin{ctable} 
[
  caption = {Apparent UBVRI magnitudes},
  cap     = {Apparent UBVRI magnitudes},
  doinside = \small,
  star,
  label   = {T:EIG_ubvri}
]
{ @{}cr@{$\,\pm\,$}rr@{$\,\pm\:$}rr@{$\,\pm\,$}rr@{$\,\pm\,$}rr@{$\,\pm\:$}r@{}
}
{
}
{
 \FL
 EIG  & \multicolumn {2}{c}{\magu} & \multicolumn {2}{c}{\magb} & \multicolumn {2}{c}{\magv} 
      & \multicolumn {2}{c}{\magr} & \multicolumn {2}{c}{\magi}
 \ML
 1s-10 & 17.48 & 0.03 & 17.91 & 0.01 & 17.448 & 0.007 & 17.155 & 0.006 & 16.81 & 0.01
 \NN
 1s-11 & 16.66 & 0.09 & 16.74 & 0.02 & 16.02 & 0.02 & 15.63 & 0.03 & 15.19 & 0.04
 \NN
 1a-04 & 14.65 & 0.01 & 14.041 & 0.003 & 12.963 & 0.002 & 12.359 & 0.004 & 11.596 & 0.004
 \NN
 2a-01 & 16.43 & 0.02 & 16.11 & 0.01 & 15.106 & 0.003 & 14.471 & 0.005 & 13.704 & 0.004
 \NN
 2a-02 & 17.01 & 0.08 & 17.14 & 0.01 & 16.33 & 0.01 & 15.79 & 0.01 & 15.21 & 0.01
 \NN
 3s-03 & 16.16 & 0.03 & 16.48 & 0.01 & 16.04 & 0.02 & 15.57 & 0.01 & 15.19 & 0.03
 \NN
 3a-01 & 15.56 & 0.03 & 15.796 & 0.007 & 15.264 & 0.006 & 14.91 & 0.01 & 14.516 & 0.006
 \NN
 3a-02 & 16.52 & 0.05 & 16.52 & 0.02 & 15.694 & 0.007 & 15.165 & 0.007 & 14.520 & 0.007
 \LL
}
\end{ctable}
 \rem{label = {T:EIG_ubvri}}


The combined R and {Net-\Halpha} ({\nHa}) images are shown in figures \ref{f:RHacolorImgEIG-1}, \ref{f:RHacolorImgEIG-2} and \ref{f:RHacolorImgEIG-3} (in negative).\footnote{\markChange{The images of EIG 2s-04 are not shown due to a bright foreground star that does not allow to clearly identify it in the image (see details in Appendix \ref{App:EIGdata}).}} Each row of images in the figures relates to a different EIG. The name of the EIG is given on the leftmost image, which shows the combined R image. The second image from the left shows the same R image using a logarithmic scale (log R). The third image from the left shows the combined {\nHa} image. The rightmost image shows the EIG in false colour; R in orange, and {\nHa} in azure (both using a negative linear scale).
The upper bar in the rightmost image shows the physical scale calculated using the distance estimate described in section \ref{s:ObsNPrc_AbsMagLum}. The lower bar in the rightmost image shows the angular size scale.

Regions of interest within some of the EIGs were measured individually. The polygonal apertures used for these measurements are drawn (along with their names) on the rightmost images of figures \ref{f:RHacolorImgEIG-1}, \ref{f:RHacolorImgEIG-2} and \ref{f:RHacolorImgEIG-3}.
Observational results of these regions of interest are described in section \ref{s:rsltsSFR}. Remarks for specific EIGs are listed in Appendix \ref{App:EIGdata}.


  \begin{figure*}
  \begin{centering}
    \includegraphics[width=17.0cm,trim=0mm 0mm 0mm 0, clip]{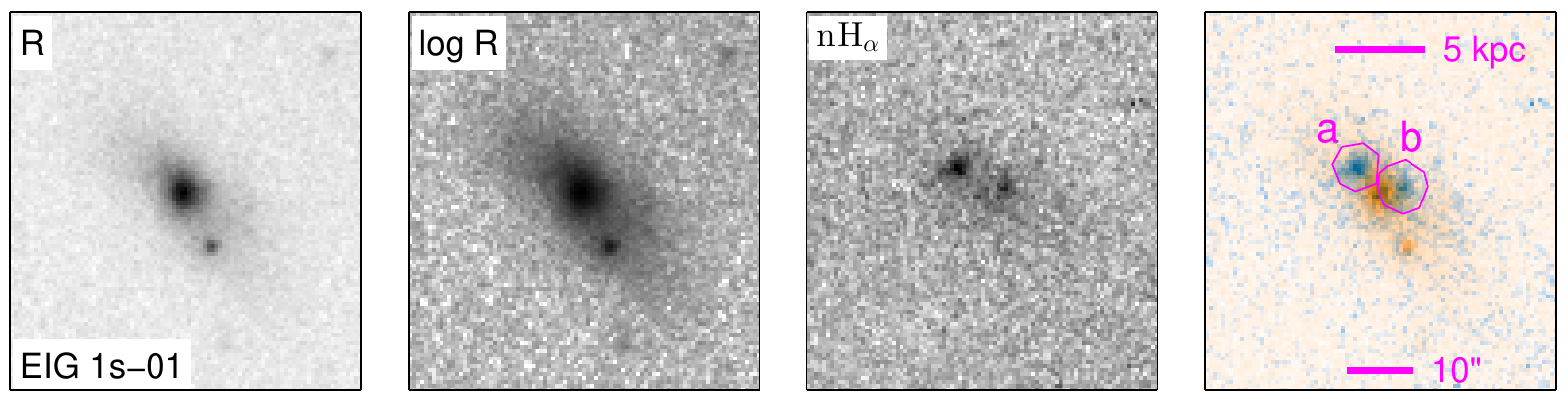}
    \includegraphics[width=17.0cm,trim=0mm 0mm 0mm 0, clip]{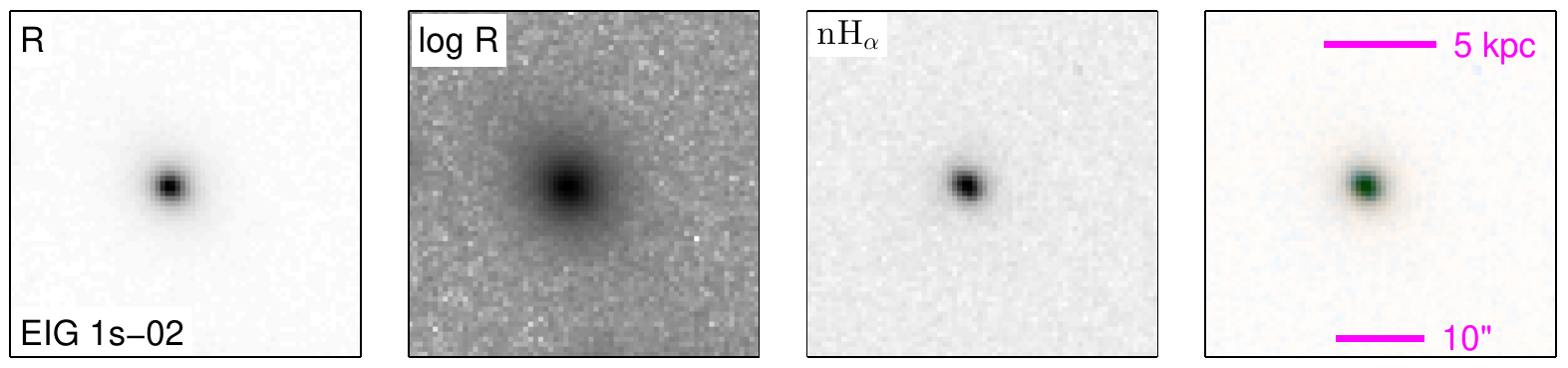}
    \includegraphics[width=17.0cm,trim=0mm 0mm 0mm 0, clip]{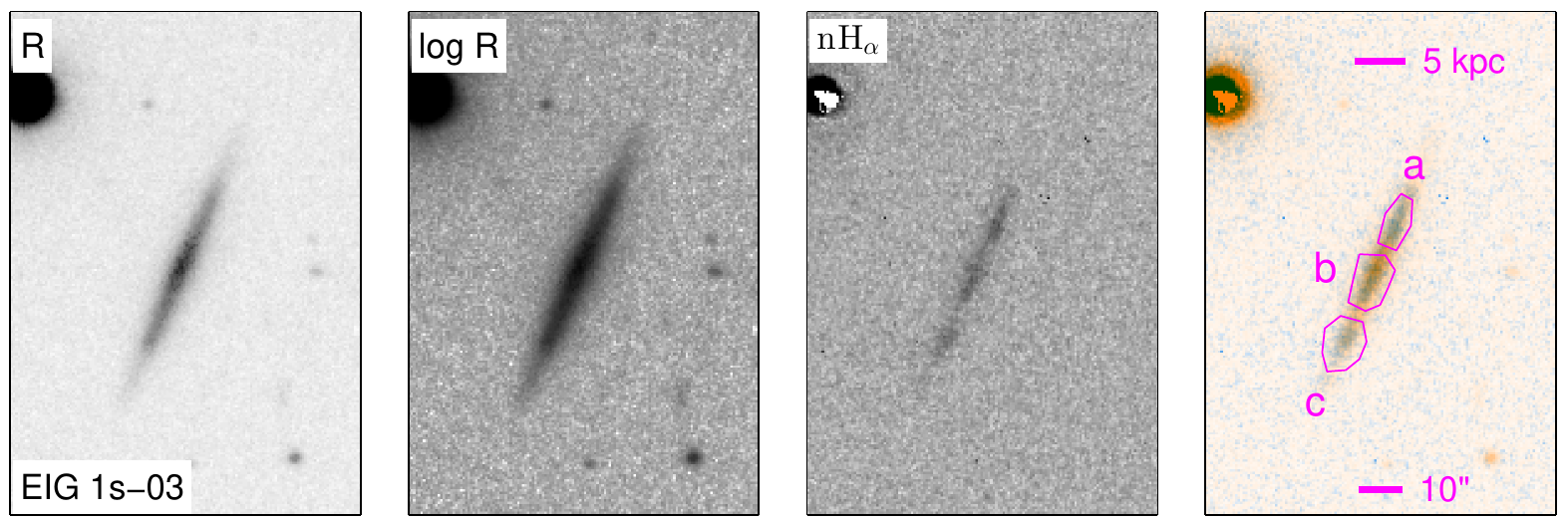}
    \includegraphics[width=17.0cm,trim=0mm 0mm 0mm 0, clip]{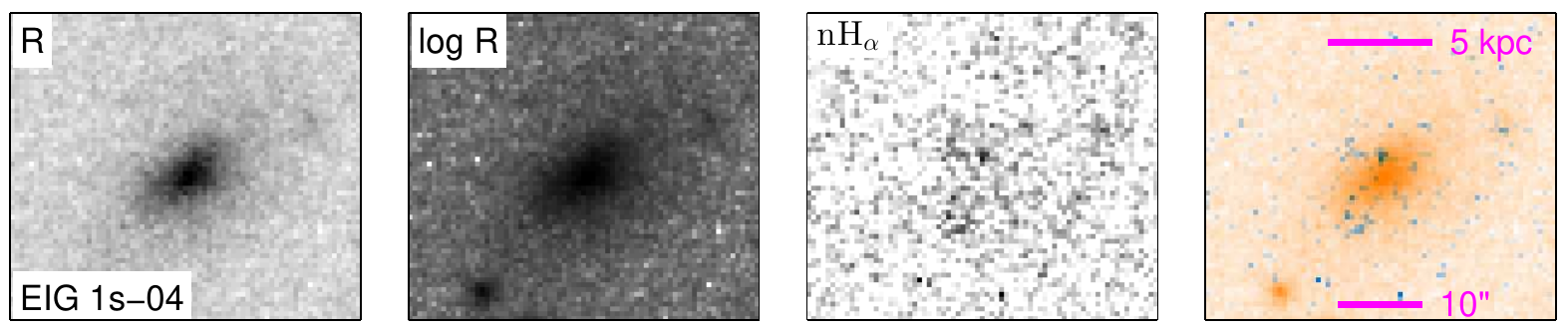}
    \includegraphics[width=17.0cm,trim=0mm 0mm 0mm 0, clip]{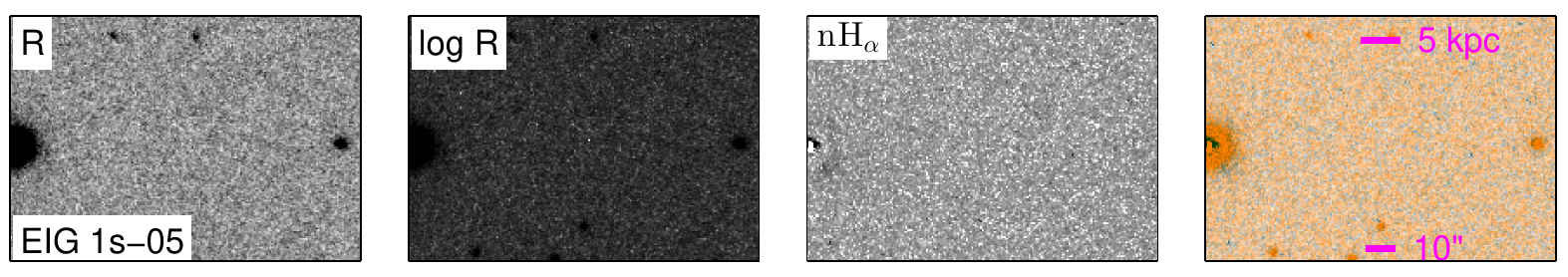}
  
    \caption [R and {\nHa} images (EIG-1)]
    {
      \markChange
      {
      R and {\nHa} images of the EIG-1 subsample (each EIG in a separate row). 
      The columns from left to right show negative images of: the combined R image (in linear scale), the combined R image in logarithmic scale, the combined {\nHa} image (linear scale), the EIG in false colour; R in orange and {\nHa} in azure (linear scale).
      The rightmost column also includes a physical distance scale, an angular size scale and where applicable the regions of interest measured individually (along with their names).\label{f:RHacolorImgEIG-1}
      }
    }
  \end{centering}
  \end{figure*}
  
  \begin{figure*}
  \begin{centering}
  
    \includegraphics[width=17.0cm,trim=0mm 0mm 0mm 0, clip]{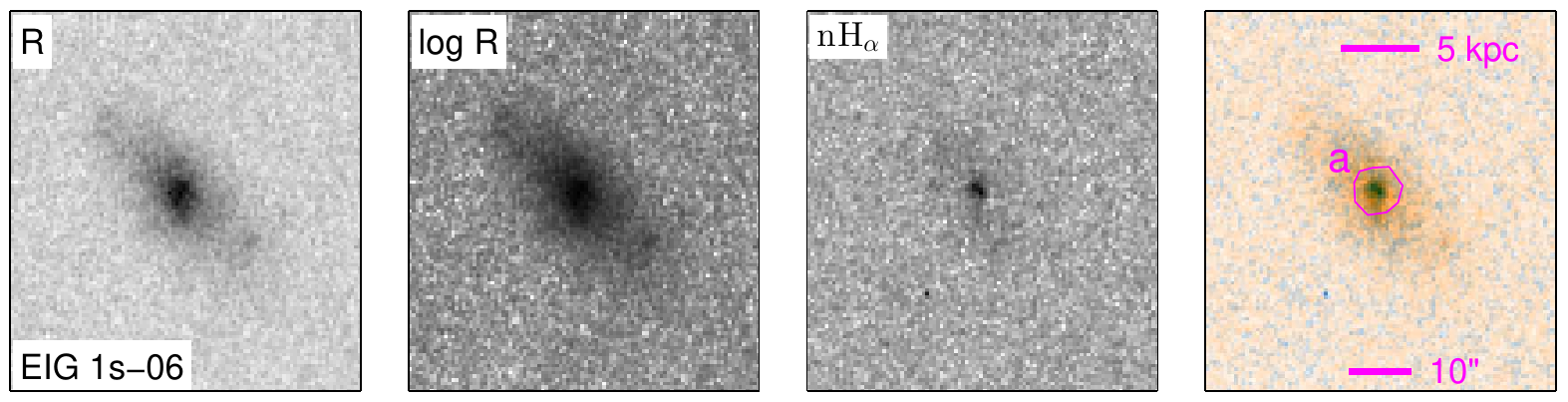}
    \includegraphics[width=17.0cm,trim=0mm 0mm 0mm 0, clip]{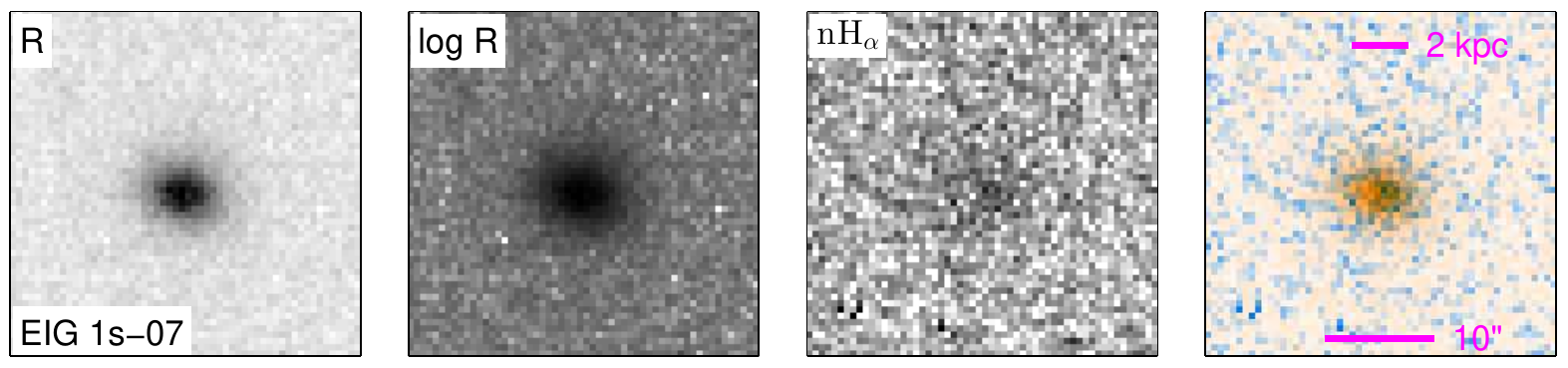}
    \includegraphics[width=17.0cm,trim=0mm 0mm 0mm 0, clip]{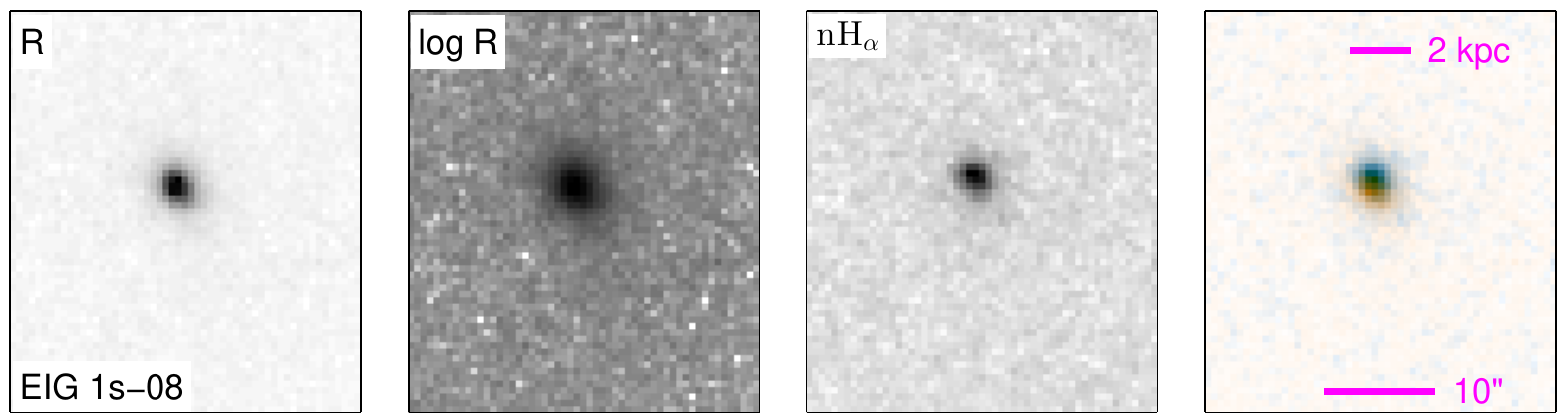}
    \includegraphics[width=17.0cm,trim=0mm 0mm 0mm 0, clip]{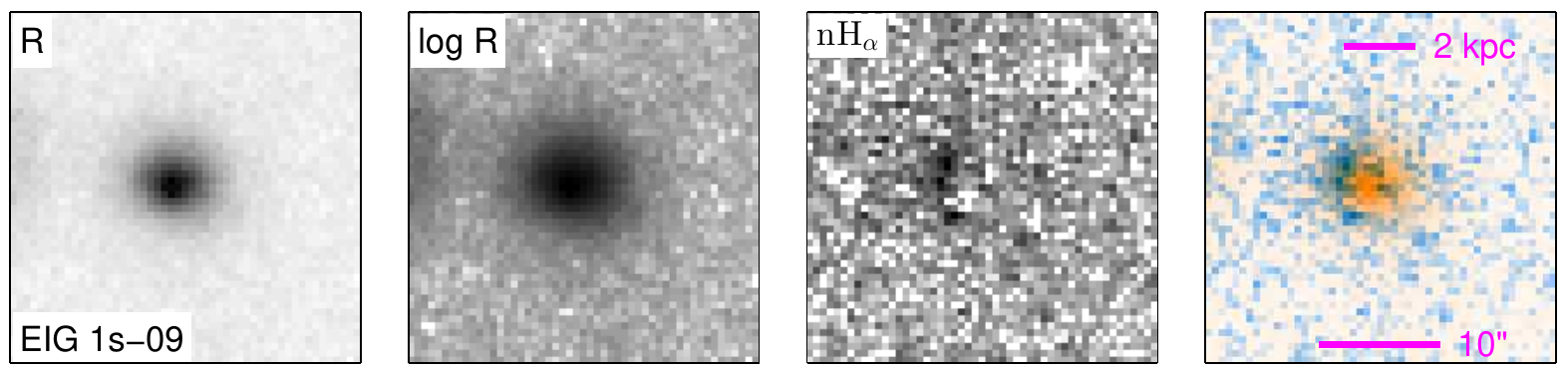}
    \includegraphics[width=17.0cm,trim=0mm 0mm 0mm 0, clip]{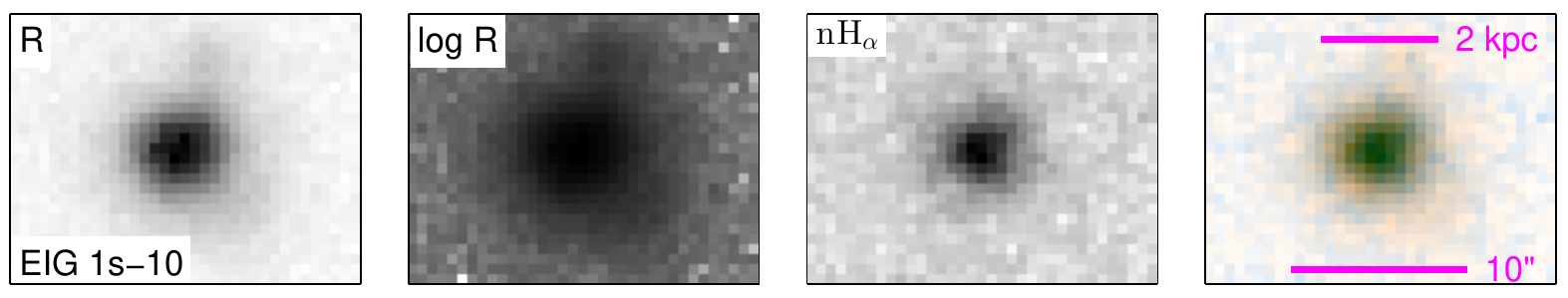}
    \includegraphics[width=17.0cm,trim=0mm 0mm 0mm 0, clip]{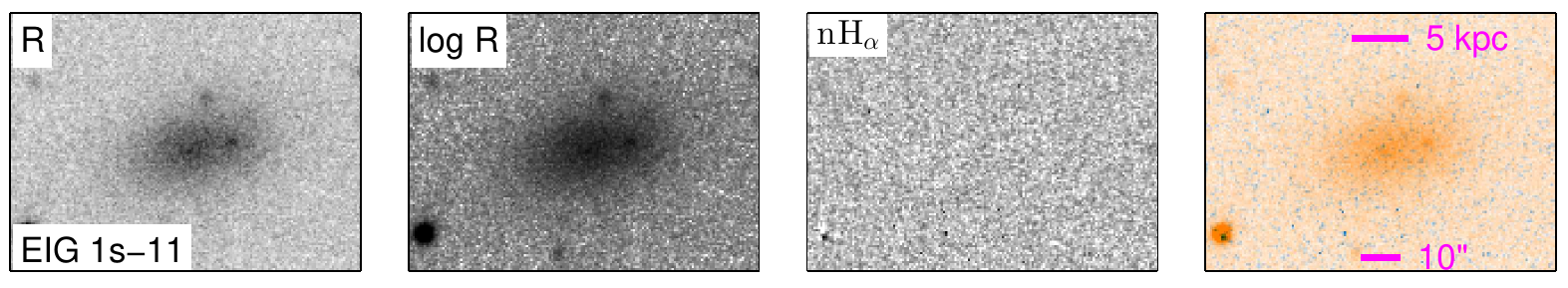}
  
    \contcaption
    {
    }
  \end{centering}
  \end{figure*}

  \begin{figure*}
  \begin{centering}
  
    \includegraphics[width=17.0cm,trim=0mm 0mm 0mm 0, clip]{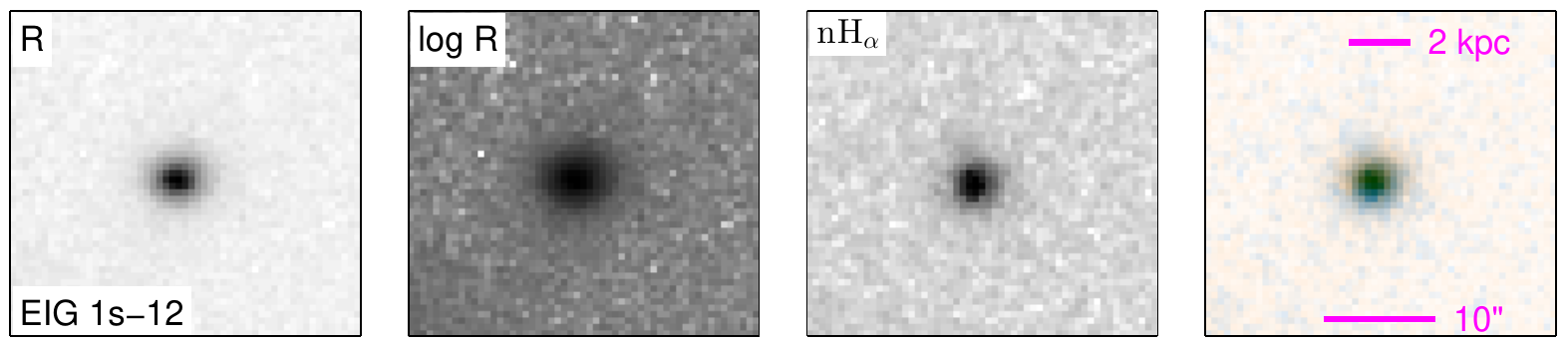}
    \includegraphics[width=17.0cm,trim=0mm 0mm 0mm 0, clip]{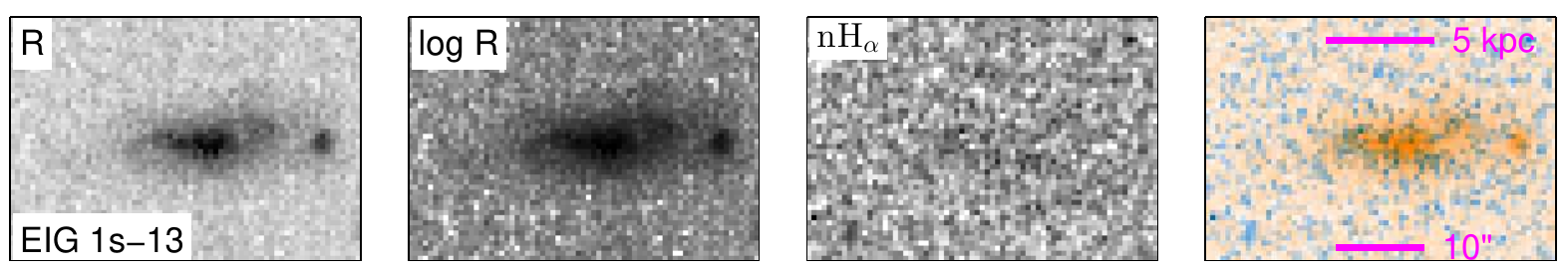}
    \includegraphics[width=17.0cm,trim=0mm 0mm 0mm 0, clip]{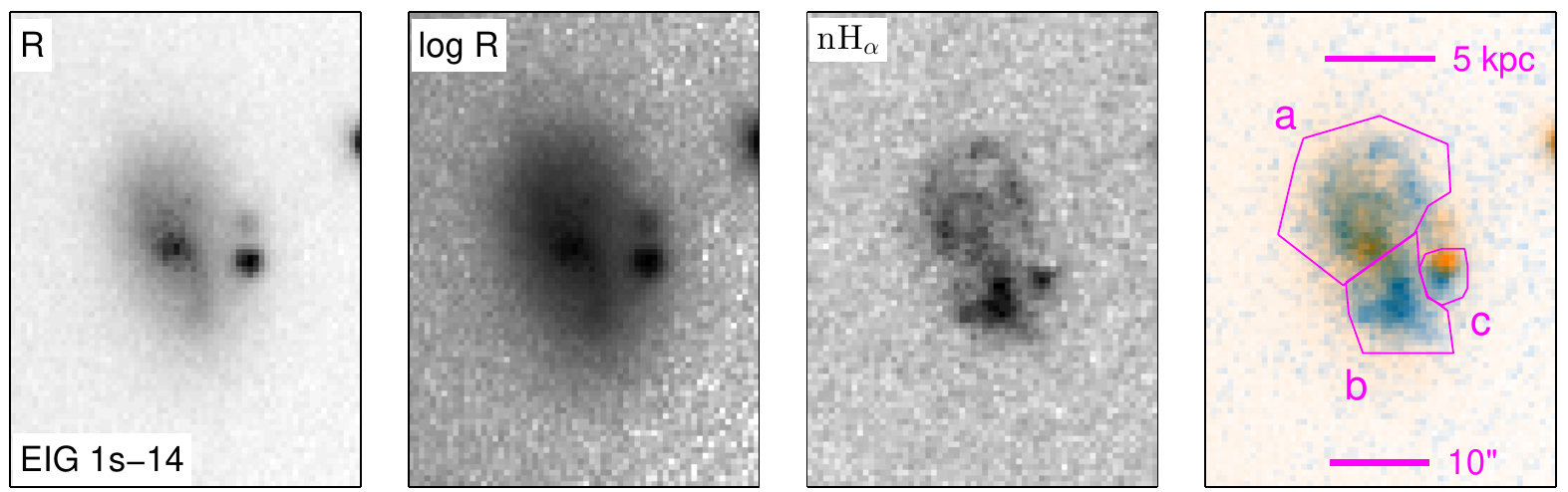}
  
    \includegraphics[width=17.0cm,trim=0mm 0mm 0mm 0, clip]{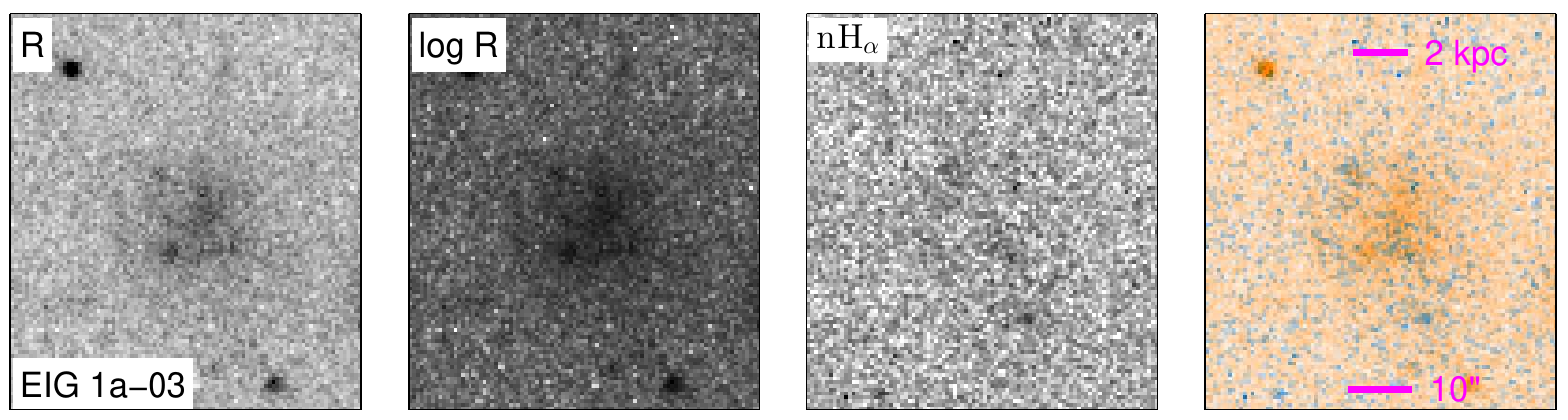}
    \includegraphics[width=17.0cm,trim=0mm 0mm 0mm 0, clip]{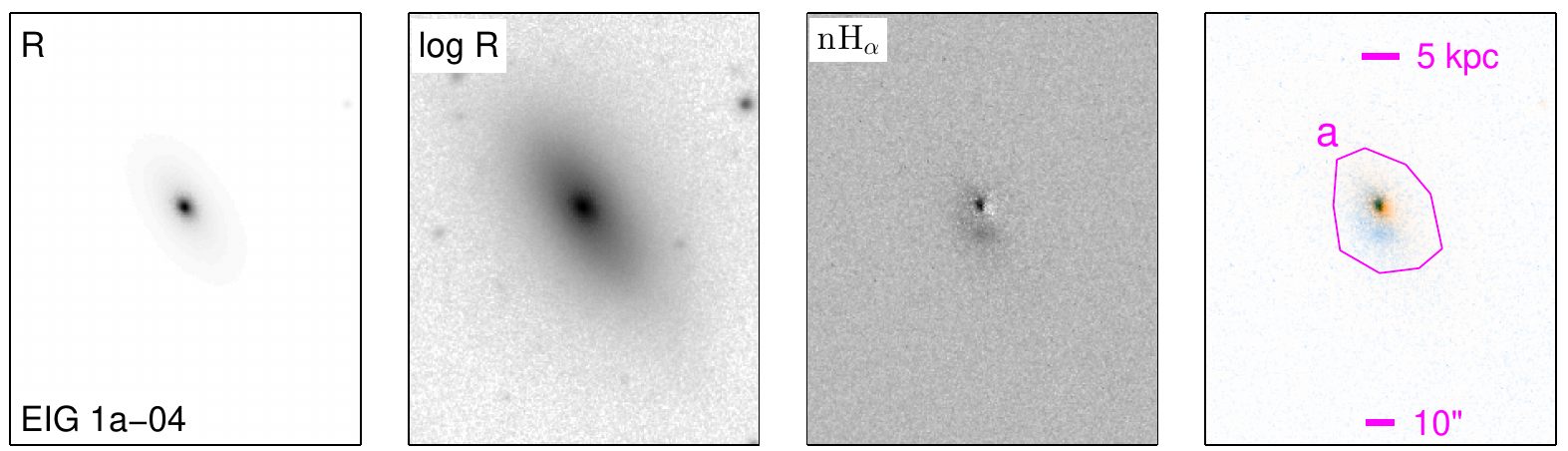}
  
    \contcaption
    {
    }
  \end{centering}
  \end{figure*}

  \begin{figure*}
  \begin{centering}
    \includegraphics[width=17.0cm,trim=0mm 0mm 0mm 0, clip]{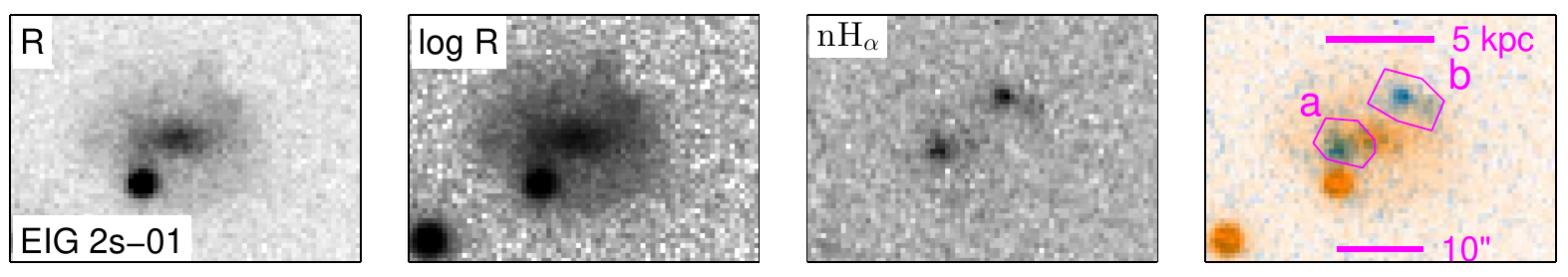}
    \includegraphics[width=17.0cm,trim=0mm 0mm 0mm 0, clip]{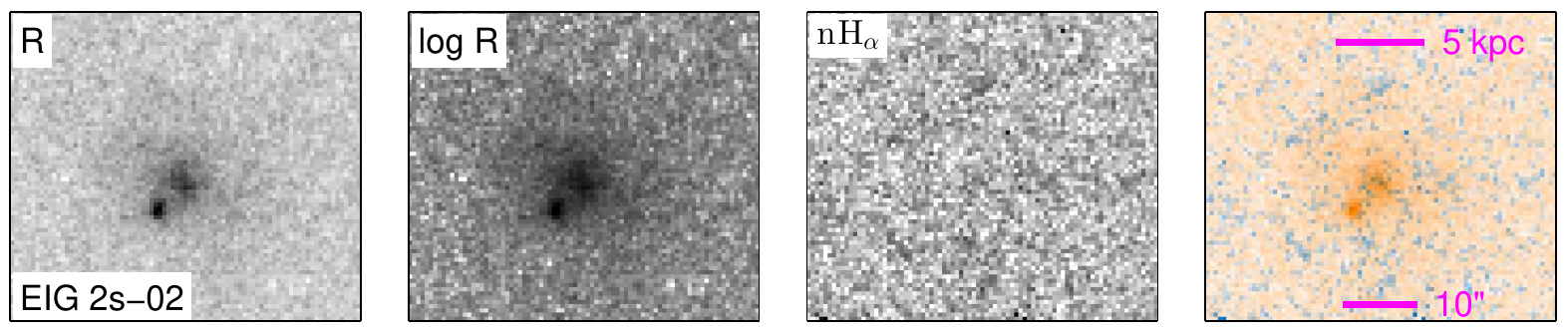}
    \rem{\includegraphics[width=17.0cm,trim=0mm 0mm 0mm 0, clip]{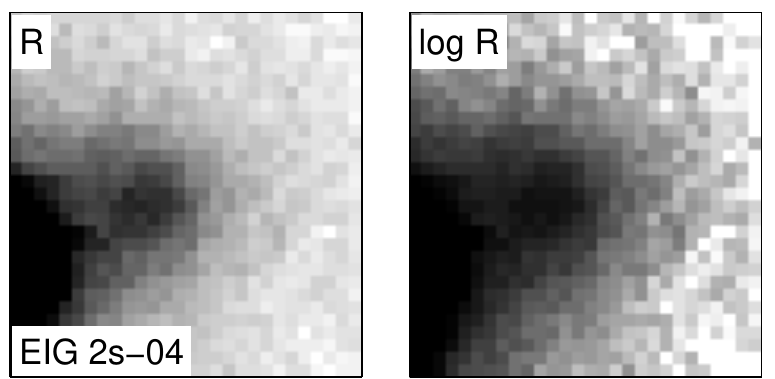}}
    \includegraphics[width=17.0cm,trim=0mm 0mm 0mm 0, clip]{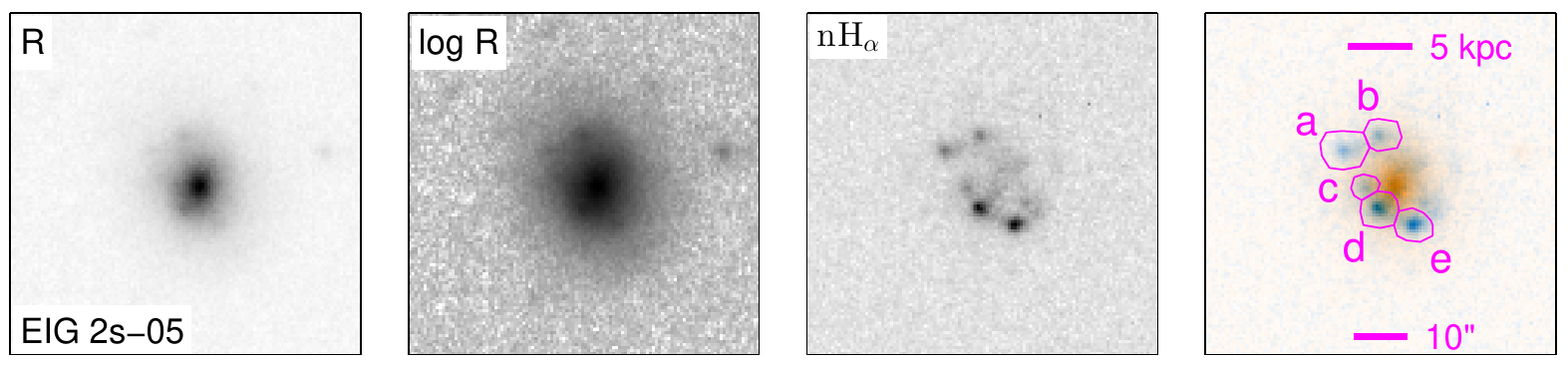}
    \includegraphics[width=17.0cm,trim=0mm 0mm 0mm 0, clip]{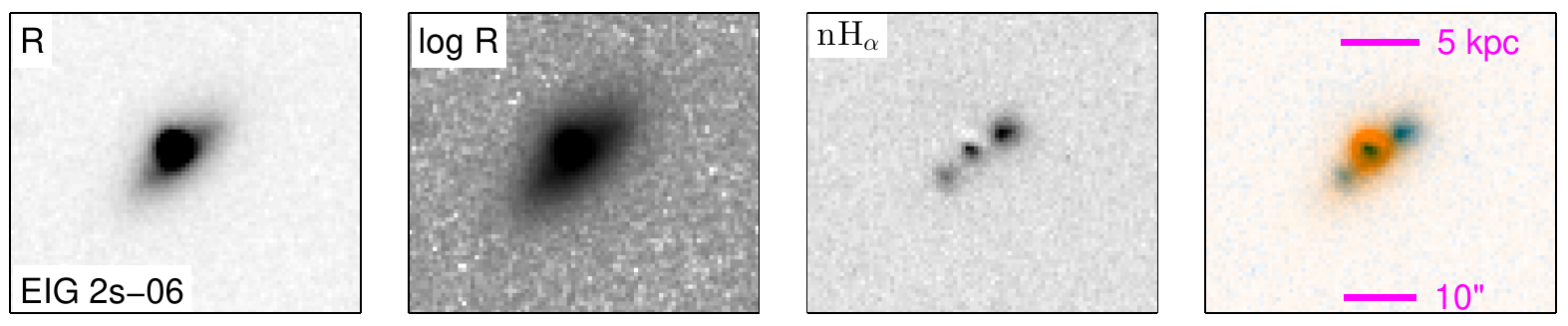}
    \includegraphics[width=17.0cm,trim=0mm 0mm 0mm 0, clip]{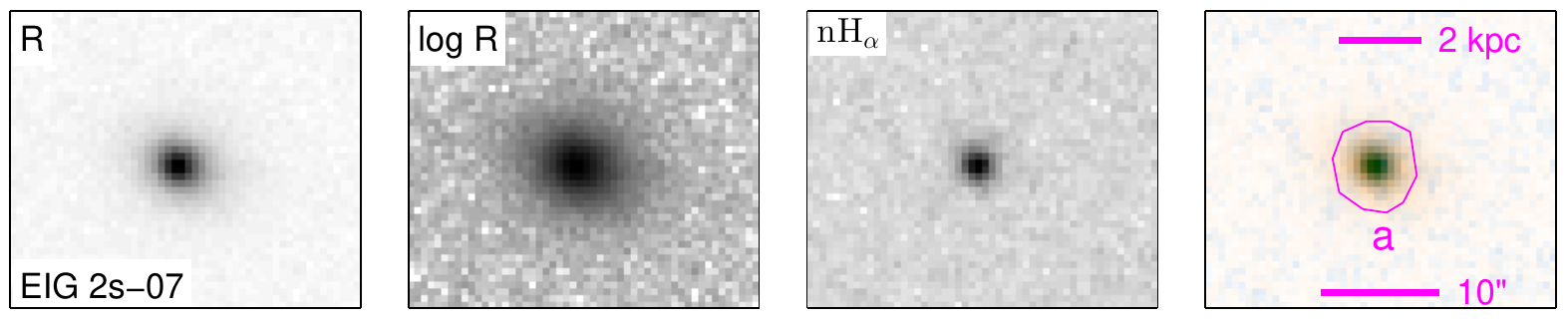}
    \includegraphics[width=17.0cm,trim=0mm 0mm 0mm 0, clip]{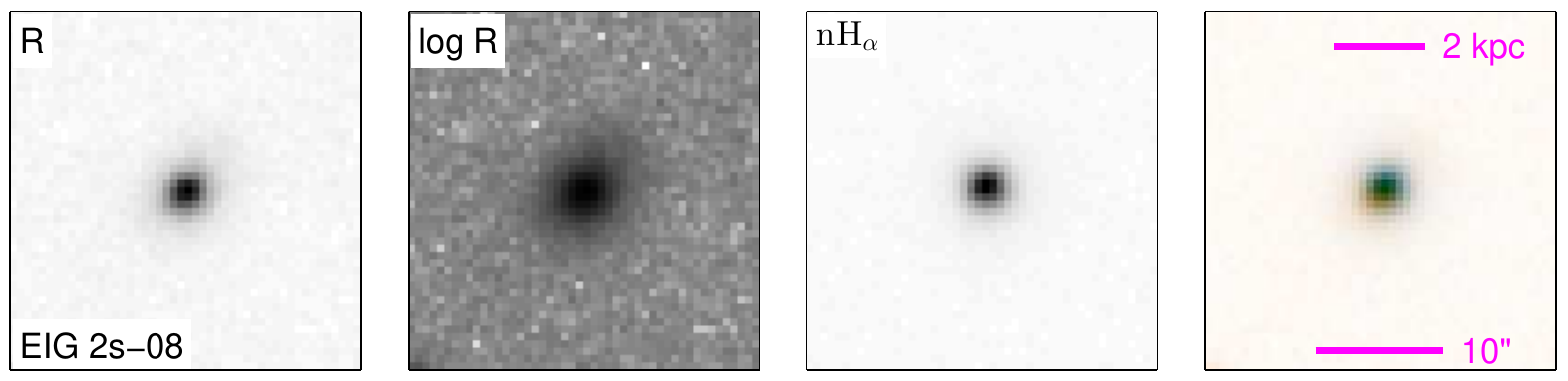}
  
    \caption [R and {\nHa} images (EIG-2)]
    {
      R and {\nHa} images of the EIG-2 subsample (each EIG in a separate row). 
      The columns from left to right show negative images of: the combined R image (in linear scale), the combined R image in logarithmic scale, the combined {\nHa} image (linear scale), the EIG in false colour; R in orange and {\nHa} in azure (linear scale).
      The rightmost column also includes a physical distance scale, an angular size scale and where applicable the regions of interest measured individually (along with their names).\label{f:RHacolorImgEIG-2}
    }
  \end{centering}
  \end{figure*}

  \begin{figure*}
  \begin{centering}
    \includegraphics[width=17.0cm,trim=0mm 0mm 0mm 0, clip]{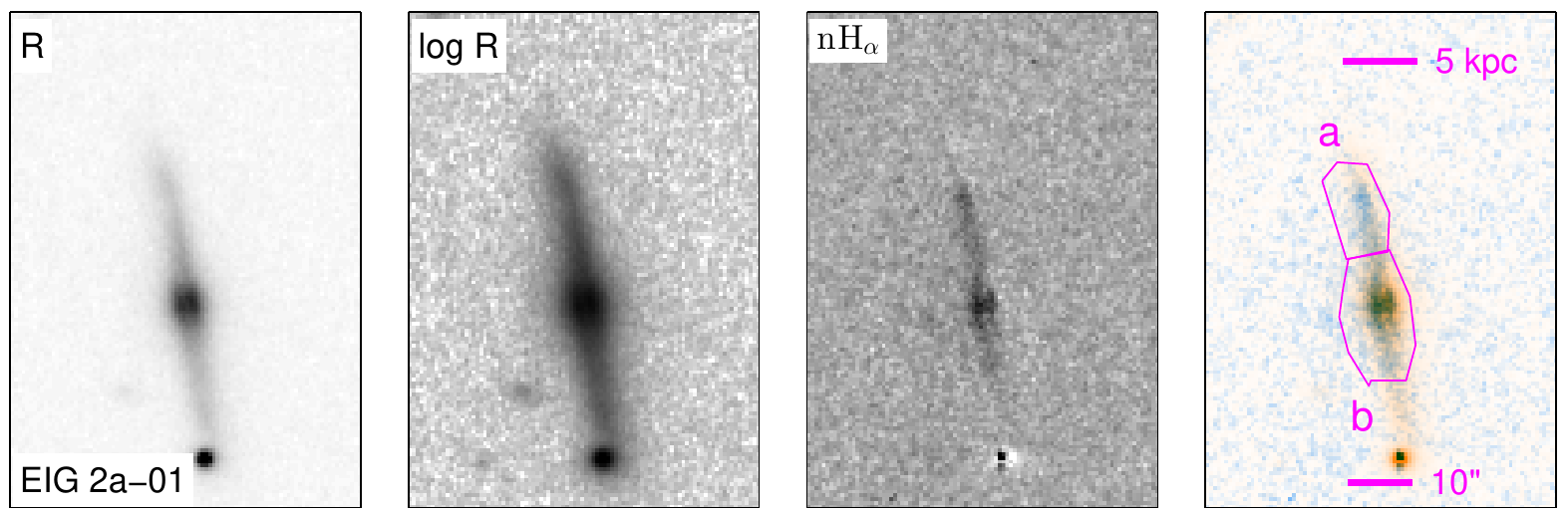}
    \includegraphics[width=17.0cm,trim=0mm 0mm 0mm 0, clip]{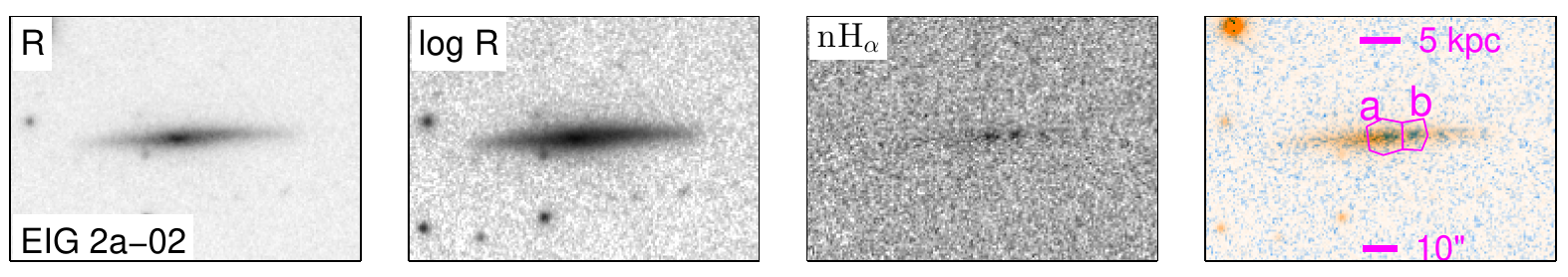}
  
    \contcaption
    {
    }
  \end{centering}
  \end{figure*}

  \begin{figure*}
  \begin{centering}
    \includegraphics[width=17.0cm,trim=0mm 0mm 0mm 0, clip]{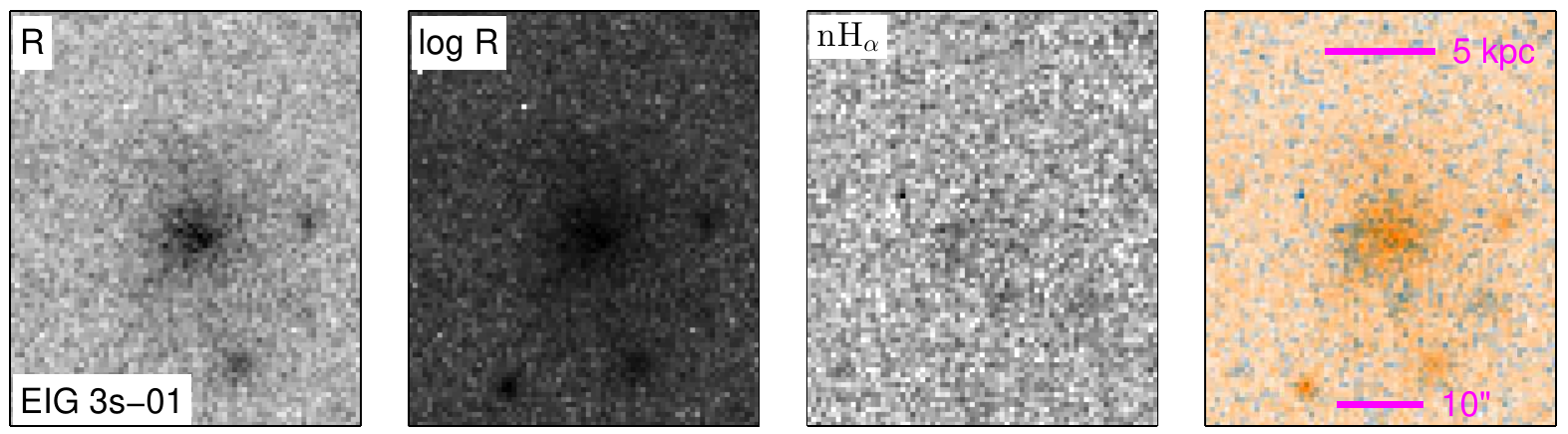}
    \includegraphics[width=17.0cm,trim=0mm 0mm 0mm 0, clip]{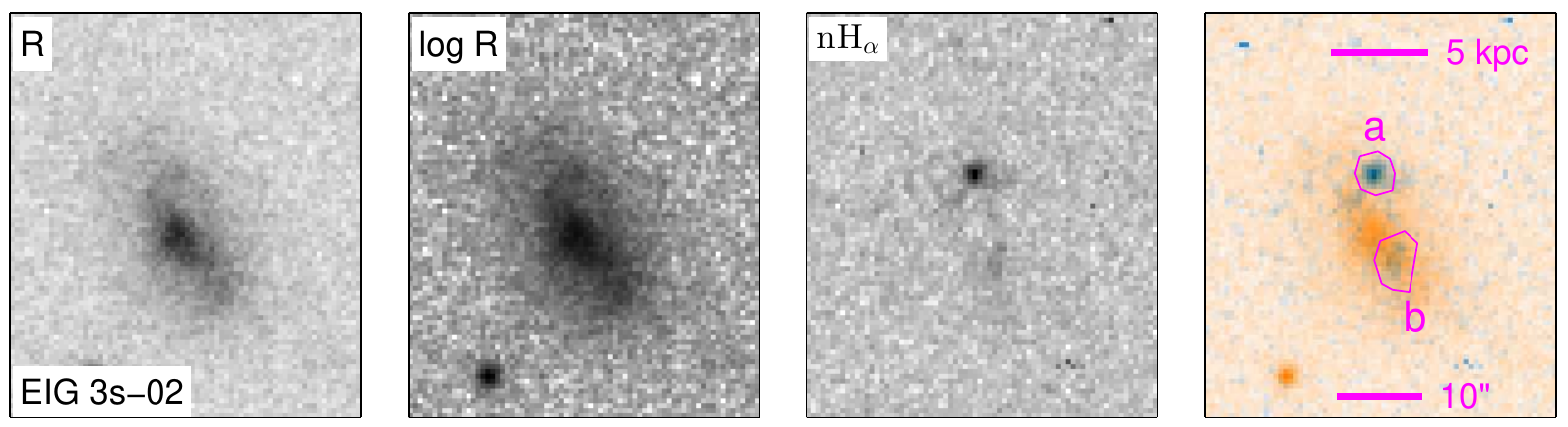}
    \includegraphics[width=17.0cm,trim=0mm 0mm 0mm 0, clip]{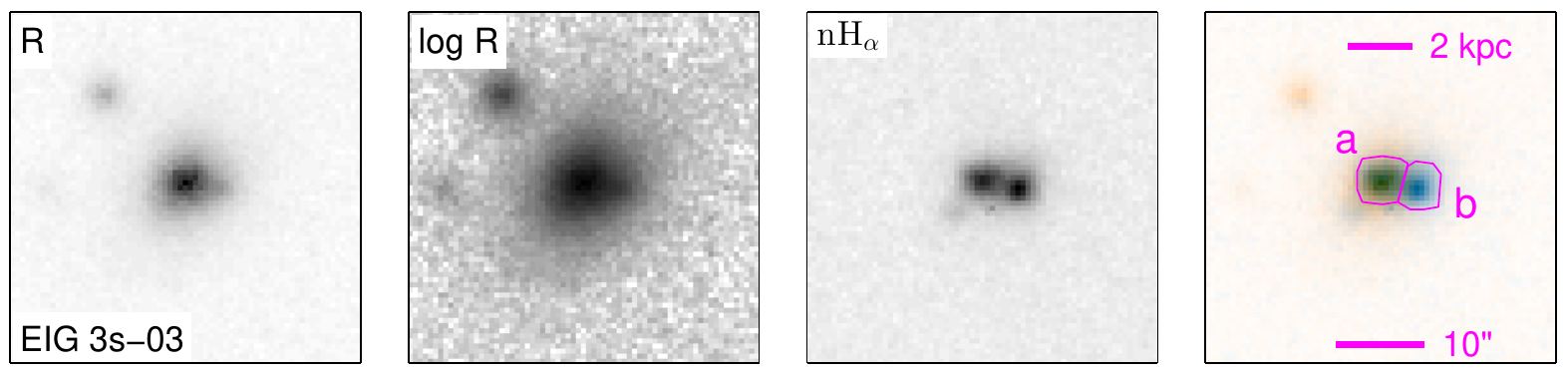}
    \includegraphics[width=17.0cm,trim=0mm 0mm 0mm 0, clip]{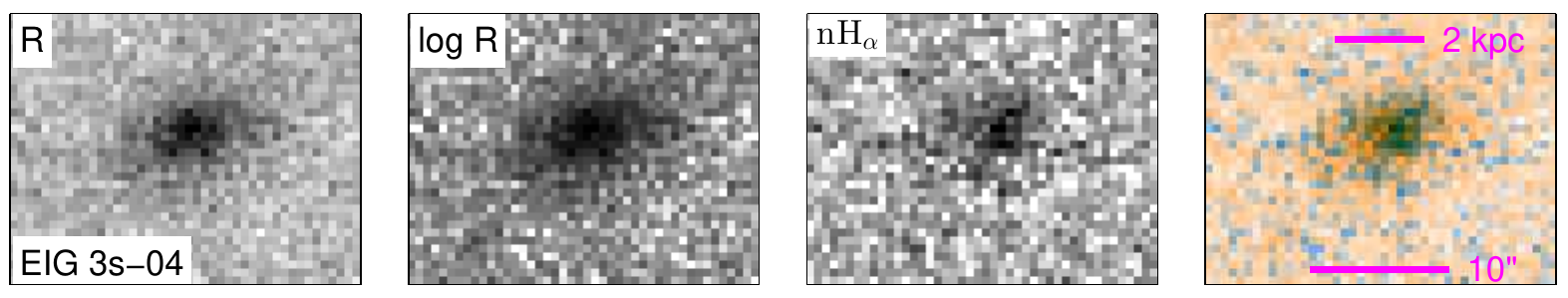}
    \includegraphics[width=17.0cm,trim=0mm 0mm 0mm 0, clip]{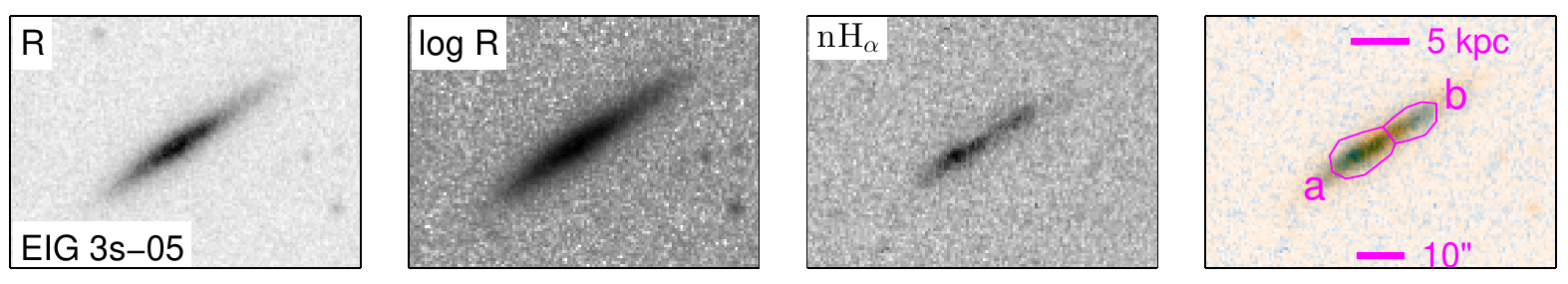}
  
    \caption [R and {\nHa} images (EIG-3)]
    {
      R and {\nHa} images of the EIG-3 subsample (each EIG in a separate row). 
      The columns from left to right show negative images of: the combined R image (in linear scale), the combined R image in logarithmic scale, the combined {\nHa} image (linear scale), the EIG in false colour; R in orange and {\nHa} in azure (linear scale).
      The rightmost column also includes a physical distance scale, an angular size scale and where applicable the regions of interest measured individually (along with their names).\label{f:RHacolorImgEIG-3}
    }
  \end{centering}
  \end{figure*}

  \begin{figure*}
  \begin{centering}
  
    \includegraphics[width=17.0cm,trim=0mm 0mm 0mm 0, clip]{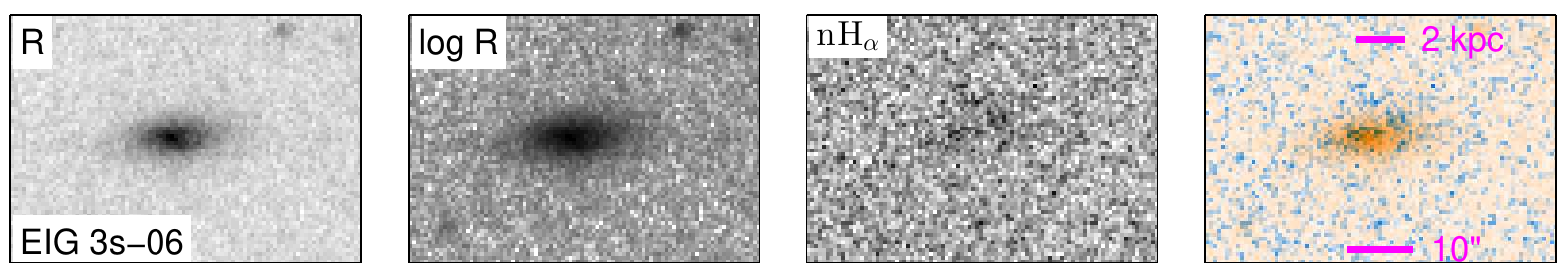}
    \includegraphics[width=17.0cm,trim=0mm 0mm 0mm 0, clip]{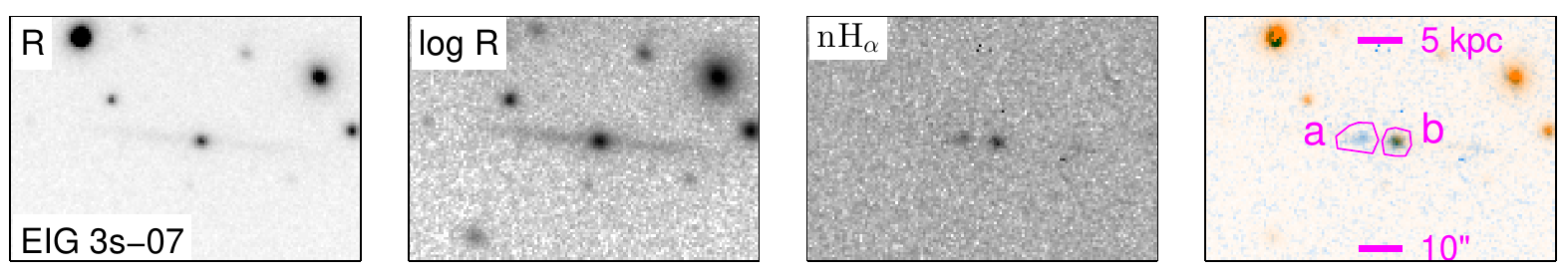}
  
    \includegraphics[width=17.0cm,trim=0mm 0mm 0mm 0, clip]{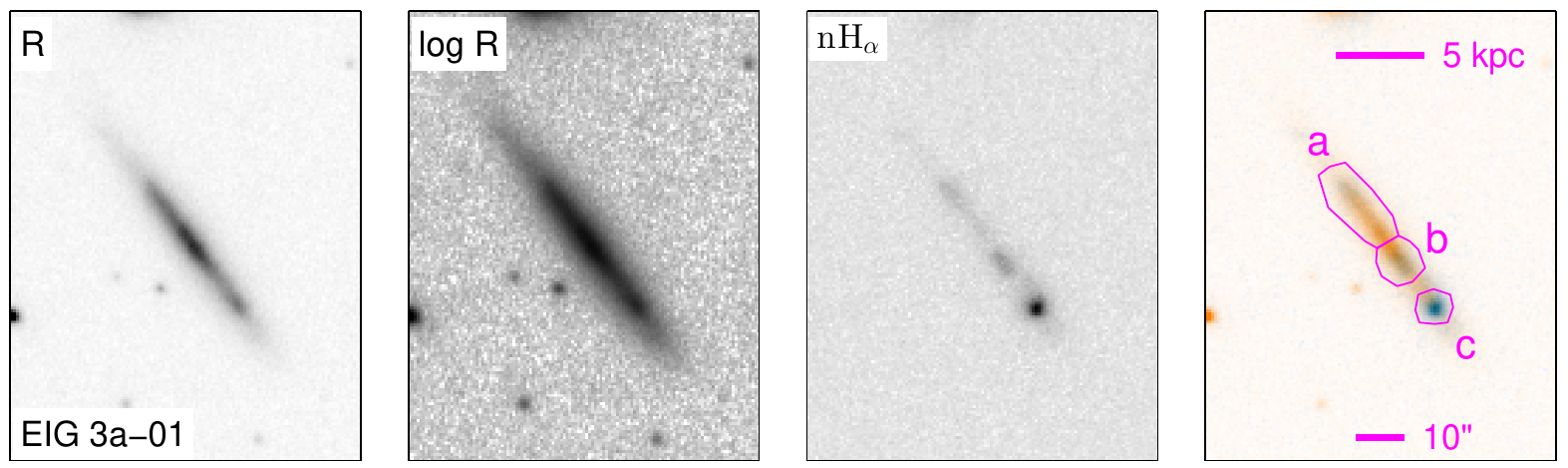}
    \includegraphics[width=17.0cm,trim=0mm 0mm 0mm 0, clip]{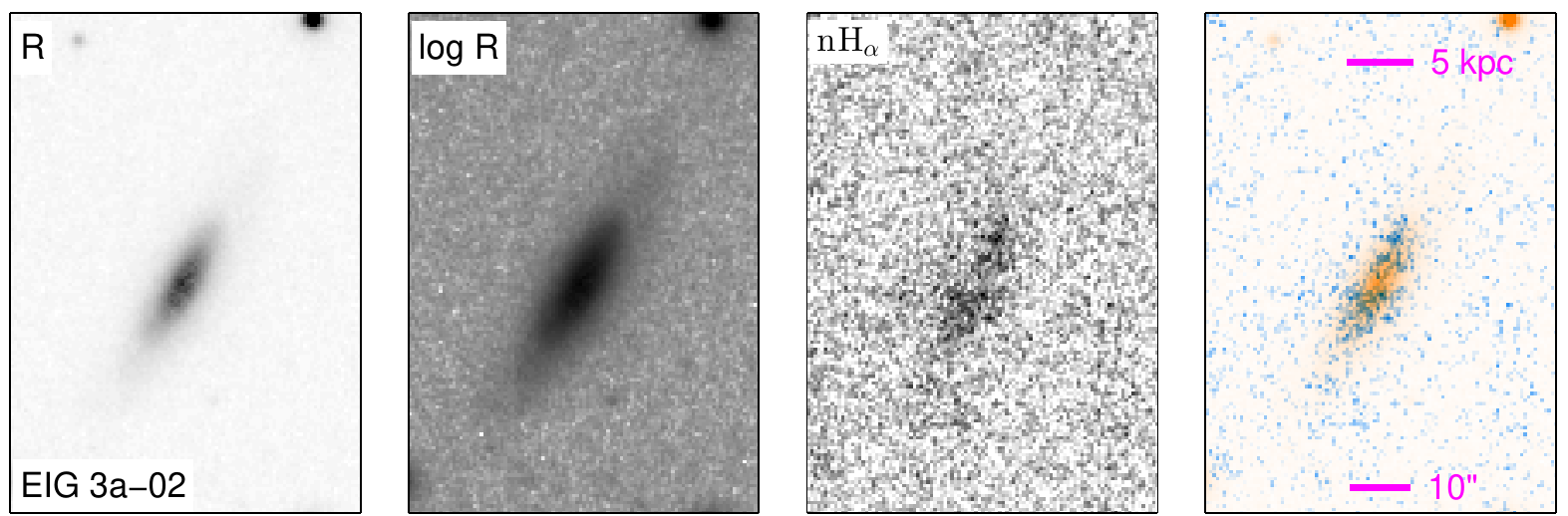}
  
    \contcaption
    {
    }
  \end{centering}
  \end{figure*}



In Figure \ref{f:RSurBrightR} the R surface brightness, $\mu_{\mbox{\scriptsize R}}$, is plotted as function of the distance from the galactic centre, $r$. The surface brightness was measured on a set of ellipses with different semi-major axes, fitted to each EIG. The $\mu_{\mbox{\scriptsize R}}$ of the innermost $2\,\arcsec$ of each galaxy is not shown to avoid confusion due to point spread function (PSF) effects. $\mu_{\mbox{\scriptsize R}}$ measurements with uncertainty $\geq 0.5\,\magAsecSq$ are also not shown.
Two profiles were fitted for each EIG's $\mu_{\mbox{\scriptsize R}}$ measurements, one typical of a late type galaxy disc (blue dashed line) and the other representing an early-type elliptical galaxy (red solid curve). The disc profile has a S\'{e}rsic's index $n=1$, and was fitted to the outskirts of the galaxy (from half of the maximum $r$ shown in the figure and further). The elliptical profile is a de Vaucouleurs relation, i.e. a S\'{e}rsic's index $n=4$. It was fitted to all the measured points shown in the figure.

\begin{figure*}
\begin{centering}
  \includegraphics[width=16.3cm,trim=0mm 0mm 0mm 0, clip]{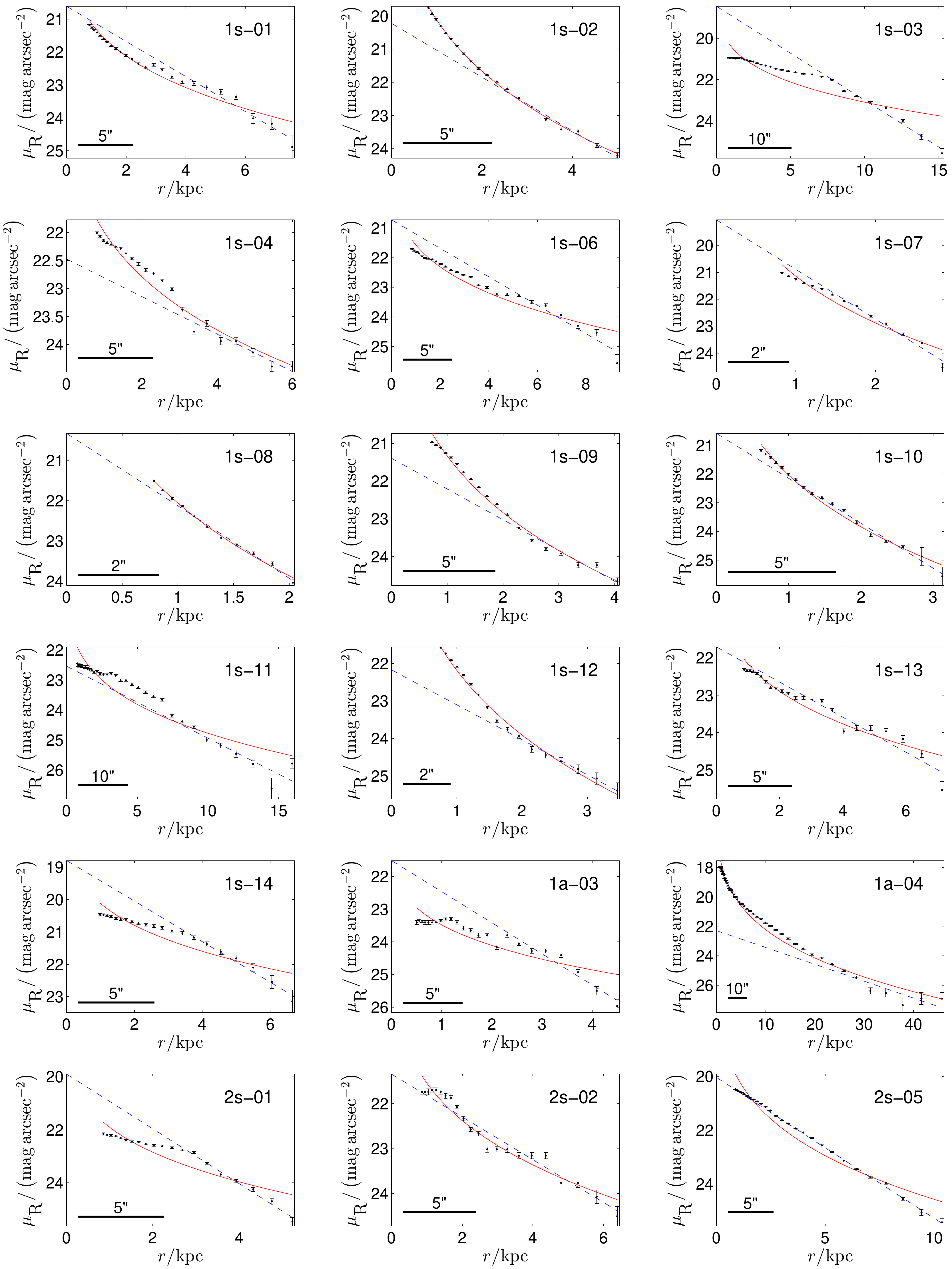}

  \caption [R surface brightness]
  {
    R surface brightness, $\mu_{\mbox{\scriptsize R}}$, as function of the distance from the galactic centre, $r$. The black horizontal bars show the angular scale.
    The blue dashed lines are linear relations, fitted to the points above half of the maximum shown $r$.
    The red solid curves show best fits to a de Vaucouleurs formula. \label{f:RSurBrightR}
  }
\end{centering}
\end{figure*}

\begin{figure*}
\begin{centering}

  \includegraphics[width=16.3cm,trim=0mm 0mm 0mm 0, clip]{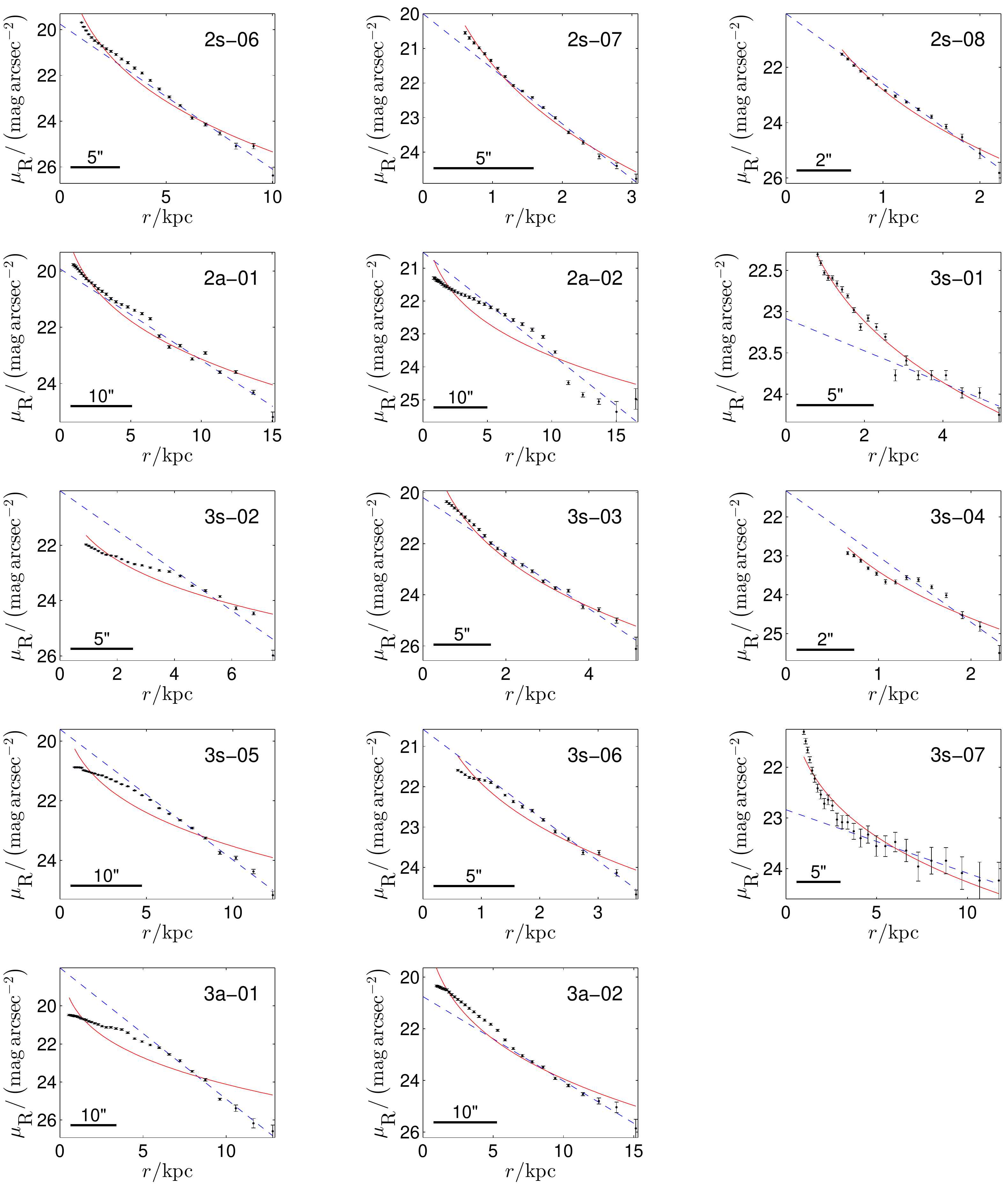}

  \contcaption
  {
  }
\end{centering}
\end{figure*}

\markChange{The EIGs were classified as early-types or late-types by visual inspection of the images of the EIGs, and the $\mu_{\mbox{\scriptsize R}}$ profiles of Figure \ref{f:RSurBrightR}.}
Whenever \markChange{the combination} of the images and $\mu_{\mbox{\scriptsize R}}$ profiles did not yield a clear identification, the EIG was classified as `unknown'. \markChange{Six out of 31 EIGs (19 per cent) were classified as `unknown'. We chose to classify such a large fraction as 'unknown' in order to reduce the probability of false identification to a minimum.
The morphological types of EIGs 1s-05, 2s-04 and 2s-06 were not classified. EIG 1s-05 was not classified because it cannot be identified in optical images, and EIGs 2s-04 and 2s-06 were not classified because of bright foreground stars projected close to them (see details in Appendix \ref{App:EIGdata})}.
The morphological classifications are listed in Table \ref{T:Res_Morphology}.

\begin{ctable} 
[
  caption = {Morphological classification of the EIGs},
  cap     = {Morphological classification of the EIGs},
  doinside = \small,
  label   = {T:Res_Morphology}
]
{ @{}ll@{}
}
{
}
{
 \FL
 Type & EIGs
 \ML
Late    & 1s-03, 1s-04, 1s-06, 1s-07, 1s-11, 1s-13, 1s-14, 1a-03, \NN 
   & 2s-01, 2s-02, 2s-05, 2s-08, 2a-01, 2a-02, 3s-02, 3s-05, \NN 
   & 3s-06, 3s-07, 3a-01, 3a-02 \NN
Unknown & 1s-01, 1s-08, 1s-10, 2s-07, 3s-03, 3s-04 \NN
Early   & 1s-02, 1s-09, 1s-12, 1a-04, 3s-01
 \LL
}
\end{ctable}
 \rem{label = {T:Res_Morphology}}


Ultraviolet data were downloaded from the GALEX \citep{2005ApJ...619L...1M} GR6/7 data release\footnote{http://galex.stsci.edu/GR6/}. The available apparent magnitudes of EIGs in the GALEX far-ultraviolet band ({\magFUV}) and near-ultraviolet band ({\magNUV}) are listed in Table \ref{T:EIG_GALEX}.

\begin{ctable} 
[
  caption = {GALEX apparent magnitudes},
  cap     = {GALEX apparent magnitudes},
  doinside = \small,
  label   = {T:EIG_GALEX}
]
{ c@{\qquad}r@{$\,\pm\,$}lr@{$\,\pm\:$}l
}
{
}
{
 \FL
 EIG  & \multicolumn {2}{c}{\magFUV} & \multicolumn {2}{c}{\magNUV}
 \ML
 1s-01 & 18.67 & 0.08 & 18.54 & 0.05
 \NN
 1s-03 & 19.50 & 0.03 & 18.88 & 0.02
 \NN
 1s-04 & 19.53 & 0.05 & 19.06 & 0.03
 \NN
 1s-06 & 18.70 & 0.02 & 18.41 & 0.01
 \NN
 1s-07 & 20.0 & 0.2 & 19.72 & 0.02
 \NN
 1s-08 & 20.4 & 0.2 & 19.84 & 0.03
 \NN
 1s-10 & 18.9 & 0.1 & 18.63 & 0.02
 \NN
 1s-11 & 23.0 & 0.4 & 19.77 & 0.06
 \NN
 1s-12 & 19.4 & 0.2 & 19.36 & 0.02
 \NN
 1s-13 & 19.72 & 0.05 & 19.43 & 0.03
 \ML
 1a-01 & 18.30 & 0.08 & 18.07 & 0.05
 \NN
 1a-02 & 19.73 & 0.09 & 19.26 & 0.05
 \NN
 1a-03 & 21.6 & 0.4 & 19.6 & 0.1
 \NN
 1a-04 & 21.5 & 0.3 & 19.35 & 0.09
 \NN
 1a-06 & 20.9 & 0.1 & 21.1 & 0.1
 \NN
 1a-07 & 18.89 & 0.09 & 18.58 & 0.05
 \ML
 2s-01 & 19.2 & 0.1 & 18.93 & 0.08
 \NN
 2s-02 & 18.75 & 0.08 & 18.36 & 0.06
 \NN
 2s-05 & 17.63 & 0.07 & 17.27 & 0.04
 \NN
 2s-07 & 19.6 & 0.2 & 19.5 & 0.1
 \NN
 2s-08 & 18.86 & 0.03 & 18.59 & 0.02
 \ML
 2a-01 & 20.8 & 0.2 & 19.30 & 0.07
 \NN
 2a-03 &  \multicolumn {2}{c}{---}  & 18.11 & 0.04
 \NN
 2a-04 & 20.53 & 0.07 & 20.18 & 0.05
 \ML
 3s-02 & 19.02 & 0.03 & 18.73 & 0.02
 \NN
 3s-03 & 18.13 & 0.07 & 17.72 & 0.03
 \NN
 3s-04 & 20.1 & 0.2 & 19.89 & 0.08
 \NN
 3s-05 & 18.9 & 0.1 & 18.45 & 0.06
 \NN
 3s-06 & 20.0 & 0.3 & 19.31 & 0.03
 \NN
 3s-07 & 19.49 & 0.04 & 19.05 & 0.03
 \ML
 3a-01 & 17.69 & 0.05 & 17.27 & 0.03
 \LL
}
\end{ctable}
 \rem{label = {T:EIG_GALEX}}


2MASS \citep{2006AJ....131.1163S} and WISE \citep{2010AJ....140.1868W} data were downloaded from the NASA/IPAC Infrared Science Archive (IRSA)\footnote{http://irsa.ipac.caltech.edu}.
The 2MASS data were taken from the All-Sky data release. Thirteen EIGs were identified in its Extended Source Catalogue. For two of these (EIGs 1a-01 and 2a-02) the quoted {\magJ}, {\magH} and {\magKs} magnitudes were not used since they translate to flux suspiciously lower than that of the {\SDSSi} band.
The {\magJ}, {\magH} and {\magKs} apparent magnitudes\rem{(wavelengths {1.24\,\um}, {1.66\,\um} and {2.16\,\um} respectively)} of the remaining eleven EIGs are listed in Table \ref{T:EIG_2MASS}.

\begin{ctable} 
[
  caption = {2MASS apparent magnitudes},
  cap     = {2MASS apparent magnitudes},
  doinside = \small,
  label   = {T:EIG_2MASS}
]
{ @{}c@{\qquad}r@{$\,\pm\,$}l@{\qquad}r@{$\,\pm\:$}l@{\qquad}r@{$\,\pm\,$}l@{}
}
{
}
{
 \FL
 EIG  & \multicolumn {2}{c}{\magJ} & \multicolumn {2}{c}{\magH} & \multicolumn {2}{c}{\magKs} 
 \ML
 1s-02 & 14.27 & 0.07 & 13.7 & 0.1 & 13.5 & 0.1
 \NN
 1s-03 & 13.76 & 0.09 & 12.63 & 0.07 & 12.9 & 0.2
 \ML
 1a-02 & 15.1 & 0.2 & 14.4 & 0.2 & 14.1 & 0.2
 \NN
 1a-04 & 10.60 & 0.02 & 9.87 & 0.02 & 9.52 & 0.03
 \NN
 1a-05 & 13.83 & 0.07 & 13.3 & 0.1 & 13.0 & 0.1
 \ML
 2s-05 & 14.11 & 0.08 & 13.08 & 0.08 & 13.2 & 0.2
 \ML
 2a-01 & 12.31 & 0.04 & 11.58 & 0.04 & 11.18 & 0.05
 \NN
 2a-03 & 13.99 & 0.09 & 13.8 & 0.2 & 13.1 & 0.2
 \ML
 3s-03 & 14.6 & 0.2 & 13.7 & 0.2 & 13.3 & 0.2
 \ML
 3a-01 & 14.2 & 0.1 & 13.4 & 0.1 & 13.7 & 0.2
 \NN
 3a-02 & 13.26 & 0.05 & 12.79 & 0.07 & 12.5 & 0.1
 \LL
}
\end{ctable}
 \rem{label = {T:EIG_2MASS}}

WISE data taken from the All-WISE catalogue are listed in Table \ref{T:EIG_StarFormation}. These include apparent magnitudes in the {\WThree}\rem{({12\,\um})} and {\WFour}\rem{({22\,\um})} bands (columns {\magWThree} and {\magWFour} respectively).
For some of the EIGs these are measurements through elliptical apertures based on the 2MASS {\Ks} isophotal apertures.
Data for EIGs, for which the elliptical aperture measurements were not possible, are profile-fitting photometry measurements, or for low SNR measurements the 0.95 confidence magnitude lower limits.

The WISE profile-fitting photometry measurements are less accurate for extended objects than the elliptical aperture measurements. To estimate their uncertainty, a comparison was made between profile-fitting photometry magnitudes and elliptical aperture based magnitudes for eight EIGs for which both were available. On average, the profile-fitting magnitudes were found to be {$0.08 \pm 0.03$\,\magnitude} ({\WThree}) and {$0.1 \pm 0.1$\,\magnitude} ({\WFour}) lower than the elliptical aperture magnitudes. The standard deviation of the difference between the two measurement methods was found to be {$\sim$0.3\,\magnitude} for {\WThree} and {$\sim$0.4\,\magnitude} for {\WFour}. These standard deviations were added to the estimated uncertainties of the profile-fitting photometry magnitudes.


Table \ref{T:EIG_ALFALFA} lists data from the ALFALFA {$\alpha$}.40 catalogue \citep{2011AJ....142..170H}. For each EIG that was detected by ALFALFA the velocity width of the HI line profile, {\Whalf}, corrected for instrumental broadening but not for disc inclination is listed. This is followed by the total HI line flux, {\FHI}, the estimated signal to noise ratio of the detection, {\SNR}, and the HI mass content, {\MHI}. Finally, the category of the HI detection, Code, is listed. Code 1 refers to a source of SNR and general qualities that make it a reliable detection. Code 2 refers to a source with $\SNR \lesssim 6.5$ that does not qualify for code 1 but was matched with a counterpart with a consistent optical redshift, and is very likely to be real.

\begin{ctable} 
[
  caption = {ALFALFA HI data: EIG name, velocity width of the HI line profile ({\Whalf}), total HI line flux ({\FHI}), estimated signal to noise ratio of the detection ({\SNR}), HI mass content ({\MHI}) and the category of the HI detection (Code).},
  cap     = {ALFALFA HI data},
  doinside = \small,
  star,
  label   = {T:EIG_ALFALFA}
]
{ @{}cr@{$\,\pm\,$}rl@{$\,\pm\,$}lcl@{$\,\pm\,$}lc@{}
}
{
}
{
 \FL
 \multirow{2}*{EIG} 
 & \multicolumn {2}{c}{\Whalf} 
 & \multicolumn {2}{c}{\FHI}   
 & \multirow{2}*{\SNR}         
 & \multicolumn {2}{c}{\multirow{2}*{$\log\frac{\MHI}{\Msun}$}} 
 & \multirow{2}*{Code} \NN
      & \multicolumn {2}{c}{$\left[ \kms \right]$} 
      & \multicolumn {2}{c}{$\left[ \Jy\,\kms \right]$} 
      &                    &  \multicolumn {2}{c}{}        &  
 \ML
 1s-01 & 176 & 6 & 1.64 & 0.07 & 12.6 & 9.35 & 0.02 & 1
 \NN
 1s-02 & 65 & 18 & 0.66 & 0.05 & 8.0 & 9.01 & 0.03 & 1
 \NN
 1s-03 & 275 & 3 & 2.47 & 0.09 & 15.3 & 9.66 & 0.02 & 1
 \NN
 1s-04 & 160 & 3 & 1.30 & 0.07 & 10.8 & 9.32 & 0.02 & 1
 \NN
 1s-05 & 32 & 11 & 0.46 & 0.05 & 8.1 & 8.84 & 0.05 & 1
 \NN
 1s-06 & 201 & 8 & 1.36 & 0.08 & 9.3 & 9.34 & 0.03 & 1
 \NN
 1s-07 & 156 & 7 & 0.59 & 0.05 & 6.4 & 8.95 & 0.04 & 2
 \NN
 1s-09 & 91 & 5 & 0.96 & 0.06 & 9.6 & 9.03 & 0.03 & 1
 \NN
 1s-10 & 109 & 9 & 0.63 & 0.06 & 6.1 & 8.76 & 0.04 & 1
 \NN
 1s-14 & 64 & 20 & 0.55 & 0.06 & 6.4 & 9.05 & 0.05 & 2
 \ML
 1a-01 & 196 & 24 & 1.78 & 0.07 & 13.4 & 9.49 & 0.02 & 1
 \NN
 1a-02 & 171 & 36 & 1.18 & 0.08 & 9.8 & 9.33 & 0.03 & 1
 \NN
 1a-03 & 29 & 1 & 0.79 & 0.03 & 16.4 & 8.65 & 0.02 & 1
 \NN
 1a-05 & 254 & 5 & 1.47 & 0.08 & 9.9 & 9.37 & 0.02 & 1
 \NN
 1a-07 & 142 & 26 & 1.86 & 0.08 & 15.7 & 8.82 & 0.02 & 1
 \ML
 2s-02 & 120 & 3 & 2.77 & 0.08 & 23.5 & 9.68 & 0.01 & 1
 \NN
 2s-04 & 153 & 8 & 1.10 & 0.08 & 9.4 & 9.30 & 0.03 & 1
 \NN
 2s-05 & 129 & 11 & 2.50 & 0.06 & 23.1 & 9.65 & 0.01 & 1
 \NN
 2s-06 & 119 & 19 & 0.77 & 0.07 & 6.8 & 9.22 & 0.04 & 1
 \ML
 2a-01 & 154 & 6 & 1.65 & 0.07 & 13.9 & 9.49 & 0.02 & 1
 \NN
 2a-02 & 222 & 17 & 2.81 & 0.08 & 19.4 & 9.70 & 0.01 & 1
 \NN
 2a-03 & 155 & 4 & 1.31 & 0.08 & 9.9 & 9.45 & 0.03 & 1
 \NN
 2a-04 & 117 & 25 & 1.07 & 0.06 & 10.4 & 8.97 & 0.02 & 1
 \ML
 3s-01 & 66 & 8 & 0.83 & 0.05 & 12.2 & 9.11 & 0.03 & 1
 \NN
 3s-02 & 199 & 66 & 0.72 & 0.08 & 5.1 & 9.14 & 0.05 & 2
 \NN
 3s-03 & 69 & 15 & 2.26 & 0.09 & 25.7 & 9.23 & 0.02 & 1
 \NN
 3s-04 & 125 & 10 & 0.97 & 0.07 & 7.6 & 8.97 & 0.03 & 1
 \NN
 3s-05 & 176 & 9 & 1.19 & 0.08 & 8.7 & 9.31 & 0.03 & 1
 \NN
 3s-06 & 113 & 10 & 0.65 & 0.06 & 6.3 & 8.69 & 0.04 & 1
 \NN
 3s-07 & 206 & 7 & 1.90 & 0.09 & 12.2 & 9.65 & 0.02 & 1
 \ML
 3a-01 & 196 & 19 & 3.68 & 0.08 & 26.5 & 9.442 & 0.009 & 1
 \NN
 3a-02 & 296 & 3 & 3.02 & 0.09 & 18.1 & 9.79 & 0.01 & 1
 \LL
}
\end{ctable}
 \rem{label = {T:EIG_ALFALFA}}

\subsection{Star formation rate}
\label{s:rsltsSFR}


SFRs of the EIGs were calculated using the WO {\Halpha} measurements and the WISE {\WThree} and {\WFour} measurements.
First, the {\Halpha} flux and the WISE {\WThree} and {\WFour} apparent magnitudes were corrected for Galactic extinction as described in section \ref{s:ObsNPrc_AbsMagLum} (the corrections for WISE were quite small, up to {0.018\,\magnitude} in W3, and up to {0.013\,\magnitude} in W4). 
Then, the {\WThree} and {\WFour} Galactic corrected magnitudes were converted to fluxes using the procedure described in section IV.4.h.i.1 of \cite{2013wise.rept....1C}\rem{\footnote{http://wise2.ipac.caltech.edu/docs/release/allsky/expsup/sec4\_4h.html\#conv2flux}} for a constant power-law spectra (same as the method used by \citealt{2014MNRAS.438...97W}). 
The {\Halpha}, {\WThree} and {\WFour} fluxes were then converted to luminosities ($\Lunimosity_{\Halpha, obs}$, $\SpecLum \left( 12\,\um \right)$ and $\SpecLum \left( 22\,\um \right)$ respectively) as described in section \ref{s:ObsNPrc_AbsMagLum}.

\cite{2014MNRAS.438...97W} found that  dust extinction-corrected {\Halpha} flux, $\Lunimosity_{\Halpha, corr}$, can be accurately derived from the observed {\Halpha} flux and either the {\WThree} or {\WFour} bands using:

\begin{equation}
\begin{IEEEeqnarraybox*}{rCl} 
  \Lunimosity_{\Halpha, corr} & = & 
     \Lunimosity_{\Halpha, obs} + 0.038 \cdot \nu \SpecLum \left( 12\,\um \right) \\
  \Lunimosity_{\Halpha, corr} & = & 
     \Lunimosity_{\Halpha, obs} + 0.035 \cdot \nu \SpecLum \left( 22\,\um \right)
\end{IEEEeqnarraybox*} 
\label{e:SFR_HaWISE_crct} 
\end{equation}

These relations are independent of the metallicity. The relation that uses $\SpecLum \left( 12\,\um \right)$ was found to be slightly more accurate.

The corrected {\Halpha} luminosity, $\Lunimosity_{\Halpha, corr}$, was calculated using equation \eqref{e:SFR_HaWISE_crct} for both {\WThree} and {\WFour}, and the average of the two results was used. If for one of the bands only a lower limit to the magnitude was given, the result of the other band was used. If both bands had only lower limits to their magnitudes, the nominal SFR was calculated assuming zero IR dust emission \markChange{(i.e., with $\SpecLum \left( 12\,\um \right) = \SpecLum \left( 22\,\um \right) = 0$)} and the effect of the possible IR dust emission was added to the uncertainty of the measurement \markChange{($\Lunimosity_{\Halpha, corr}$ was calculated using both options of \eqref{e:SFR_HaWISE_crct} with uncertainties in $\SpecLum$ calculated using the 0.95 confidence lower limit of the {\WThree} and {\WFour} fluxes, and the option with the lower uncertainty was used)}.

Finally, the dust-extinction-corrected {\Halpha} luminosity, $\Lunimosity_{\Halpha, corr}$, was used to calculate the SFR using the following equation adapted from \cite{2012ARA&A..50..531K}, \cite{2011ApJ...737...67M} and \cite{2011ApJ...741..124H}:

\begin{equation}
  \log\, \frac{SFR}{\MsunPerYr} = \Lunimosity_{\Halpha, corr} - 41.27
  \label{e:SFR_GenCalibKen2009} 
\end{equation}

\vspace{12pt}

Table \ref{T:EIG_StarFormation} lists measured star formation properties of the sample galaxies. For each EIG, the equivalent width, EW, and the {\Halpha} flux, $F_{\Halpha}$, are listed. These are followed by the WISE magnitudes, {\magWThree} and {\magWFour}, used for calculating the {\Halpha} flux that was extinguished within the EIG. Listed finally, are the fraction of the total {\Halpha} flux extinguished within the EIG, $\mbox{frac}_{{\Halpha}, \mbox{\scriptsize ext}}$, and the SFR.

\begin{ctable} 
[
  caption = {Star formation measurements: EIG name, equivalent width (EW), {\Halpha} flux ($F_{\Halpha}$), WISE magnitudes ({\magWThree} and {\magWFour}), fraction of the total {\Halpha} flux extinguished within the EIG ($\mbox{frac}_{{\Halpha}, \mbox{\scriptsize ext}}$) and the SFR.},
  cap     = {Star formation measurements},
  doinside = \small,
  star,
  label   = {T:EIG_StarFormation}
]
{ @{}cr@{$\,\pm\,$}rr@{$\,\pm\:$}rr@{$\,\pm\:$}lr@{$\,\pm\:$}ll@{$\,\pm\:$}ll@{$\,\pm\:$}l@{}
}
{
  \tnote[a]
  {
    `$<$' and `$>$' indicate 0.95 confidence level upper and lower limits (respectively).
  }
}
{
 \FL
 \multirow{2}*{EIG}
  & \multicolumn {2}{c}{EW}
  & \multicolumn {2}{c}{$F_{\Halpha}$}
  & \multicolumn {2}{c}{\multirow{2}*{{\magWThree}\tmark[a]}}
  & \multicolumn {2}{c}{\multirow{2}*{{\magWFour}\tmark[a]}}
  & \multicolumn {2}{c}{\multirow{2}*{$\mbox{frac}_{{\Halpha}, \mbox{\scriptsize ext}}$\tmark[a]}}
  & \multicolumn {2}{c}{SFR}
 \NN
  & \multicolumn {2}{c}{$\left[ \Angst \right]$}
  & \multicolumn {2}{c}{$\left[ \ergcms \right]$}
  & \multicolumn {2}{c}{}
  & \multicolumn {2}{c}{}
  & \multicolumn {2}{c}{}
  & \multicolumn {2}{c}{$\left[ \MsunPerYr \right]$}
 \ML
 1s-01 & 50 & 5 & (4.0 & 0.9)\,$10^{-14}$ & \multicolumn {2}{c}{$>$11.9} & \multicolumn {2}{c}{$>$8.4} & \multicolumn {2}{c}{$<$0.12} & 0.16 & 0.04
 \NN
 1s-02 & 40 & 2 & (6.3 & 1.3)\,$10^{-14}$ & 9.9 & 0.3 & 7.4 & 0.4 & 0.3 & 0.1 & 0.45 & 0.07
 \NN
 1s-03 & 27 & 3 & (4.4 & 1.0)\,$10^{-14}$ & 10.25 & 0.05 & 7.9 & 0.2 & 0.34 & 0.05 & 0.38 & 0.06
 \NN
 1s-04 & 26 & 4 & (1.0 & 0.3)\,$10^{-14}$ & \multicolumn {2}{c}{$>$12.0} & \multicolumn {2}{c}{$>$9.0} & \multicolumn {2}{c}{$<$0.43} & 0.05 & 0.01
 \NN
 1s-06 & 39 & 5 & (3.0 & 0.7)\,$10^{-14}$ & \multicolumn {2}{c}{$>$11.6} & \multicolumn {2}{c}{$>$8.1} & \multicolumn {2}{c}{$<$0.23} & 0.14 & 0.04
 \NN
 1s-07 & 40 & 12 & (1.7 & 0.6)\,$10^{-14}$ & 11.4 & 0.4 & \multicolumn {2}{c}{$>$8.4} & 0.3 & 0.1 & 0.11 & 0.03
 \NN
 1s-08 & 137 & 16 & (3.9 & 0.9)\,$10^{-14}$ & 11.4 & 0.4 & \multicolumn {2}{c}{$>$7.9} & 0.16 & 0.07 & 0.20 & 0.04
 \NN
 1s-09 & 43 & 6 & (2.4 & 0.6)\,$10^{-14}$ & \multicolumn {2}{c}{$>$12.4} & \multicolumn {2}{c}{$>$9.0} & \multicolumn {2}{c}{$<$0.13} & 0.08 & 0.02
 \NN
 1s-10 & 75 & 3 & (1.9 & 0.1)\,$10^{-14}$ & 12.3 & 0.5 & 8.6 & 0.5 & 0.16 & 0.09 & 0.063 & 0.007
 \NN
 1s-11 & 28 & 4 & (2.0 & 0.7)\,$10^{-14}$ & \multicolumn {2}{c}{$>$12.6} & \multicolumn {2}{c}{$>$9.0} & \multicolumn {2}{c}{$<$0.13} & 0.08 & 0.03
 \NN
 1s-12 & 170 & 24 & (3.2 & 0.2)\,$10^{-14}$ & \multicolumn {2}{c}{$>$12.1} & \multicolumn {2}{c}{$>$9.1} & \multicolumn {2}{c}{$<$0.13} & 0.13 & 0.01
 \NN
 1s-13 & 48 & 9 & (1.3 & 0.2)\,$10^{-14}$ & \multicolumn {2}{c}{$>$12.3} & \multicolumn {2}{c}{$>$9.0} & \multicolumn {2}{c}{$<$0.27} & 0.07 & 0.01
 \NN
 1s-14 & 34 & 7 & (8.2 & 2.5)\,$10^{-14}$ & 9.3 & 0.3 & 7.3 & 0.4 & 0.4 & 0.1 & 0.9 & 0.2
 \ML
 1a-03 & 39 & 10 & (1.6 & 0.5)\,$10^{-14}$ & \multicolumn {2}{c}{$>$12.0} & \multicolumn {2}{c}{$>$8.9} & \multicolumn {2}{c}{$<$0.24} & 0.03 & 0.01
 \NN
 1a-04 & 4.6 & 0.4 & (13.8 & 1.7)\,$10^{-14}$ & 9.21 & 0.09 & 7.5 & 0.4 & 0.27 & 0.04 & 0.95 & 0.09
 \ML
 2s-01 & 51 & 4 & (1.5 & 0.1)\,$10^{-14}$ & \multicolumn {2}{c}{$>$12.2} & \multicolumn {2}{c}{$>$8.8} & \multicolumn {2}{c}{$<$0.26} & 0.070 & 0.008
 \NN
 2s-02 & 23 & 5 & (1.4 & 0.4)\,$10^{-14}$ & \multicolumn {2}{c}{$>$11.7} & \multicolumn {2}{c}{$>$8.2} & \multicolumn {2}{c}{$<$0.40} & 0.07 & 0.02
 \NN
 2s-05 & 51 & 3 & (12.2 & 2.6)\,$10^{-14}$ & 10.5 & 0.3 & 8.7 & 0.4 & 0.11 & 0.03 & 0.7 & 0.1
 \NN
 2s-06 & 58 & 3 & (7.1 & 0.5)\,$10^{-14}$ & 9.5 & 0.3 & 6.3 & 0.4 & 0.4 & 0.2 & 0.8 & 0.1
 \NN
 2s-07 & 58 & 4 & (3.1 & 0.2)\,$10^{-14}$ & 9.3 & 0.3 & 5.9 & 0.4 & 0.7 & 0.4 & 0.23 & 0.07
 \NN
 2s-08 & 460 & 39 & (11.4 & 2.5)\,$10^{-14}$ & 11.3 & 0.4 & 7.6 & 0.4 & 0.07 & 0.04 & 0.30 & 0.06
 \ML
 2a-01 & 13.1 & 0.7 & (4.2 & 0.5)\,$10^{-14}$ & 8.44 & 0.01 & 6.52 & 0.06 & 0.72 & 0.07 & 0.94 & 0.06
 \NN
 2a-02 & 19 & 1 & (1.7 & 0.3)\,$10^{-14}$ & 11.3 & 0.4 & \multicolumn {2}{c}{$>$7.9} & 0.3 & 0.1 & 0.16 & 0.02
 \ML
 3s-01 & 37 & 6 & (1.4 & 0.4)\,$10^{-14}$ & 12.6 & 0.6 & 9.2 & 0.6 & 0.2 & 0.1 & 0.08 & 0.02
 \NN
 3s-02 & 28 & 4 & (1.3 & 0.3)\,$10^{-14}$ & \multicolumn {2}{c}{$>$12.2} & \multicolumn {2}{c}{$>$8.8} & \multicolumn {2}{c}{$<$0.28} & 0.07 & 0.02
 \NN
 3s-03 & 71 & 3 & (13.4 & 1.5)\,$10^{-14}$ & 11.1 & 0.4 & 8.1 & 0.4 & 0.08 & 0.03 & 0.33 & 0.04
 \NN
 3s-04 & 73 & 14 & (0.5 & 0.2)\,$10^{-14}$ & \multicolumn {2}{c}{$>$12.6} & \multicolumn {2}{c}{$>$9.1} & \multicolumn {2}{c}{$<$0.46} & 0.016 & 0.005
 \NN
 3s-05 & 35 & 3 & (4.4 & 0.9)\,$10^{-14}$ & 11.2 & 0.4 & \multicolumn {2}{c}{$>$8.6} & 0.18 & 0.07 & 0.27 & 0.05
 \NN
 3s-06 & 26 & 5 & (0.9 & 0.2)\,$10^{-14}$ & \multicolumn {2}{c}{$>$12.7} & \multicolumn {2}{c}{$>$9.1} & \multicolumn {2}{c}{$<$0.27} & 0.021 & 0.004
 \NN
 3s-07 & 35 & 7 & (1.6 & 0.5)\,$10^{-14}$ & \multicolumn {2}{c}{$>$12.3} & \multicolumn {2}{c}{$>$9.3} & \multicolumn {2}{c}{$<$0.20} & 0.12 & 0.04
 \ML
 3a-01 & 48 & 2 & (13.9 & 0.9)\,$10^{-14}$ & 12.2 & 0.5 & \multicolumn {2}{c}{$>$8.9} & 0.03 & 0.01 & 0.33 & 0.02
 \NN
 3a-02 & 33 & 2 & (6.8 & 1.0)\,$10^{-14}$ & 10.06 & 0.03 & 8.1 & 0.2 & 0.25 & 0.03 & 0.67 & 0.07
 \LL
}
\end{ctable}
 \rem{label = {T:EIG_StarFormation}}


The {\Halpha} flux and EW were measured for each region of interest within the EIGs (plotted in figures \ref{f:RHacolorImgEIG-1}, \ref{f:RHacolorImgEIG-2} and \ref{f:RHacolorImgEIG-3}).  
Table \ref{T:EIG_PlgnHaFlux} lists for each region the {\Halpha} flux as a fraction of the total EIGs {\Halpha} flux.
Table \ref{T:EIG_PlgnEW} lists the EWs of the regions.

The regions of interest were defined with some additional area around the star-forming regions so that all the {\Halpha} flux will be measured. Therefore, the EWs of star-forming regions may be larger than those listed in the Table \ref{T:EIG_PlgnEW}. Similarly, the {\Halpha} flux fractions of the star-forming regions may be smaller than those listed in Table \ref{T:EIG_PlgnHaFlux}, since some diffuse {\Halpha} flux from the surrounding area may have been included in the measurement.

\begin{ctable} 
[
  caption = {Fraction of the EIG's total {\Halpha} flux within regions of interest \markChange{shown in the last column of figures \ref{f:RHacolorImgEIG-1}, \ref{f:RHacolorImgEIG-2} and \ref{f:RHacolorImgEIG-3}}},
  cap     = {Fraction of the EIG's total {\Halpha} flux within regions of interest},
  doinside = \small,
  star,
  label   = {T:EIG_PlgnHaFlux}
]
{ @{}cl@{$\,\pm\,$}ll@{$\,\pm\,$}ll@{$\,\pm\,$}ll@{$\,\pm\,$}ll@{$\,\pm\,$}l@{}
}
{
}
{
 \FL
 EIG & \multicolumn {2}{c}{a} & \multicolumn {2}{c}{b} & \multicolumn {2}{c}{c}
     & \multicolumn {2}{c}{d} & \multicolumn {2}{c}{e}
 \ML
 1s-01 & 0.13 & 0.04 & 0.12 & 0.04 &  \multicolumn {2}{c}{-}  &  \multicolumn {2}{c}{-}  &  \multicolumn {2}{c}{-} 
 \NN
 1s-03 & 0.11 & 0.04 & 0.13 & 0.05 & 0.13 & 0.04 &  \multicolumn {2}{c}{-}  &  \multicolumn {2}{c}{-} 
 \NN
 1s-06 & 0.13 & 0.04 &  \multicolumn {2}{c}{-}  &  \multicolumn {2}{c}{-}  &  \multicolumn {2}{c}{-}  &  \multicolumn {2}{c}{-} 
 \NN
 1s-14 & 0.5 & 0.2 & 0.4 & 0.1 & 0.07 & 0.03 &  \multicolumn {2}{c}{-}  &  \multicolumn {2}{c}{-} 
 \ML
 1a-04 & 0.6 & 0.1 &  \multicolumn {2}{c}{-}  &  \multicolumn {2}{c}{-}  &  \multicolumn {2}{c}{-}  &  \multicolumn {2}{c}{-} 
 \ML
 2s-01 & 0.27 & 0.03 & 0.28 & 0.05 &  \multicolumn {2}{c}{-}  &  \multicolumn {2}{c}{-}  &  \multicolumn {2}{c}{-} 
 \NN
 2s-05 & 0.07 & 0.02 & 0.06 & 0.02 & 0.04 & 0.01 & 0.13 & 0.04 & 0.10 & 0.03
 \NN
 2s-07 & 0.90 & 0.09 &  \multicolumn {2}{c}{-}  &  \multicolumn {2}{c}{-}  &  \multicolumn {2}{c}{-}  &  \multicolumn {2}{c}{-} 
 \ML
 2a-01 & 0.22 & 0.04 & 0.6 & 0.1 &  \multicolumn {2}{c}{-}  &  \multicolumn {2}{c}{-}  &  \multicolumn {2}{c}{-} 
 \NN
 2a-02 & 0.33 & 0.07 & 0.15 & 0.04 &  \multicolumn {2}{c}{-}  &  \multicolumn {2}{c}{-}  &  \multicolumn {2}{c}{-} 
 \ML
 3s-02 & 0.21 & 0.07 & 0.11 & 0.04 &  \multicolumn {2}{c}{-}  &  \multicolumn {2}{c}{-}  &  \multicolumn {2}{c}{-} 
 \NN
 3s-03 & 0.32 & 0.05 & 0.29 & 0.05 &  \multicolumn {2}{c}{-}  &  \multicolumn {2}{c}{-}  &  \multicolumn {2}{c}{-} 
 \NN
 3s-05 & 0.3 & 0.1 & 0.20 & 0.06 &  \multicolumn {2}{c}{-}  &  \multicolumn {2}{c}{-}  &  \multicolumn {2}{c}{-} 
 \NN
 3s-07 & 0.20 & 0.08 & 0.3 & 0.1 &  \multicolumn {2}{c}{-}  &  \multicolumn {2}{c}{-}  &  \multicolumn {2}{c}{-} 
 \ML
 3a-01 & 0.18 & 0.02 & 0.16 & 0.02 & 0.19 & 0.06 &  \multicolumn {2}{c}{-}  &  \multicolumn {2}{c}{-} 
 \LL
}
\end{ctable}
 \rem{label = {T:EIG_PlgnHaFlux}}
\begin{ctable} 
[
  caption = {EW [\Angst] within regions of interest \markChange{shown in the last column of figures \ref{f:RHacolorImgEIG-1}, \ref{f:RHacolorImgEIG-2} and \ref{f:RHacolorImgEIG-3}}},
  cap     = {},
  doinside = \small,
  star,
  label   = {T:EIG_PlgnEW}
]
{ @{}cr@{$\,\pm\,$}rr@{$\,\pm\,$}rr@{$\,\pm\,$}rr@{$\,\pm\,$}rr@{$\,\pm\,$}r@{}
}
{
}
{
 \FL
 EIG & \multicolumn {2}{c}{a} & \multicolumn {2}{c}{b} & \multicolumn {2}{c}{c}
     & \multicolumn {2}{c}{d} & \multicolumn {2}{c}{e}
 \ML
 1s-01 & 71 & 6 & 45 & 3 &  \multicolumn {2}{c}{-}  &  \multicolumn {2}{c}{-}  &  \multicolumn {2}{c}{-} 
 \NN
 1s-03 & 31 & 2 & 23 & 1 & 30 & 3 &  \multicolumn {2}{c}{-}  &  \multicolumn {2}{c}{-} 
 \NN
 1s-06 & 45 & 4 &  \multicolumn {2}{c}{-}  &  \multicolumn {2}{c}{-}  &  \multicolumn {2}{c}{-}  &  \multicolumn {2}{c}{-} 
 \NN
 1s-14 & 30 & 5 & 49 & 5 & 26 & 3 &  \multicolumn {2}{c}{-}  &  \multicolumn {2}{c}{-} 
 \ML
 1a-04 & 3.2 & 0.3 &  \multicolumn {2}{c}{-}  &  \multicolumn {2}{c}{-}  &  \multicolumn {2}{c}{-}  &  \multicolumn {2}{c}{-} 
 \ML
 2s-01 & 67 & 7 & 112 & 17 &  \multicolumn {2}{c}{-}  &  \multicolumn {2}{c}{-}  &  \multicolumn {2}{c}{-} 
 \NN
 2s-05 & 100 & 6 & 70 & 4 & 48 & 2 & 61 & 2 & 104 & 4
 \NN
 2s-07 & 73 & 3 &  \multicolumn {2}{c}{-}  &  \multicolumn {2}{c}{-}  &  \multicolumn {2}{c}{-}  &  \multicolumn {2}{c}{-} 
 \ML
 2a-01 & 17 & 1 & 11.8 & 0.5 &  \multicolumn {2}{c}{-}  &  \multicolumn {2}{c}{-}  &  \multicolumn {2}{c}{-} 
 \NN
 2a-02 & 18 & 1 & 21 & 2 &  \multicolumn {2}{c}{-}  &  \multicolumn {2}{c}{-}  &  \multicolumn {2}{c}{-} 
 \ML
 3s-02 & 119 & 28 & 22 & 5 &  \multicolumn {2}{c}{-}  &  \multicolumn {2}{c}{-}  &  \multicolumn {2}{c}{-} 
 \NN
 3s-03 & 75 & 2 & 172 & 6 &  \multicolumn {2}{c}{-}  &  \multicolumn {2}{c}{-}  &  \multicolumn {2}{c}{-} 
 \NN
 3s-05 & 39 & 2 & 38 & 3 &  \multicolumn {2}{c}{-}  &  \multicolumn {2}{c}{-}  &  \multicolumn {2}{c}{-} 
 \NN
 3s-07 & 56 & 15 & 26 & 3 &  \multicolumn {2}{c}{-}  &  \multicolumn {2}{c}{-}  &  \multicolumn {2}{c}{-} 
 \ML
 3a-01 & 26 & 1 & 37 & 1 & 140 & 5 &  \multicolumn {2}{c}{-}  &  \multicolumn {2}{c}{-} 
 \LL
}
\end{ctable}
     \rem{label = {T:EIG_PlgnEW}}

As can be seen, the {\Halpha} flux fractions of the distinct star-forming regions of EIGs do not account for the entire {\Halpha} flux. In most EIGs the diffuse {\Halpha} is a significant component.
In many of the EIGs (e.g., 1S-13) there are no detectable star-forming regions at all. In some of these the diffused {\Halpha} component is not detectable in the {\nHa} images, even though the total {\Halpha} EW, shown in table \ref{T:EIG_StarFormation}, is considerable. This is a result of noise in the {\nHa} images, which was reduced in the EW measurements by averaging over the whole galaxy. This noise is significantly less considerable in the {\R} images, because the spectral width of the {\R} filter is more than an order of magnitude larger than the {\nHa} EW of the EIGs for which {\Halpha} emission is not easily detectable in the {\nHa} images. 

The fact that diffused {\Halpha} is the dominant component in most EIGs indicates that a possible active galactic nucleus (AGN) contribution to the {\Halpha} flux is insignificant. This is supported by the fact that SDSS did not classify any of the EIGs' measured spectra as containing detectable AGN emission lines.
Since the {\Halpha} measurements utilize narrow bands they include a contribution of the [NII] lines flanking the {\Halpha} line. For the central parts of the 22 EIGs which have SDSS spectra this correction was found to be: $\log \left( F_{\Halpha} / F_{\Halpha+[NII]} \right) = -0.07 \pm 0.05$ (with 0.95 confidence level). The correction factor for a whole galaxy is expected to be significantly lower than this value, since central parts of galaxies typically have high metal abundance and high [NII] to {\Halpha} flux ratio compared to those measured for whole galaxies. This difference in the correction factor is expected to be more significant for EIGs in which the diffused {\Halpha} is the dominant component. In light of this we chose not to correct for these effects.

\subsection{Model fitting}
\label{sec:ResultsModel}

The SFH, dust attenuation and stellar mass of the EIGs were estimated by fitting a five-parameter model to their UV-to-near-IR spectral energy distributions (SEDs). The model assumes that the SFH can be described by a first population of stars with an exponentially decreasing or increasing star formation rate (SFR), and a possible addition to recent star formation (a second population). The second population is modelled as an instantaneous star formation that occurred $1\,\Myr$ ago and is meant to compensate for a possible recent deviation from an exponential SFH, which may have a significant effect on the emission from the galaxy (especially in the UV and {\Halpha}).
The model can also describe a scenario of a constant star formation or a sudden star formation (first population) with or without a recent star formation burst (second population). 
This model is obviously a simplification of the actual SFH and is limited in the diversity of possible scenarios. However, not much more can be done given the available SED measured points.

\vspace{12pt}
The five free parameters of the model are:
\begin{description}
  \item[$\Mass_{1}$] - The mass of the first population of stars, created along the history of the galaxy (integral over time of the SFR of the first population). 

  \item[$Age_{1}$] - The look-back time of the beginning of star formation of the first population.

  \item[$\tau$] - The e-folding time of the exponentially decreasing (positive $\tau$) or increasing (negative $\tau$) SFH of the first population. $\tau \ll Age_{1}$ indicates a sudden star formation, while $\tau \gg  Age_{1}$ indicates an almost constant star formation.

  \item[\EBV] - The {\BMinV} colour excess that results from dust within the galaxy.

  \item[$\Mass_{2}$] - The mass of the second population of stars that was created.

  \item[]
\end{description}

The model fitting procedure used Bayesian statistical inference with uniform prior probability distributions of $\log \left( \Mass_{1} \right)$, $\log \left( Age_{1} \right)$, $\tau^{-1}$, {\EBV} and $\log \left( \Mass_{2} \right)$. The parameters were restricted to the following ranges:
$10^{7}\,\Msun < \Mass_{1} < 1.6 \cdot 10^{15}\,\Msun$, \; 
$10^{8}\,\yr < Age_{1} < $ age of the Universe at the redshift of the galaxy, \; 
$0 < \EBV < 2$, \;
$2.7\,\Msun < \Mass_{2} < 10^{7}\,\Msun$.
The metallicities of the first and second populations were set to {0.4\,\Zsun} and {1\,\Zsun} respectively.

The model fitting was computed using the GalMC software \citep{IAU:8669162, 2011ApJ...737...47A}\footnote{http://ctp.citytech.cuny.edu/$\sim$vacquaviva/web/GalMC.html}.
GalMC is a Markov Chain Monte Carlo (MCMC) algorithm designed to fit the SEDs of galaxies to infer physical properties such as age, stellar mass, dust reddening, metallicity, redshift, and star formation rate.
The Markov chains produced by GalMC were analysed using the GetDist software, a part of the CosmoMC software \citep{2002PhRvD..66j3511L}\footnote{http://cosmologist.info/cosmomc/}.

The stellar emission was calculated using the Charlot \& Bruzual 2007 stellar population synthesis model (Charlot \& Bruzual, private communication, CB07\footnote{http://www.bruzual.org/}) assuming a Salpeter initial mass function.
Nebular emission was calculated following \cite{1998ApJ...497..618S} and \cite{2009A&A...502..423S}, as described in section 2.2.4 of \cite{2011ApJ...737...47A}.
Dust extinction within the EIG was calculated from the {\EBV} parameter using the \cite{1994ApJ...429..582C}\rem{Calzetti} law with $\R_{v} = 4.05$ \citep{2000ApJ...533..682C}.
The emission of the EIGs were also corrected for absorption by neutral hydrogen in the intergalactic
medium (IGM) using \cite{1995ApJ...441...18M}.

The input to the model-fitting software included the EIG's redshift and the data from measured bands with wavelengths shorter than {3\,\um} (the CB07 model does not estimate correctly the emission beyond the first PAH feature at {$\sim$3\,\um}). Each band measurement was corrected for Galactic extinction and then converted to flux, which was used as input to the GalMC software.
The calibrated Bessell U, B, V, R and I magnitudes, measured at the Wise Observatory, were translated to AB magnitudes (and then to flux) using the relations listed in Table 1 of \cite{2007AJ....133..734B}.
The 2MASS magnitudes were converted to fluxes using the zero magnitude isophotal monochromatic
intensities listed in Table 2 of \cite{2003AJ....126.1090C}.
Foreground Galactic extinction was corrected as described in section \ref{s:ObsNPrc_AbsMagLum}.

For each EIG four MCMC runs were made, each with a different set of free parameters as a starting point. Best-fitting parameters and covariance matrices of these four runs were then used as inputs for continued runs. Each run included 50000 sampled parameter sets.
Only one of each EIG's MCMC runs (chains) was used for analysis. This run was chosen based on the speed of convergence of the chain, on its average likelihood, on its multiplicity (number of trial steps before moving to the next location in parameter space), and on how well it covered the parameter space. Chains that probed the parameter space with $Age_{1} <0.2\,\Gyr$ for a large fraction of their length were disfavoured (if another good chain existed it was selected instead of them).

Models were not fitted to EIG 1s-05 and EIG 2s-04, because these do not have the necessary SED data. EIG 1s-05 has only {21\,cm} data, and EIG 2s-04 has only {\SDSSg}, {\SDSSr}, {\SDSSi} and {\SDSSz} measurements (due to a bright foreground star close to it).
The model that was fitted to EIG 1s-11 did not reproduce its {\Halpha} emission successfully ($\mbox{EW} = 28 \pm 4\,\Angst$). The best-fitting parameters of all MCMC runs of EIG 1s-11 yielded lower EW values.

\vspace{12pt}

Marginalized posterior distributions (the predicted probability distribution functions, PDFs) of the free parameters and of the calculated total mass of stars, {\Mstar}, considering mass-loss mechanisms, were calculated from the selected chain of each EIG.
Figure \ref{f:ResGalMC_1D} in Appendix \ref{App:ModelledEIG_Properties} shows, for each of the modelled EIGs, the marginalized posterior distributions of $Age_{1}$, $\tau^{-1}$, \EBV, $\Mass_{2}$ and {\Mstar}.
Table \ref{T:EIG_MStar} lists the {\Mstar} values predicted by the model.

\begin{ctable} 
[
  caption = {Modelled Stellar Mass},
  cap     = {},
  doinside = \small,
  star,
  label   = {T:EIG_MStar}
]
{ @{}cc@{$\,\pm\,$}clcc@{$\,\pm\,$}clcc@{$\,\pm\,$}clcc@{$\,\pm\,$}c@{}
}
{
}
{
 \FL
 {EIG} 
 & \multicolumn {2}{c}{$\log\frac{\Mstar}{\Msun}$} & {\qquad}
 & {EIG}   
 & \multicolumn {2}{c}{$\log\frac{\Mstar}{\Msun}$} & {\qquad}
 & {EIG}   
 & \multicolumn {2}{c}{$\log\frac{\Mstar}{\Msun}$} & {\qquad}
 & {EIG}   
 & \multicolumn {2}{c}{$\log\frac{\Mstar}{\Msun}$} 
 \NN
 \cmidrule(r){1-3}\cmidrule(rl){5-7}\cmidrule(l){9-11}\cmidrule(l){13-15}
 1s-01 & 8.9 & 0.1 && 1s-12 & 8.0 & 0.1 && 2s-05 & 9.7 & 0.2 && 3s-03 & 8.7 & 0.1
 \NN
 1s-02 & 9.3 & 0.1 && 1s-13 & 8.3 & 0.1 && 2s-06 & 9.3 & 0.3 && 3s-04 & 7.4 & 0.1
 \NN
 1s-03 & 9.5 & 0.2 && 1s-14 & 9.7 & 0.2 && 2s-07 & 8.52 & 0.05 && 3s-05 & 9.3 & 0.2
 \NN
 1s-04 & 8.6 & 0.1 && 1a-01 & 9.6 & 0.2 && 2s-08 & 7.3 & 0.1 && 3s-06 & 8.20 & 0.08
 \NN
 1s-06 & 9.0 & 0.1 && 1a-02 & 9.1 & 0.2 && 2a-01 & 10.5 & 0.2 && 3s-07 & 8.9 & 0.2
 \NN
 1s-07 & 9.0 & 0.2 && 1a-03 & 8.3 & 0.1 && 2a-02 & 9.7 & 0.2 && 3a-01 & 9.3 & 0.1
 \NN
 1s-08 & 8.5 & 0.1 && 1a-04 & 11.40 & 0.07 && 2a-03 & 9.7 & 0.2 && 3a-02 & 10.1 & 0.2
 \NN
 1s-09 & 8.41 & 0.07 && 1a-07 & 8.1 & 0.1 && 2a-04 & 8.3 & 0.2 &&  & \multicolumn {2}{c}{}
 \NN
 1s-10 & 8.05 & 0.08 && 2s-01 & 8.6 & 0.2 && 3s-01 & 8.8 & 0.2 &&  & \multicolumn {2}{c}{}
 \NN
 1s-11 & 9.1 & 0.2 && 2s-02 & 9.0 & 0.2 && 3s-02 & 8.67 & 0.06 &&  & \multicolumn {2}{c}{}
 \LL
}
\end{ctable}
 \rem{label = {T:EIG_MStar}}

Two-dimensional marginalized PDFs of pairs of the model parameters were also analysed. It was found that in most cases there is some dependence between pairs of the free parameters. The $\Mass_{1}$ and $Age_{1}$ parameters were found to be highly correlated.
The $\left( Age_{1}, \Mass_{1} \right)$ two-dimensional marginalized PDFs do not seem to depend on whether they are part of the EIG-1, EIG-2 or EIG-3 subsample, i.e. the $\left( Age_{1}, \Mass_{1} \right)$ space is filled similarly by the \markChange{three subsamples}.

\subsection{Dynamic mass}

We estimated dynamic mass lower limits for the EIGs using the ALFALFA measured HI rotation, the elliptical isophotes fitted to the combined {\R} images (described in section \ref{sec:Wise_Photometry}) and the surface brightness measurements, $\mu_{\mbox{\scriptsize R}}$. The calculations were based on the methods described by \cite{2014RvMP...86...47C}.
First, we estimated the inclination of the galaxies using eq.~6 of \cite{2014RvMP...86...47C} for the $\mu_{\mbox{\scriptsize R}} = 24\,\magAsecSq$ elliptical isophote:

\begin{equation}
  i \cong \cos ^{-1} \sqrt{\frac{\left( b_{24}/a_{24} \right)^2 - q_0^2}{1 - q_0^2}}
  \label{e:rsltsInclination} 
\end{equation}
 where:
 \begin{description}
  \item[$i$] is the estimated inclination of the galaxy;
  \item[$a_{24}$, $b_{24}$] are the semi-major axis and semi-minor axis (respectively) at $\mu_{\mbox{\scriptsize R}} = 24\,\magAsecSq$;
  \item[$q_0$] is the axial ratio of a galaxy viewed edge on.
  \item[]
 \end{description} 

The inclination of EIGs classified as early-types was not measured, because their $q_0$ is unknown.
For a sample of 13482 spiral galaxies \cite{1992MNRAS.258..404L} found $q_0 \cong 0.2$ by applying statistical techniques to explore triaxial models. \cite{2012MNRAS.425.2741H} found $q_0 \cong 0.13$ for spirals using SDSS data on a sample of 871 edge-on galaxies. 
Here we adopted $q_0 = 0.17 \pm 0.05$ for the galaxies classified as late-types or `unknown' (see table \ref{T:Res_Morphology}).

we measured $a_{24}$ using the linear fit of figure \ref{f:RSurBrightR}. 
The semi-minor to semi-major axes ratio, $b_{24}/a_{24}$, was calculated by linear interpolation of its values for the EIG's ellipse isophotes just below and just above $a_{24}$.
The speed of rotation of the HI gas, $v_{rot}$, was calculated using the HI velocity width, {\Whalf}, listed in Table \ref{T:EIG_ALFALFA} and the inclination, $i$, using: 

\begin{equation}
  v_{rot} = \frac{\Whalf}{2 \cdot \sin i}
  \label{e:v_rot}
\end{equation}

The dynamic mass lower limit, $M_{dyn,24}$, was then calculated using:

\begin{equation}
  \Mass_{dyn,24} = \frac{v_{rot}^2 \cdot a_{24}}{G}
\end{equation}

$M_{dyn,24}$ is a lower limit to the galactic total mass, since the extent of the neutral gas in spiral galaxies can often exceed twice that of the stars \citep{2014RvMP...86...47C}, and the dark matter (DM) halo may extend even further.
An additional source of uncertainty in $M_{dyn,24}$ comes from the assumption behind \eqref{e:v_rot} that all of the HI velocity width, {\Whalf}, is due to the rotational velocity, $v_{rot}$. This may somewhat increase the $M_{dyn,24}$ estimate, but probably by much less than it is decreased due to the underestimation of the dark mass diameter.
Table \ref{T:EIG_DynamicMass} lists $\Mass_{dyn,24}$ of EIGs along with the values used for its calculation. It also lists the ratio of the measured dynamic mass to stellar plus HI mass, $\Mass_{dyn,24} / \left( \Mstar + \MHI \right)$.

\begin{ctable} 
[
  caption = {Dynamic mass},
  cap     = {Dynamic mass},
  doinside = \small,
  star,
  label   = {T:EIG_DynamicMass}
]
{ @{}cr@{$\,\pm\,$}lr@{$\,\pm\:$}lr@{$\,\pm\,$}rr@{$\,\pm\,$}rr@{$\,\pm\,$}lr@{$\,\pm\,$}l@{}
}
{
  \tnote[a]
  {
    Due to stars/galaxies projected by chance on EIG 1a-03 and interfering with the
    measurement $b_{24}/a_{24}$ was measured on the third largest
    ellipse isophote ($\mu_{\mbox{\scriptsize R}} = 24.93 \pm 0.08\,\magAsecSq$).
  }
}
{
 \FL
 \multirow{2}*{EIG} 
 & \multicolumn {2}{c}{$a_{24}$} 
 & \multicolumn {2}{c}{\multirow{2}*{$b_{24}/a_{24}$}} 
 & \multicolumn {2}{c}{$i$}      
 & \multicolumn {2}{c}{$v_{rot}$} 
 & \multicolumn {2}{c}{\multirow{2}*{$\log \frac{\Mass_{dyn,24}}{\Msun}$}} 
 & \multicolumn {2}{c}{\multirow{2}*{$\frac{\Mass_{dyn,24}}{\Mstar + \MHI}$}} \NN
      & \multicolumn {2}{c}{$\left[ \kpc \right]$} 
 & \multicolumn {2}{c}{} 
 & \multicolumn {2}{c}{$[deg]$}     
 & \multicolumn {2}{c}{$\left[ \kms \right]$} 
 & \multicolumn {2}{c}{} 
 & \multicolumn {2}{c}{} 
 \ML
 1s-01 & 6.4 & 0.4 & 0.46 & 0.02 & 64 & 2 & 98 & 4 & 10.15 & 0.04 & 3.8 & 0.4
 \NN
 1s-03 & 12.2 & 0.4 & 0.231 & 0.009 & 81 & 3 & 139 & 2 & 10.74 & 0.02 & 6.0 & 0.5
 \NN
 1s-04 & 4.6 & 0.4 & 0.57 & 0.03 & 56 & 2 & 96 & 3 & 9.99 & 0.04 & 3.1 & 0.3
 \NN
 1s-06 & 6.8 & 0.5 & 0.48 & 0.02 & 63 & 2 & 113 & 5 & 10.31 & 0.05 & 5.2 & 0.7
 \NN
 1s-07 & 2.69 & 0.09 & 0.92 & 0.08 & 24 & 11 & 191 & 79 & 10.4 & 0.4 & 10 & 8
 \NN
 1s-10 & 2.18 & 0.08 & 0.95 & 0.04 & 19 & 7 & 166 & 57 & 10.1 & 0.3 & 16 & 11
 \NN
 1s-14 & 8.3 & 0.3 & 0.81 & 0.04 & 37 & 4 & 54 & 17 & 9.7 & 0.3 & 0.9 & 0.6
 \ML
 \;\;1a-03 \tmark[a] & 2.6 & 0.2 & 0.90 & 0.05 & 27 & 7 & 32 & 7 & 8.8 & 0.2 & 0.8 & 0.4
 \ML
 2s-02 & 5.6 & 0.4 & 0.85 & 0.07 & 33 & 8 & 111 & 22 & 10.2 & 0.2 & 2.2 & 0.9
 \NN
 2s-05 & 7.6 & 0.2 & 0.82 & 0.02 & 35 & 2 & 112 & 10 & 10.35 & 0.08 & 2.1 & 0.5
 \NN
 2s-06 & 6.7 & 0.4 & 0.58 & 0.02 & 56 & 1 & 72 & 11 & 9.9 & 0.1 & 1.8 & 0.7
 \ML
 2a-01 & 13 & 1 & 0.30 & 0.02 & 75 & 2 & 80 & 3 & 10.27 & 0.05 & 0.6 & 0.2
 \NN
 2a-02 & 11 & 2 & 0.29 & 0.01 & 76 & 2 & 114 & 9 & 10.53 & 0.09 & 3.0 & 0.7
 \ML
 3s-02 & 5.5 & 0.5 & 0.56 & 0.02 & 57 & 2 & 119 & 39 & 10.3 & 0.3 & 8 & 5
 \NN
 3s-03 & 3.5 & 0.2 & 0.78 & 0.02 & 39 & 2 & 55 & 12 & 9.4 & 0.2 & 0.9 & 0.4
 \NN
 3s-04 & 1.6 & 0.1 & 0.49 & 0.02 & 62 & 2 & 71 & 6 & 9.27 & 0.08 & 1.5 & 0.3
 \NN
 3s-05 & 10.0 & 0.3 & 0.24 & 0.01 & 80 & 3 & 89 & 5 & 10.27 & 0.05 & 4.1 & 0.6
 \NN
 3s-06 & 3.1 & 0.1 & 0.43 & 0.02 & 67 & 2 & 62 & 5 & 9.44 & 0.08 & 3.4 & 0.7
 \NN
 3s-07 & 9 & 1 & 0.24 & 0.02 & 80 & 3 & 105 & 4 & 10.37 & 0.05 & 3.5 & 0.5
 \ML
 3a-01 & 8.7 & 0.3 & 0.217 & 0.006 & 82 & 4 & 99 & 10 & 10.30 & 0.09 & 3.6 & 0.7
 \NN
 3a-02 & 9.9 & 0.4 & 0.367 & 0.009 & 71 & 2 & 157 & 2 & 10.75 & 0.02 & 2.9 & 0.4
 \LL
}
\end{ctable}
 \rem{label = {T:EIG_DynamicMass}}

\section{Analysis}
\label{ch:Analysis}

\subsection{Colours of the EIGs}
%
%

Colour-mass and colour-colour diagrams of large scale surveys show that galaxies tend to populate two main regions, the `blue cloud' of star-forming galaxies and the `red sequence' of quiescent galaxies, with a small fraction of galaxies in a `green valley' range in between \citep{2001AJ....122.1861S, 2006MNRAS.373..469B, 2014MNRAS.440..889S}.
Star-forming main sequence galaxies populate the blue cloud, whether their star formation started recently or a long time ago. When star formation is quenched, galaxies leave the main sequence and their changing colours can be interpreted  as a reflection of the quenching process \citep{2014MNRAS.440..889S}.

\begin{figure*}
\begin{centering}
\includegraphics[width=14cm,trim=0mm 0mm 0mm 0, clip]{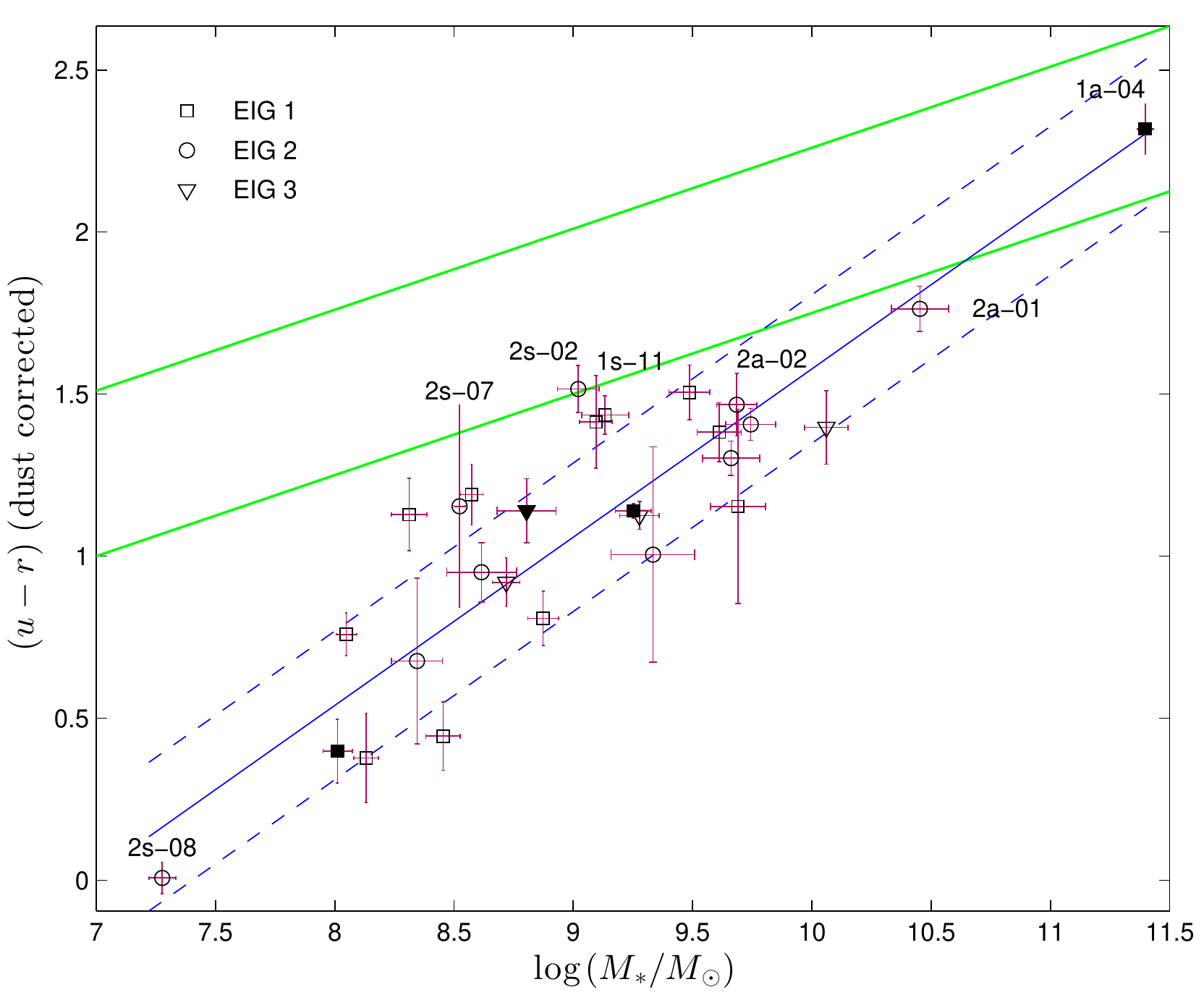}
\caption [Colour to stellar mass diagram]
{
  Colour-to-stellar mass diagram. {\uMinr}, corrected for both Galactic extinction and dust extinction within the EIG, as function of stellar mass, {\Mstar}. The green thick lines show the limits of the `green valley', based on equations (1) and (2) of \cite{2014MNRAS.440..889S}. The thin blue line shows a linear fit to the EIG data. The dashed blue lines show the $\pm 1\sigma$ deviation from this fit.
  Filled symbols indicate EIGs classified as early-types. \label{f:Res_ColorMass}
}
\end{centering}
\end{figure*}

Figure \ref{f:Res_ColorMass} shows a {\uMinr} to {\Mstar} colour-mass diagram of the EIGs, with the approximate edges of the `green valley' marked in green bold lines. The {\uMinr} colour was corrected both for Galactic extinction (as described in section \ref{s:ObsNPrc_AbsMagLum}), and for dust within the EIG using the \cite{2000ApJ...533..682C} extinction law (with $R'_{V} = 4.05$). 

It is evident from Figure \ref{f:Res_ColorMass} that most EIGs are `blue cloud' galaxies. There are no EIGs in the `red sequence' and only one that is certainly within the `green valley' (EIG 1a-04). Based on comparison between the measurements and the `green valley' limit shown in Figure \ref{f:Res_ColorMass}, we can conclude with 0.95 confidence that the probability for an EIG to be in the `blue cloud' is $>$0.76. The probability for an EIG to be in the `red sequence' is $<$0.12.

A linear relation was fitted to the measured EIG points of Figure \ref{f:Res_ColorMass} (the thin blue line in the figure): $\uMinr_{\mbox{\scriptsize dust corrected}} = 0.52 \cdot \log \left( \Mstar / \Msun \right) - 3.61$ . The expected standard deviation in {\uMinr} around this fit is {0.23\,\magnitude} (marked in the figure by dashed blue lines).
The 0.52 slope of this fit is significantly larger than the slope of the `green valley' limits, 0.25 \citep{2014MNRAS.440..889S}. It can be concluded from this that the EIGs will be closer to the `red sequence', the higher their stellar mass, {\Mstar}, is. EIGs with stellar mass smaller than $10^{(10.6 \pm 0.9)}\,\Msun$ are typically `blue cloud' galaxies.

A similar colour-mass relation probably holds also for less isolated galaxies, as indicated by the results of \cite{2012A&A...540A..47F} who studied the AMIGA sample. They have measured colour-luminosity correlation in different morphological subtypes, and found that the more massive spirals show redder colours, and that there is little evidence for `green valley' galaxies in the AMIGA sample.

\vspace{12pt}

\begin{figure*}
\begin{centering}
\includegraphics[width=14cm,trim=0mm 0mm 0mm 0, clip]{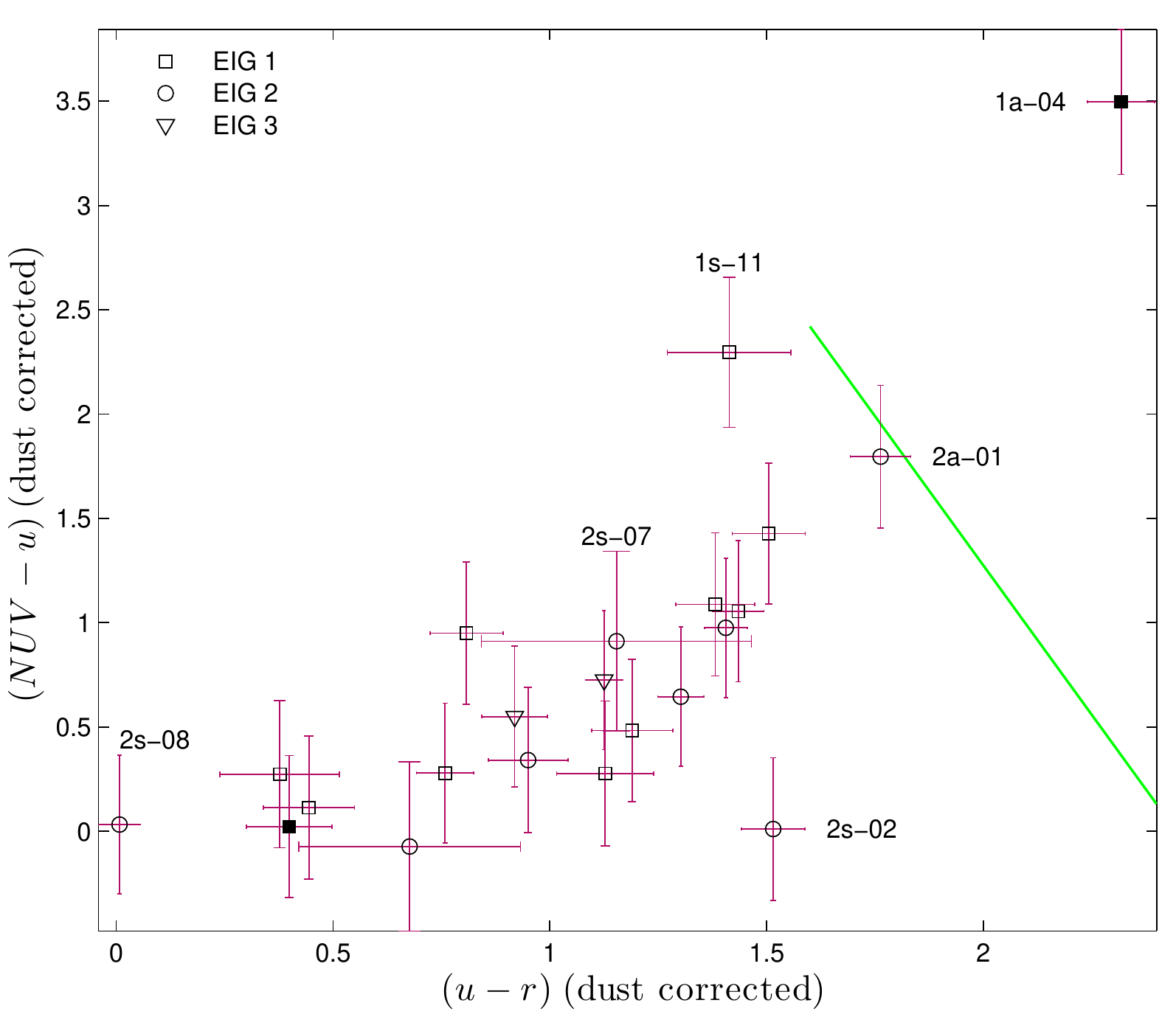}
\caption [{\NUVMinu} vs.~{\uMinr} colour-colour diagram of the EIG]
{
  Dust corrected {\NUVMinu} vs.~{\uMinr} colour-colour diagram of the EIG. The green line shows the approximated limit between the `blue cloud' and the `green valley', based on Fig.~7 of \cite{2014MNRAS.440..889S} (the `blue cloud' is to the left of the line, and the `green valley' is to the right).
  Filled symbols indicate EIGs classified as early-types. \label{f:Res_NUVu_ur}
}
\end{centering}
\end{figure*}

Figure \ref{f:Res_NUVu_ur} shows an {\NUVMinu} vs.~{\uMinr} colour-colour diagram of the EIGs, with the approximate limit between the `blue cloud' and the `green valley' marked with a green line (the `blue cloud' is to the left of the line, and the `green valley' is to the right). These colours were corrected for Galactic extinction and dust within the EIG, as in Figure \ref{f:Res_ColorMass}.
Other than distinguishing between blue and red galaxies, this diagram is useful for diagnosing how rapidly star formation quenches in `green valley' galaxies. The faster the star formation quenching is, the higher the {\NUVMinu} colour of galaxies would be as their {\uMinr} is gradually increased \citep[Fig.~7]{2014MNRAS.440..889S}.

The `green valley' galaxy, EIG 1a-04, is significantly redder in both colours, ${\uMinr} = 2.32 \pm 0.08$ and ${\NUVMinu} = 3.5 \pm 0.3$, compared to the other EIGs as Figure \ref{f:Res_NUVu_ur} shows.
A comparison to the simulated SFH scenarios from Fig.~7 of \cite{2014MNRAS.440..889S} shows that EIG 1a-04 fits a scenario of a galaxy that passed, more than {1\,\Gyr} ago, a rapid star formation quenching (with e-folding time significantly shorter than {1\,\Gyr}). The model fitted in this work (see Figure \ref{f:ResGalMC_1D}, page \pageref{f:ResGalMC_1D_2}, row 8) matches this scenario.
Another evidence supporting this scenario is that EIG 1a-04 was not detected by ALFALFA. This indicates that it might have lost most of its HI content which, given its extremely high stellar mass, perhaps was once very high.
It should be noted in this context that EIG 1a-04 is possibly not an extremely isolated galaxy (a possible false positive) as discussed in Appendix \ref{App:EIGdata}. Its {\Halpha} images show that LEDA 213033, a galaxy separated by {$\sim$2\arcmin} from it, may be a neighbour less than {300\,\kms} away.

Other EIGs with less than 0.90 confidence of being in the `blue cloud' are 2s-02, 2s-07, and 1s-11 (calculated from data shown in Figure \ref{f:Res_ColorMass}). From Figure \ref{f:Res_NUVu_ur} it is evident that EIG 1s-11 (as well as EIG 2a-01) deviate towards the red section. EIG 2s-02 is somewhat redder in {\uMinr} or bluer in {\NUVMinu} than the bulk. EIG 2s-07 seems to be well within the bulk of EIGs in Figure \ref{f:Res_NUVu_ur}, which indicates that it is a regular `blue cloud' galaxy (data of Figure \ref{f:Res_ColorMass} indicates 0.77 probability that it is in the `blue cloud').

\subsection{Comparison to the main sequence of star-forming galaxies}
%
%

Most star-forming galaxies exhibit a tight, nearly linear correlation between galaxy stellar mass and SFR (on a log-log scale; \citealt{2007ApJ...660L..43N}). This correlation is termed `the main sequence of star-forming galaxies' (or simply `the main sequence'). Up to redshifts \z$\sim$2 the correlation changes only in its normalization \citep{2012ApJ...752...66L, 2014MNRAS.440..889S}. 
Models of \cite{2010ApJ...718.1001B} and \cite{2013ApJ...772..119L} suggest that the main sequence is a result of an equilibrium between galaxy inflows and outflows. 

For a specific range of redshift, \z, and stellar mass, \Mstar, the main sequence can be expressed as:

\begin{equation}
  \log \left( \frac{\SFR}{\MsunPerYr} \right) = \alpha \cdot \log \left( \frac{\Mstar}{\Msun} \right) + \beta
  \label{e:Intr_MS} 
\end{equation}
 where:
 \begin{description}
  \item[$\alpha$, $\beta$] are the free parameters fitted to the observed data.
  \item[]
 \end{description}

The $\alpha$ and $\beta$ parameters somewhat vary with redshift. \cite{2007ApJS..173..267S} and \cite{2012ApJ...756..113H} indicated that below $\log \left( \Mstar \right) \sim 9.5$ the slope, $\alpha$, of the main sequence increases. \cite{2012ApJ...756..113H} have studied a sample of local Universe ALFALFA galaxies with SDSS and GALEX photometry and have found the following main sequence relation:

\begin{equation}
\begin{IEEEeqnarraybox*}{rCl} 
  \alpha & = &
    \begin{cases} 
       0.851 & \text{ for\quad} \log \left( \Mstar / \Msun \right) \leq 9.5  \\
       0.241 & \text{ for\quad} \log \left( \Mstar / \Msun \right) > 9.5
     \end{cases} \\
  \vspace{6pt} \\
  \beta & = &
    \begin{cases} 
       -8.207  & \text{ for\quad} \log \left( \Mstar / \Msun \right) \leq 9.5  \\
       -2.402  & \text{ for\quad} \log \left( \Mstar / \Msun \right) > 9.5
     \end{cases}
  \label{e:Intr_MS_ALFALFA}\rem{based on equation. 8 of \cite{2012ApJ...756..113H}}
\end{IEEEeqnarraybox*} 
\end{equation}

\begin{figure*}
\begin{centering}
\includegraphics[width=13.8cm,trim=0mm 0mm 0mm 0, clip]{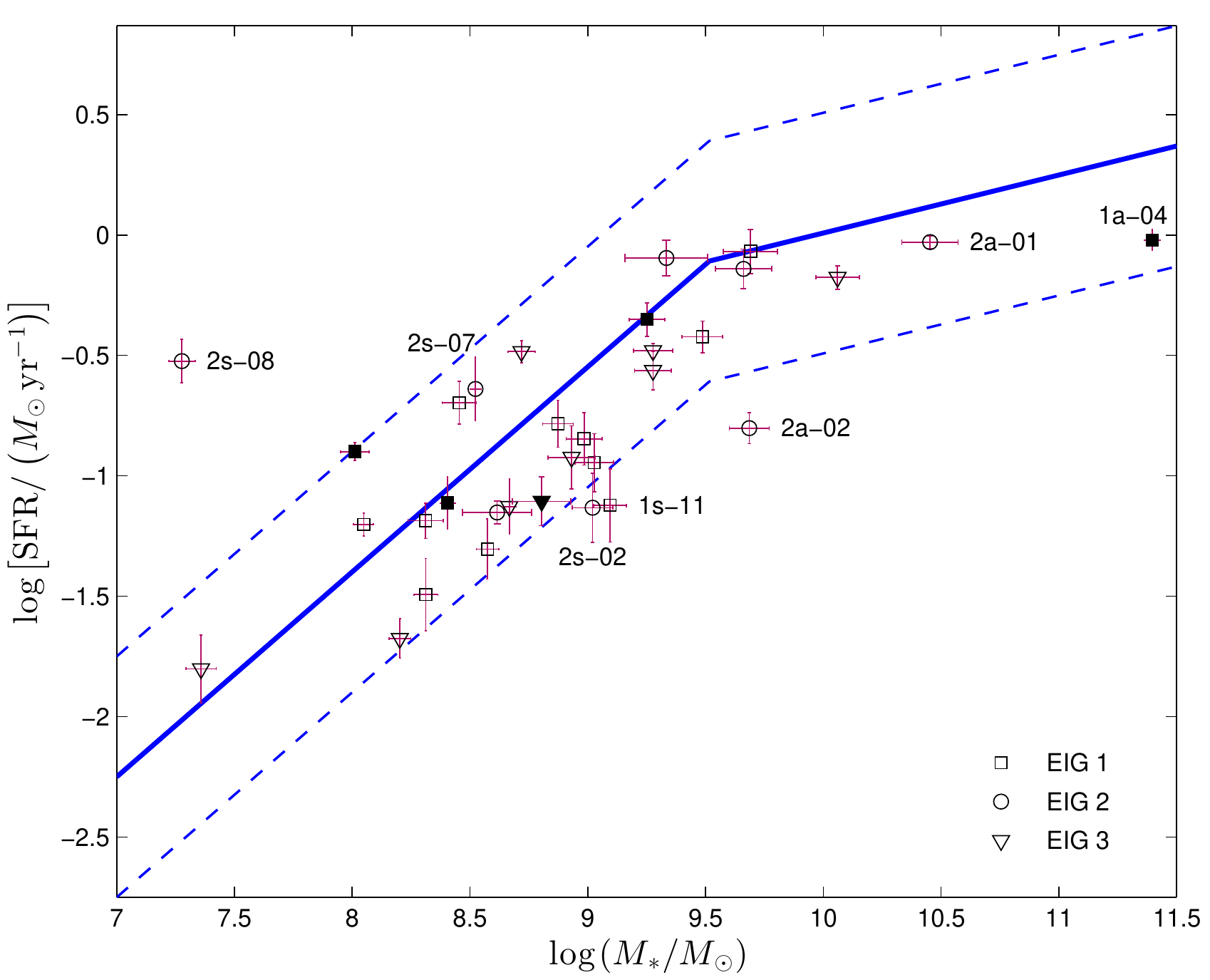}
\includegraphics[width=13.8cm,trim=0mm 0mm 0mm 0, clip]{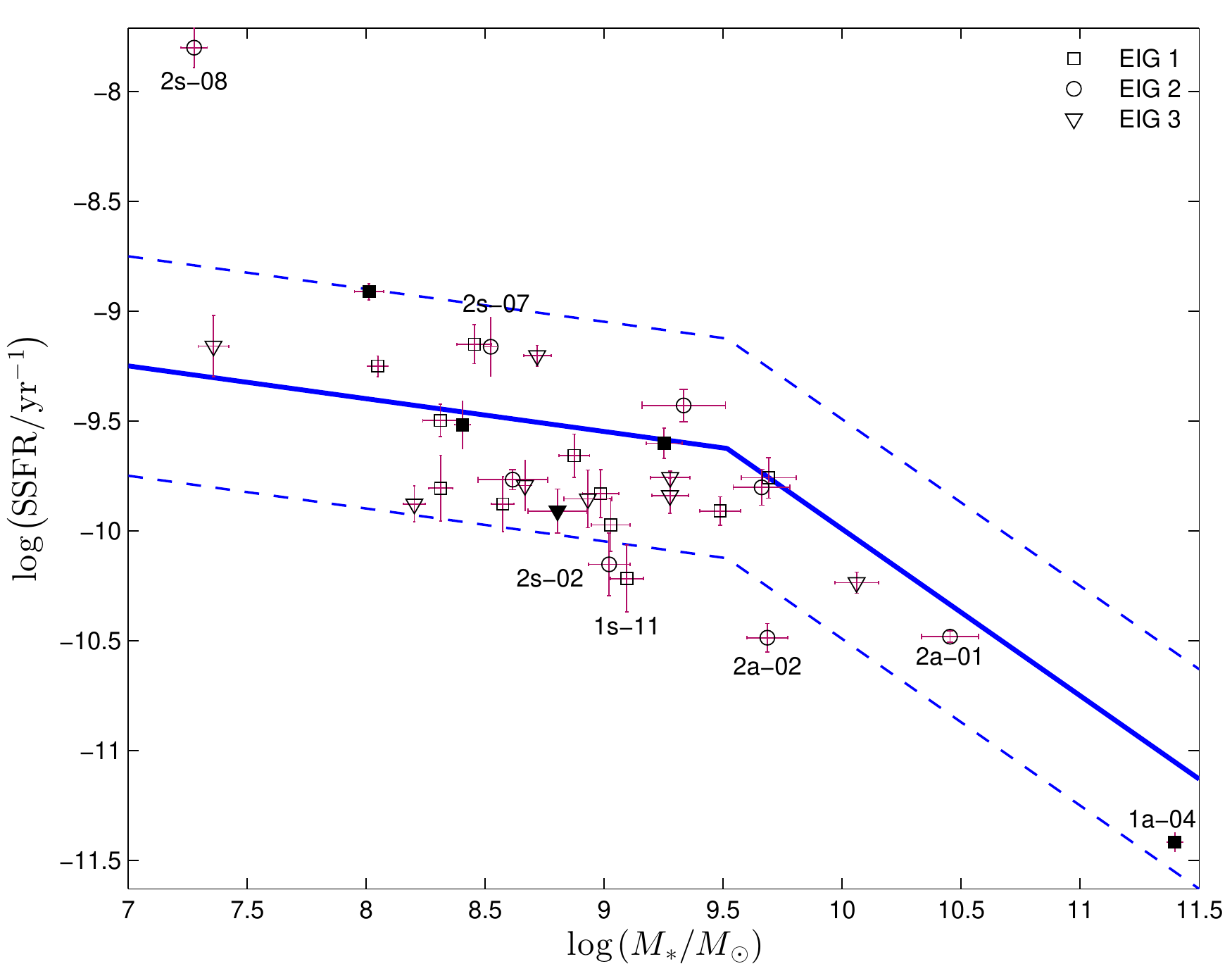}
\caption [EIGs compared to the main sequence of star-forming galaxies]
{
  EIGs compared to the main sequence of star-forming galaxies \markChange{({\SFR} vs. {\Mstar} in the top chart and {\SSFR} vs. {\Mstar} in the bottom chart)}. The solid blue lines are the best fit main sequence found by \cite{2012ApJ...756..113H} and described by \eqref{e:Intr_MS_ALFALFA}. The dashed blue lines indicate deviations of $\pm\,0.5\,\dex$ from the main sequence (a typical $1\,\sigma$ deviation of main sequence fits).
  Filled symbols indicate EIGs classified as early-types. \label{f:Res_MainSeq}
}
\end{centering}
\end{figure*}

Figure \ref{f:Res_MainSeq} shows the current SFR (upper plot) and SSFR (lower plot) of the EIGs as function of stellar mass, \Mstar, compared to the `main sequence' of \cite{2012ApJ...756..113H}.
It indicates that, in general, EIGs fit the `main sequence of star-forming galaxies'. A fraction of $0.88^{+0.08}_{-0.16}$ of the EIGs fit the `main sequence' to within $\pm 0.5\,\dex$ (assuming that whether an EIG deviates by more than $\pm 0.5\,\dex$ or not follows a binomial distribution, and using a Wilson score interval with 0.95 confidence level). 
On average, the SFR of the EIGs is $0.1\,\dex$ lower than the main sequence, with a standard deviation of $0.4\,\dex$. This deviation from the main sequence is similar for all EIG sub-samples (1, 2 and 3).
It may indicate a tendency of the EIGs to have slightly lower SFRs compared to main sequence galaxies. However, it could also result from differences between the SFR and {\Mstar} estimation methods used by \cite{2012ApJ...756..113H} and the ones used here.
\markChange{\cite{2012ApJ...756..113H} derived stellar masses and SFRs by SED fitting of the seven
GALEX and SDSS bands\rem{ (as described in section 4.1 of \citealt{2012AJ....143..133H})}, while in this work the {\Halpha} emission line and 2MASS data were also used when available for the SED fitting. Furthermore, in this work the SED fitting was used only for {\Mstar} estimation. The SFR was estimated based on {\Halpha} and WISE fluxes.}

The EIGs that deviate by more than $0.5\,\dex$ in {\SFR} from the main sequence are EIG 1s-11 ($-0.7 \pm 0.2\,\dex$), EIG 2s-02 ($-0.6 \pm 0.1\,\dex$), EIG 2s-08 ($1.49 \pm 0.09\,\dex$) and EIG 2a-02 ($-0.73 \pm 0.06\,\dex$).
EIG 2s-08 is, therefore, the only EIG known to deviate significantly from the main sequence. It has the highest {\SSFR} of all the measured EIGs, $\log \left( \SSFR/ \yr^{-1} \right) = -7.80  \pm 0.09$, as well as the highest {\Halpha} EW ($460 \pm 39\,\Angst$). It also has the lowest stellar mass, $\log \left( \Mstar / \Msun \right) = 7.28  \pm  0.06$, as well as the lowest model estimated age, $\log(Age_{1}/\yr) =  8.4  \pm  0.2$.
Based on this we conclude (with 0.95 confidence) that the probability for an EIG to have SFR that deviates significantly from the main sequence is $<0.16$.

For the LOG catalogue of isolated galaxies \cite{2013AstBu..68..243K} measured {\SSFR} values as function of {\Mstar} lower than those measured here for the EIGs and lower than the main sequence as measured by \cite{2012ApJ...756..113H}. This may be a result of their different method of estimating {\Mstar}. \cite{2013AstBu..68..243K} used {\Ks} band measurements assuming a {\Ks} luminosity to stellar mass ratio as that of the Sun, as opposed to fitting a model to measurements in several bands as was done here and by \cite{2012ApJ...756..113H}.
They also observed that almost all of the LOG galaxies have $\log \left( \SSFR / \yr^{-1} \right)$ lower than $-9.4$ \citep[Fig. 4]{2013AstBu..68..243K}. As the lower plot of Figure \ref{f:Res_MainSeq} indicates, the $\log \left( \SSFR / \yr^{-1} \right)$ limit we measured for the EIGs is $-8.9$ with the exception of EIG 2s-08 that is above this value.

\subsection{Mass histograms}
\label{s:Res_MassHisgrms}

\begin{figure*}
\begin{centering}
\includegraphics[width=14cm,trim=0mm 0mm 0mm 0, clip]{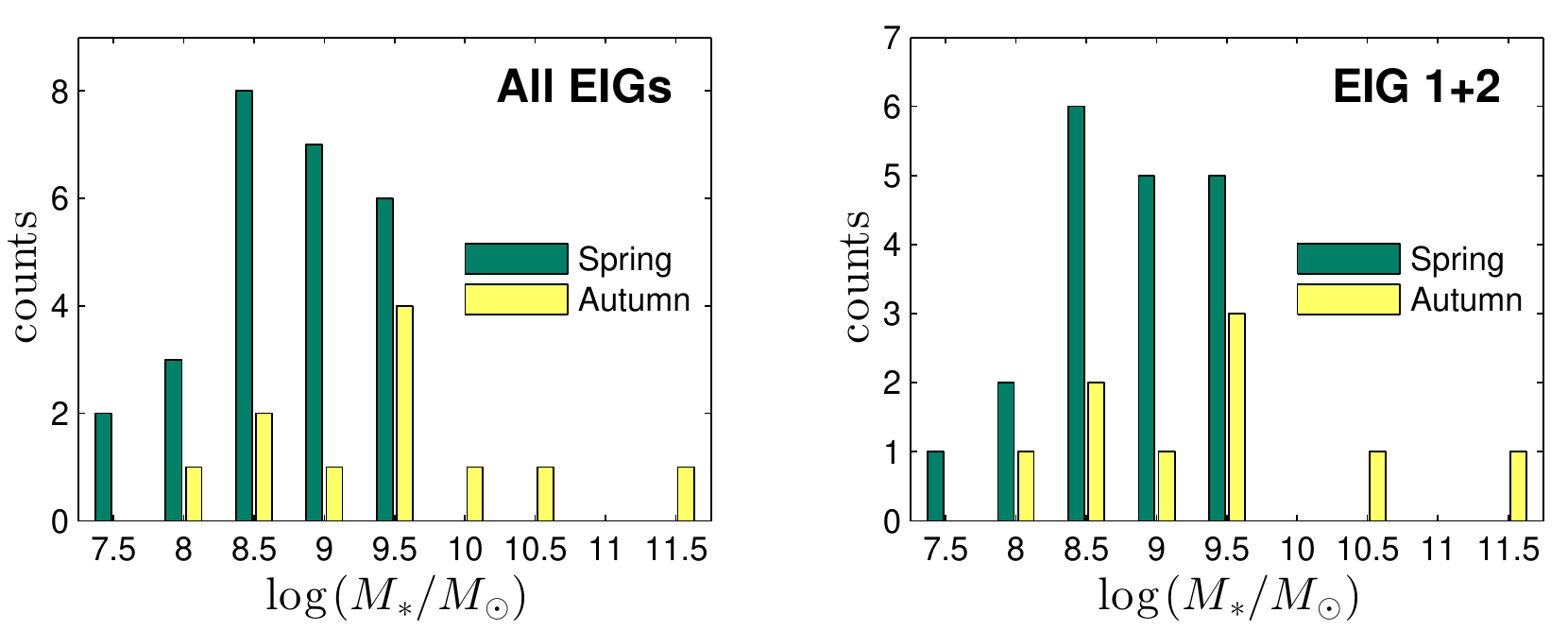}
\caption [Stellar mass histograms for EIGs]
{
  Stellar mass, {\Mstar}, histograms for Spring and Autumn EIGs. The left chart shows the histograms for all the EIGs. The right chart shows the histograms of subsamples EIG-1 and EIG-2. \label{f:Res_MstarHist}
}
\end{centering}
\end{figure*}

\begin{figure*}
\begin{centering}
\includegraphics[width=14cm,trim=0mm 0mm 0mm 0, clip]{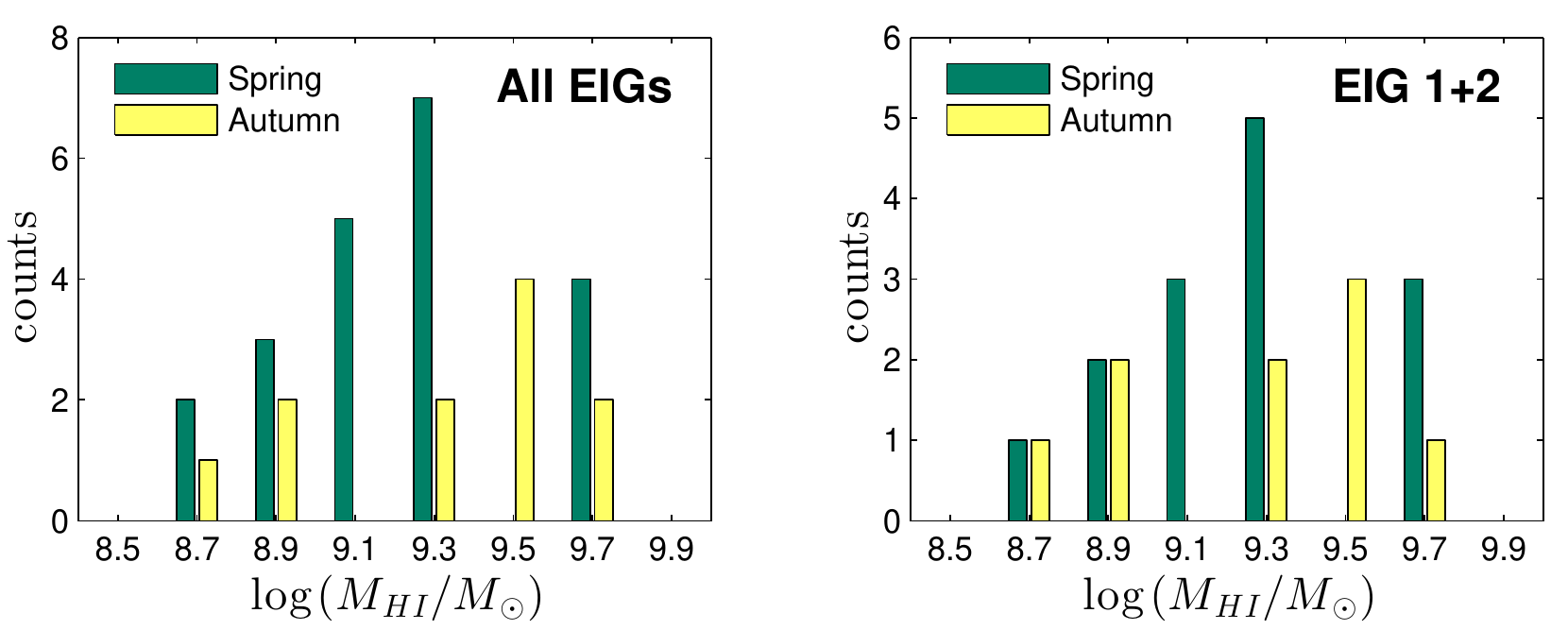}
\caption [HI mass histograms for EIGs]
{
  HI mass, {\MHI}, histograms for Spring and Autumn EIGs. The left chart shows the histograms for all the EIGs. The right chart shows the histograms of subsamples EIG-1 and EIG-2. \label{f:Res_M_HI_Hist}
}
\end{centering}
\end{figure*}

\begin{figure*}
\begin{centering}
\includegraphics[width=14cm,trim=0mm 0mm 0mm 0, clip]{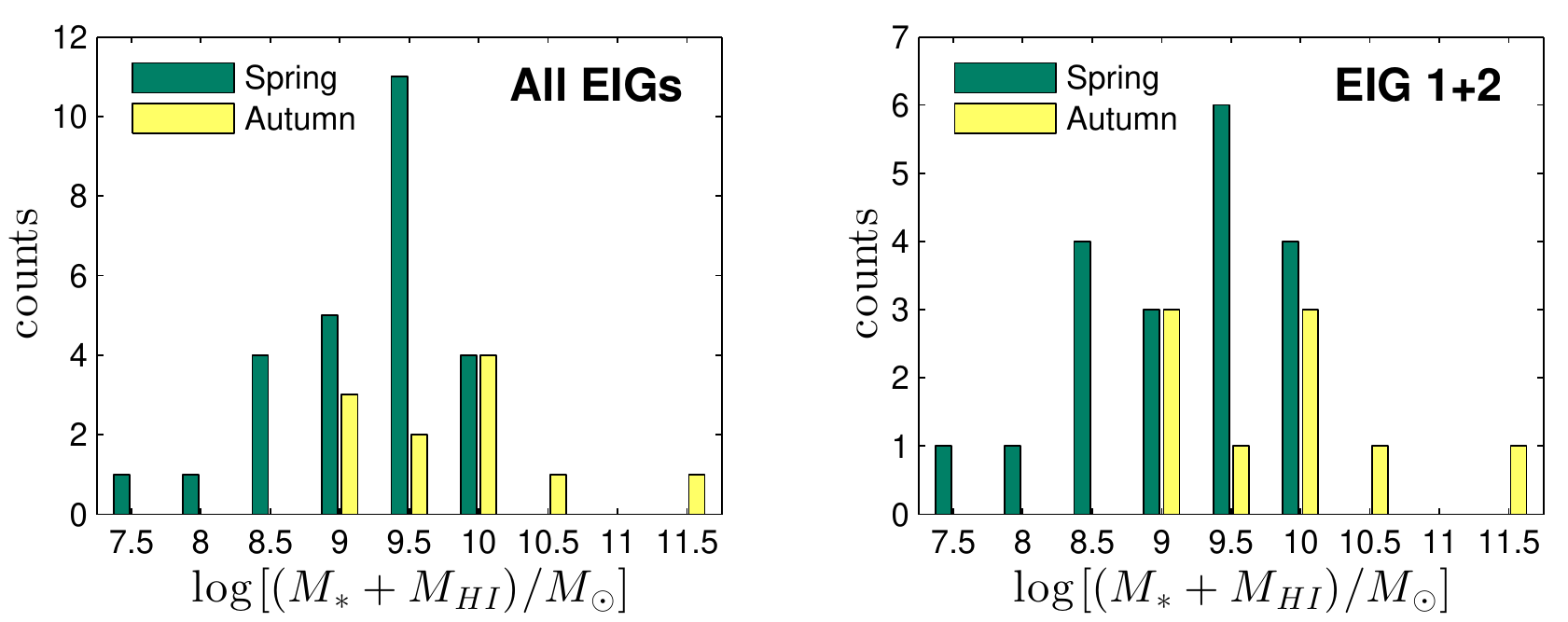}
\caption [Stellar plus HI mass histograms for EIGs]
{
  Stellar plus HI mass, $\left( \Mstar + \MHI \right)$, histograms for Spring and Autumn EIGs. The left chart shows the histograms for all the EIGs. The right chart shows the histograms of subsamples EIG-1 and EIG-2. \label{f:Res_MStarHI_Hist}
}
\end{centering}
\end{figure*}

\begin{figure*}
\begin{centering}
\includegraphics[width=14cm,trim=0mm 0mm 0mm 0, clip]{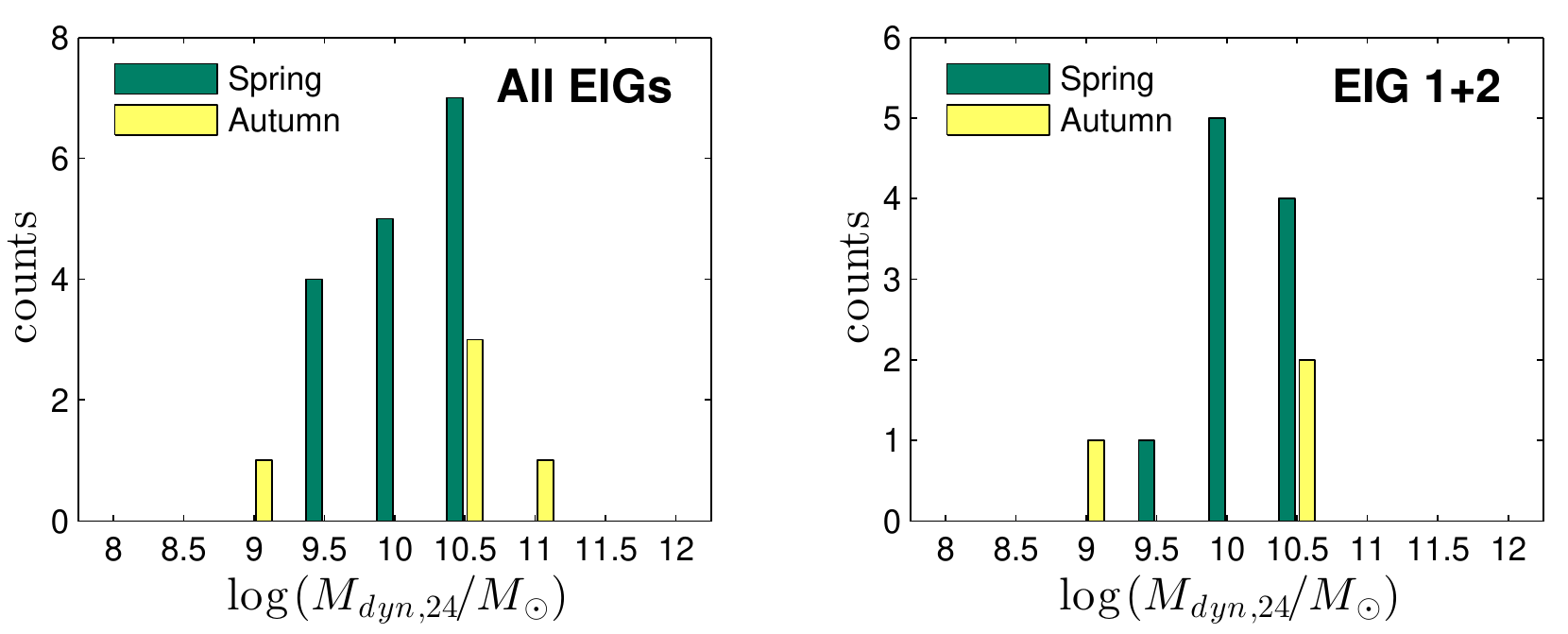}
\caption [Dynamic mass histograms for EIGs]
{
  $\Mass_{dyn,24}$ histograms for Spring and Autumn EIGs. The left chart shows the histograms for all the EIGs. The right chart shows the histograms of subsamples EIG-1 and EIG-2. \label{f:Res_Mdyn24Hist}
}
\end{centering}
\end{figure*}

\begin{figure*}
\begin{centering}
\includegraphics[width=6.8cm,trim=0mm 0mm 0mm 0, clip]{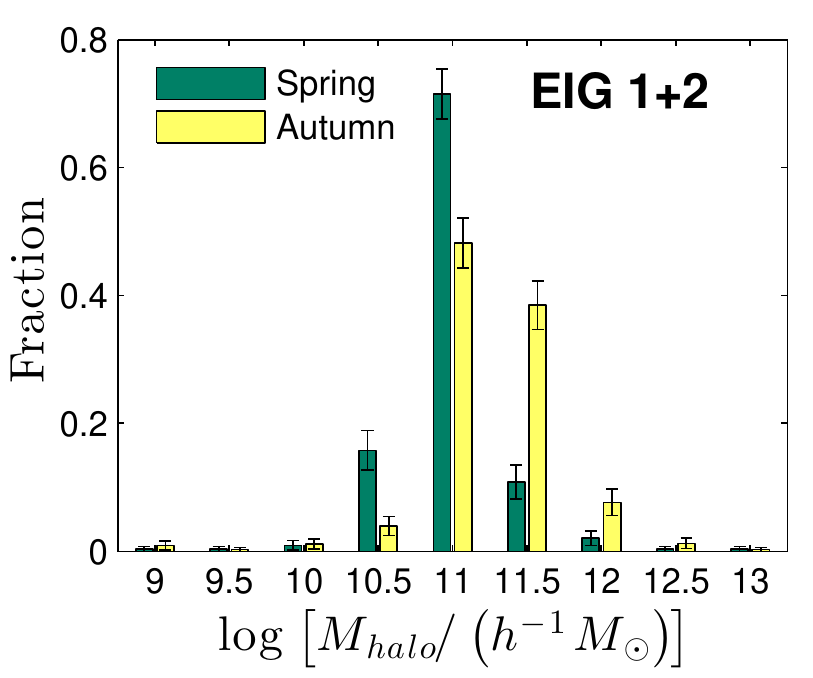}
\caption [Halo mass histogram for mock EIGs (Mill2 simulation)]
{
  \markChange{Simulation-based halo mass, {\Mhalo}, histogram calculated for the combination of subsamples EIG-1 and EIG-2. Adapted from Figure 6 of SB16.} \label{f:MhaloFromEIG_I}
}
\end{centering}
\end{figure*}

The stellar mass, {\Mstar}, HI mass, {\MHI}, and dynamic mass, $\Mass_{dyn,24}$, of EIGs were analysed in a fashion similar to that of the analysis of the dark matter (DM) subhalo mass, {\Mhalo}, of the Millennium II-SW7 simulation (Mill2; \citealt{2013MNRAS.428.1351G}) described in section 3.5 of \markChange{SB16}.
Figures \ref{f:Res_MstarHist}, \ref{f:Res_M_HI_Hist} and \ref{f:Res_Mdyn24Hist} show histograms of {\Mstar}, {\MHI} and $\Mass_{dyn,24}$ (respectively) for the EIGs of the Spring and Autumn sky regions.
Figure \ref{f:Res_MStarHI_Hist} shows histograms of $\left( \Mstar + \MHI \right)$ of all EIGs for which {\Mstar} was estimated. For EIGs not detected by ALFALFA, $\Mstar$ was used as representing $\left( \Mstar + \MHI \right)$. Thus the $\left( \Mstar + \MHI \right)$ statistics presented here, is expected to be slightly biased towards lower masses and a wider distribution.
Similarly, Figure \ref{f:Res_M_HI_Hist} includes only EIGs that were detected by ALFALFA, and is, therefore, expected to be slightly biased towards higher masses and a narrower distribution.
The right-hand side charts of figures \ref{f:Res_MstarHist}, \ref{f:Res_M_HI_Hist}, \ref{f:Res_MStarHI_Hist} and \ref{f:Res_Mdyn24Hist} may be compared to \markChange{Figure \ref{f:MhaloFromEIG_I}, an adaptation of Figure 6 of SB16,} which shows the simulation-based estimates of {\Mhalo} calculated for the combination of subsamples EIG-1 and EIG-2.

As can be seen in these figures, the distributions of both {\Mstar} and $\left( \Mstar + \MHI \right)$ are more scattered than those of {\Mhalo}. For EIGs in the Spring sky region the standard deviation of $\log \left[ \Mhalo / \left( \Msunh \right) \right]$ is $\sim$0.3, compared to $\sim$0.6 for $\log \left( \Mstar / \Msun \right)$ and $\sim$0.7 for $\log \left[ \left( \Mstar + \MHI \right) / \Msun \right]$. For the Autumn EIGs the standard deviation of $\log \left[ \Mhalo / \left( \Msunh \right) \right]$ is $\sim$0.4, compared to $\sim$1.1 for $\log \left( \Mstar / \Msun \right)$ and $\sim$0.8 for $\log \left[ \left( \Mstar + \MHI \right) / \Msun \right]$.
\markChange{Such higher scatter of {\Mstar} compared to {\Mhalo} is expected from the stellar mass to halo mass (SMHM) relation derived from simulations (e.g., Fig. 2 of \citealt{2009ApJ...696..620C},\rem{ Fig. 5 of \citealt{2010ApJ...717..379B},} Fig. 7 of \citealt{2013ApJ...770...57B}, Fig. 2 of \citealt{2015A&A...576L...7D}, Fig. 5 of \citealt{2015ApJ...799..130R} and Fig. 1 of \citealt{2017MNRAS.465.2381M}). The vast majority of EIGs have masses $\log \left( \Mstar / \Msun \right) < 10.5$ and $\log \left[ \Mhalo / \left( \Msunh \right) \right] < 12$ in which the SMHM relation's slope is large, i.e. in which {\Mstar} varies faster than {\Mhalo}.}
The distribution of the HI mass, {\MHI}, seems to have a width similar to that of {\Mhalo} (the standard deviation of $\log \left( \MHI / \Msun \right)$ is $\sim$0.3 for the Spring EIGs and $\sim$0.4 for the Autumn EIGs).

The distribution of the simulation predicted {\Mhalo} \markChange{(shown in Figure \ref{f:MhaloFromEIG_I})} is very different from the distribution of $\Mass_{dyn,24}$ (shown in Figure \ref{f:Res_Mdyn24Hist}). The average $\log \left( \Mass_{dyn,24} / \Msun \right)$ is $\sim$10.2 for the Spring and $\sim$9.9 for the Autumn. This is about an order of magnitude lower than the average $\log \left[ \Mhalo / \left( \Msunh \right) \right]$ (11.0 for Spring and 11.3 for Autumn).
The standard deviation of $\log \left( \Mass_{dyn,24} / \Msun \right)$ is $\sim$0.3 for the Spring EIGs and $\sim$0.9 for the Autumn EIGs (compared to $\sim$0.3 for the Spring and $\sim$0.4 for the Autumn $\log \left[ \Mhalo / \left( \Msunh \right) \right]$).
This difference between the simulated distribution of {\Mhalo} and the measured distribution of $\Mass_{dyn,24}$ may be the result of a large discrepancy between $\Mass_{dyn,24}$ and the actual dynamic mass (had it been measured using HI rotation curves), a large discrepancy of the simulation results from reality, or both.

The average $\log \left( \Mstar / \Msun \right)$ for EIG-1 and EIG-2 is $\sim$8.8 (Spring) and $\sim$9.4 (Autumn). From a comparison to the average $\log \left[ \Mhalo / \left( \Msunh \right) \right]$ (11.0 for Spring, and 11.3 for Autumn) we conclude that the stellar masses, {\Mstar}, of EIGs are {$\sim$2.4\,\dex} (Spring) and {$\sim$2.0\,\dex} (Autumn) lower on average than the EIGs' DM masses. This, compared to a {$\sim$0.7\,\dex} difference between the baryonic to DM average densities in the Universe (according to WMAP7).
If the dark-to-baryonic matter ratio of isolated galaxies is similar to the Universal average, then the geometric average of the fraction of baryonic mass turned into stars is  {$\sim$0.02} (Spring) and {$\sim$0.05} (Autumn).


\vspace{12pt}

It is interesting to compare the stellar and HI content of galaxies from subsample EIG-1 with those of subsample EIG-2. As described in section \ref{sec:Introduction}, the EIG-1 subsample contains galaxies that passed the isolation criterion using both NED and ALFALFA data. The EIG-2 subsample contains galaxies that passed the criterion using NED data, but did not pass it using ALFALFA data (have neighbours closer than {3\,\Mpch} with sufficient HI content to be detected by ALFALFA).
Therefore, the distance to the closest ALFALFA neighbour for EIG-1 galaxies is {$>$3\,\Mpch} by definition. For EIG-2 galaxies this distance is in the range {0.66--2.74\,\Mpch} ({0.9--3.9\,\Mpc}; see Table 8 of \citealt{2016MNRAS.456..885S}).

\begin{figure*}
\begin{centering}
\includegraphics[width=14cm,trim=0mm 0mm 0mm 0, clip]{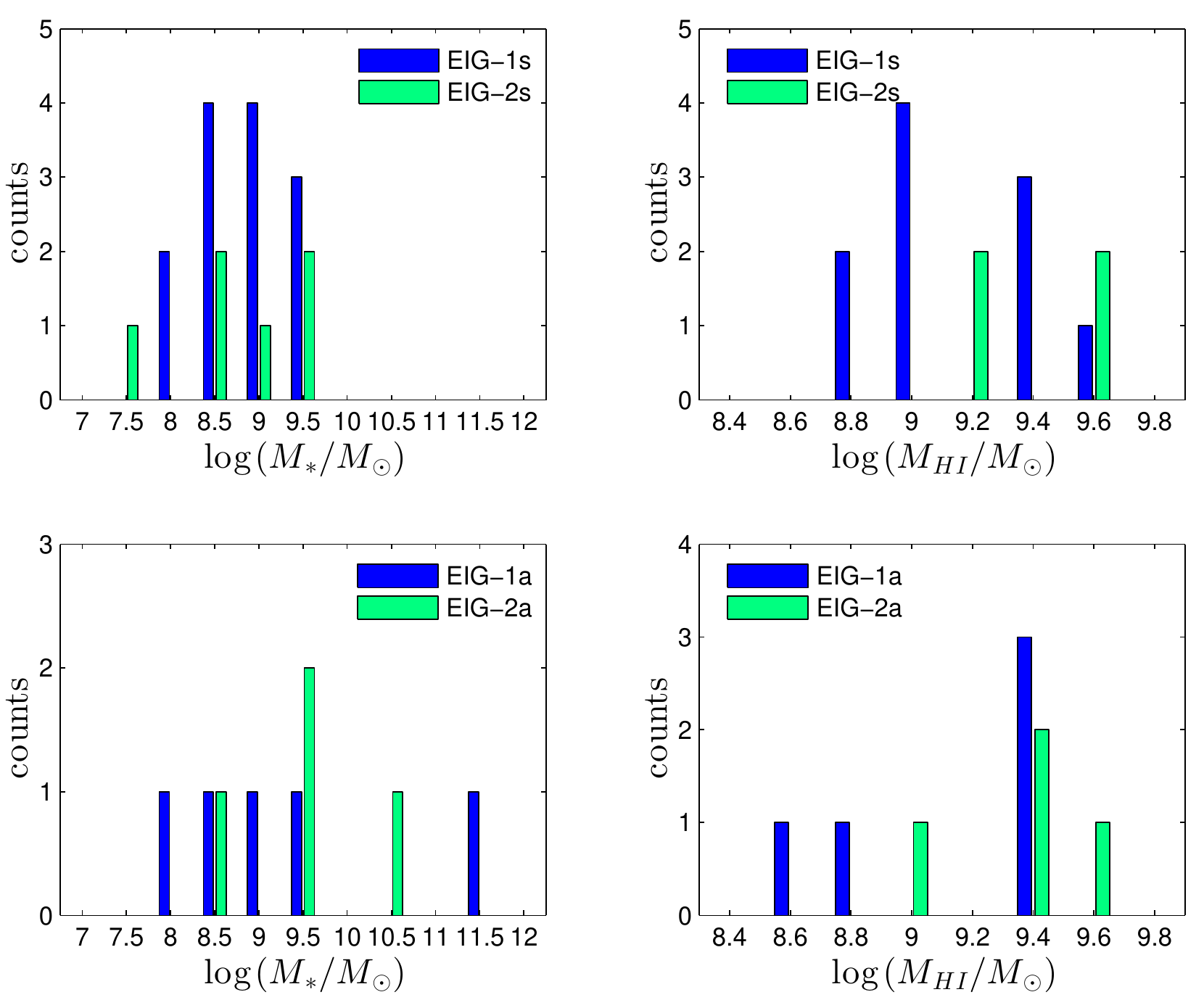}
\caption [Stellar and HI content histograms comparing subsamples EIG-1 and EIG-2]
{
  {\Mstar} (left charts), and {\MHI} (right charts) histograms comparing subsample EIG-1s with EIG-2s (upper charts) and EIG-1a with EIG-2a (lower charts).
  One EIG-1s galaxy, one EIG-2s galaxy and two EIG-1a galaxies are not included in the left charts, because they do not have {\Mstar} data. Four EIG-1s galaxies, three EIG-2s galaxies and two EIG-1a galaxies are not included in the right charts, because they were not detected by ALFALFA. \label{f:Res_EIG-1_2_Hist}
}
\end{centering}
\end{figure*}

Figure \ref{f:Res_EIG-1_2_Hist} shows histograms of {\Mstar} (left charts), and {\MHI} (right charts) comparing subsample EIG-1s with EIG-2s (upper charts) and subsample EIG-1a with EIG-2a (lower charts).
As can be seen, the {\Mstar} distribution of EIG-1s is similar to that of EIG-2s, and the distribution of EIG-1a is similar to that of EIG-2a.
The average $\log \left( \Mstar / \Msun \right)$ of the measured EIG-1s galaxies is $8.8 \pm 0.1$ with a standard deviation of {0.5}. This, compared to an average of $8.7 \pm 0.3$ with standard deviation {0.8} for the measured EIG-2s galaxies.
The average $\log \left( \Mstar / \Msun \right)$ of the measured EIG-1a galaxies is $9.3 \pm 0.6$ with a standard deviation of {1.3}. This, compared to an average of $9.6 \pm 0.4$ with standard deviation {0.9} for the measured EIG-2a galaxies. These differences are not statistically significant.

In contrast to this, the {\MHI} distributions differ significantly between the EIG-1 and the EIG-2 subsamples.
The average $\log \left( \MHI / \Msun \right)$ of the EIG-1s galaxies detected by ALFALFA is $9.13 \pm 0.09$ (with standard deviation {0.3}). This, compared to an average of $9.5 \pm 0.1$ (with standard deviation {0.2}) for the ALFALFA-detected EIG-2s galaxies; a $2.3\,\sigma$ difference.

The difference in the average $\log \left( \MHI / \Msun \right)$ between the Autumn subsamples is the same as for the Spring subsamples ({$\sim$0.3}), but with lower statistical significance ($1.2\,\sigma$) due to the smaller numbers of measured galaxies.
The average $\log \left( \MHI / \Msun \right)$ of the EIG-1a galaxies detected by ALFALFA is $9.1 \pm 0.2$ (with standard deviation {0.4}). This, compared to an average of $9.4 \pm 0.2$ (with standard deviation {0.3}) for the ALFALFA-detected EIG-2a galaxies.

Combining the differences measured for the Spring and Autumn, gives an expected difference between $\log \left( \MHI / \Msun \right)$ of EIG-2 and EIG-1 of $0.3 \pm 0.1\,\dex$.
Therefore, from the data of the ALFALFA-detected galaxies we conclude with $2.5\,\sigma$ significance that EIG-2 galaxies have higher {\MHI}, on average, than EIG-1 galaxies (by a factor of $2.1 \pm 0.6$).

The actual average $\log \left( \MHI / \Msun \right)$ of all EIGs of these subsamples are expected to be lower than the above values, since the EIGs not detected by ALFALFA are expected to have low {\MHI} values. However, adding the {\MHI} values of those EIGs is not expected to make the distributions of the EIG-1 and EIG-2 subsamples significantly more similar.

We can therefore conclude that extremely isolated galaxies that have neighbours with significant {\MHI} content at distances {$<$3\,\Mpch} tend to have higher {\MHI} compared to extremely isolated galaxies lacking such neighbours. The HI content of galaxies, therefore, seems to be environmentally dependent even in extremely isolated regions.

\subsection{Scaling relations}

\begin{figure*}
\begin{centering}
\includegraphics[width=14cm,trim=0mm 0mm 0mm 0, clip]{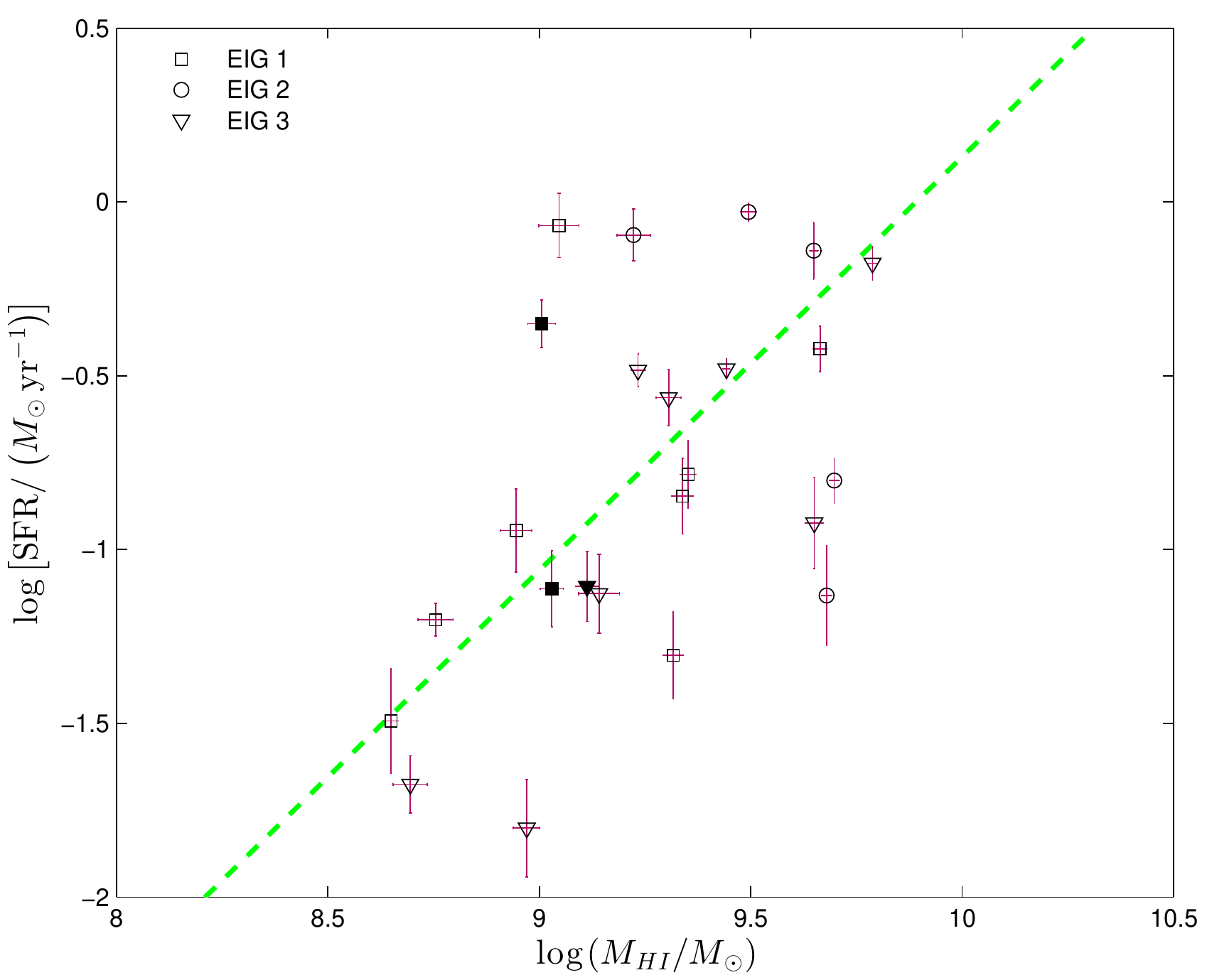}
\includegraphics[width=14cm,trim=0mm 0mm 0mm 0, clip]{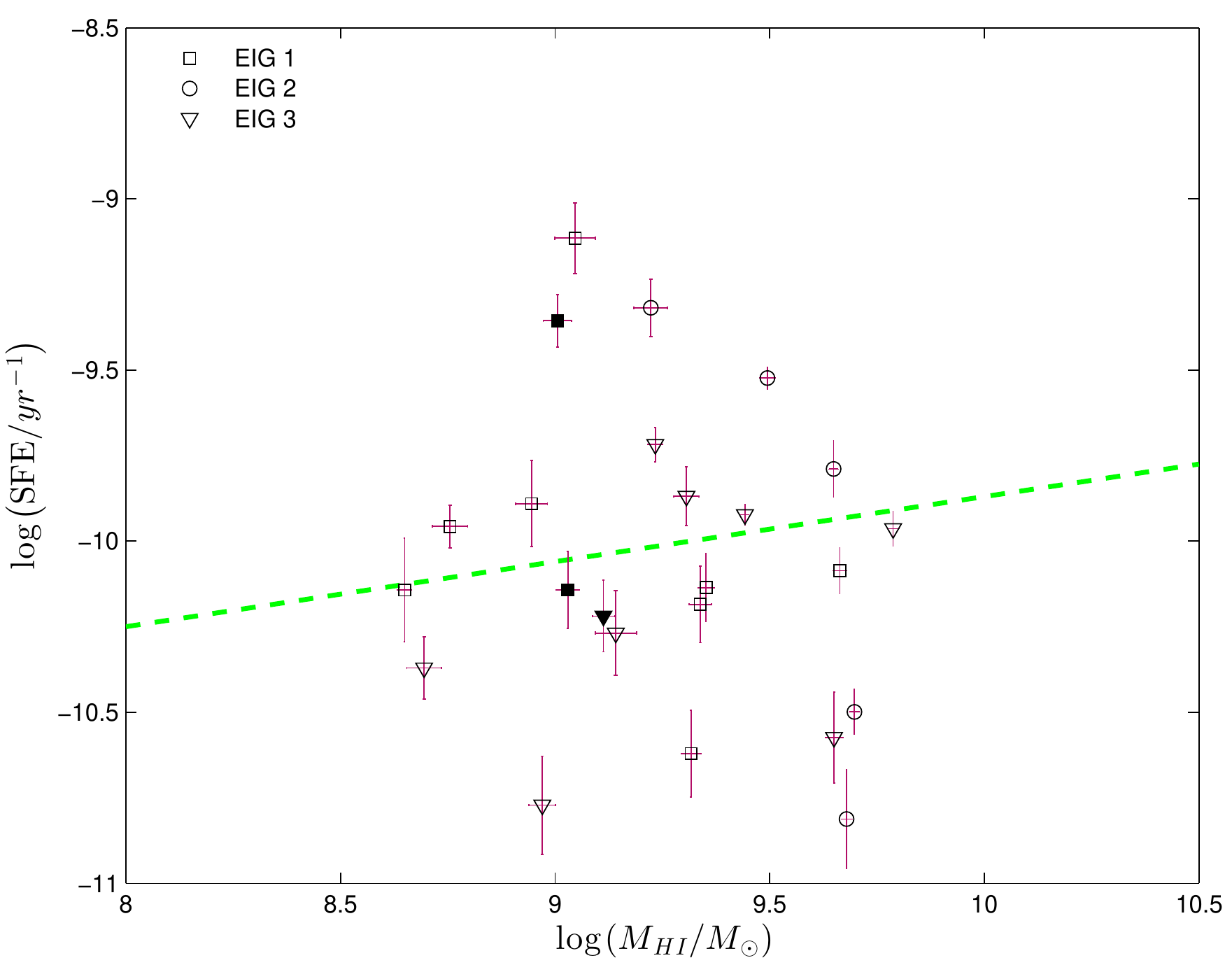}
\caption [Star formation vs.~HI mass]
{
  Star formation ({\SFR} in the upper chart, and {\SFE} in the lower chart) vs.~HI mass, {\MHI}, of the EIGs.
  The green thick dashed lines show the fit found by \cite[Fig.~4.b]{2012ApJ...756..113H} for {\SFR} to {\MHI} of star-forming ALFALFA galaxies.
  Filled symbols indicate EIGs classified as early-types.
  \label{f:Res_SF_MHI}
}
\end{centering}
\end{figure*}

\begin{figure*}
\begin{centering}
\includegraphics[width=14cm,trim=0mm 0mm 0mm 0, clip]{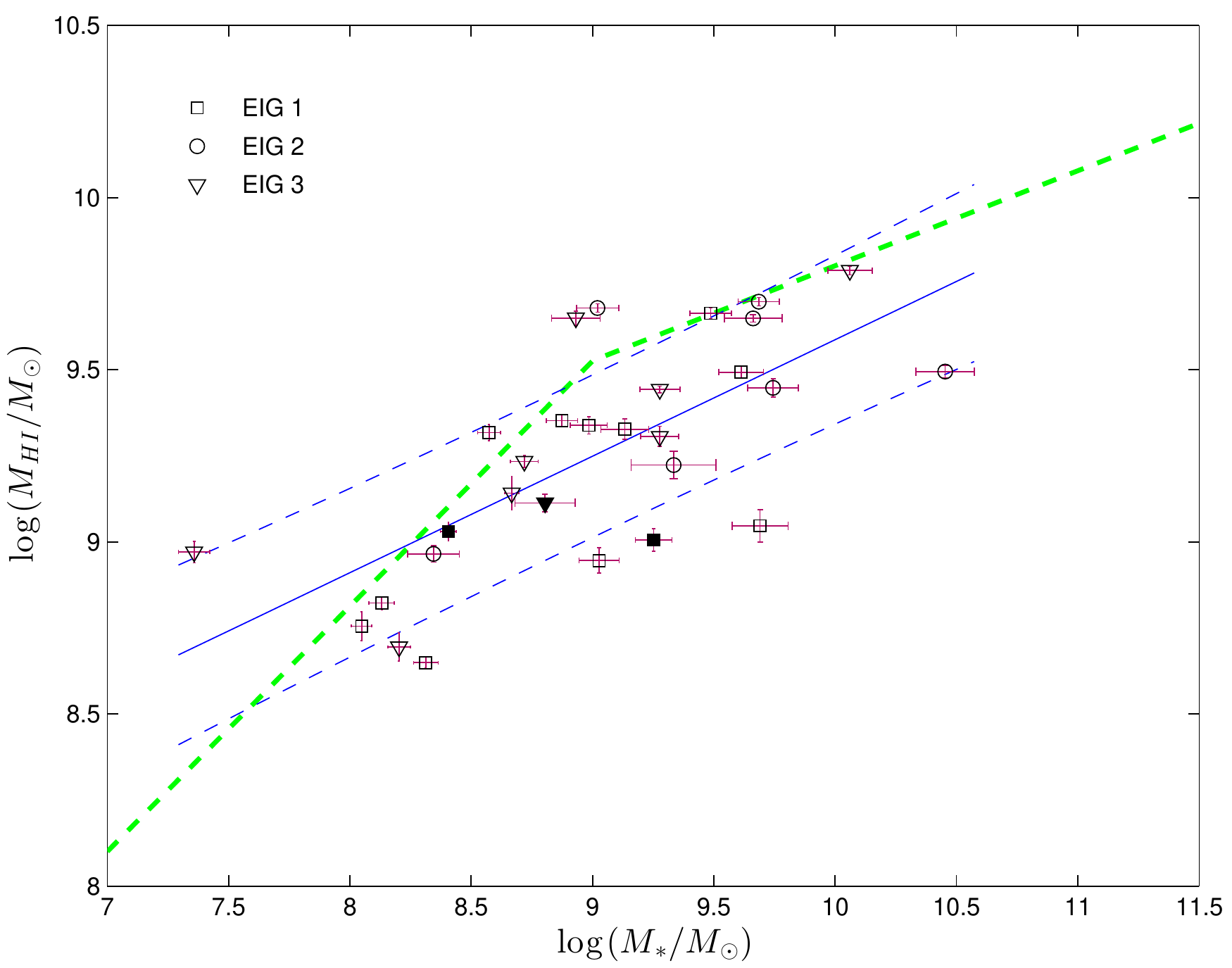}
\includegraphics[width=14cm,trim=0mm 0mm 0mm 0, clip]{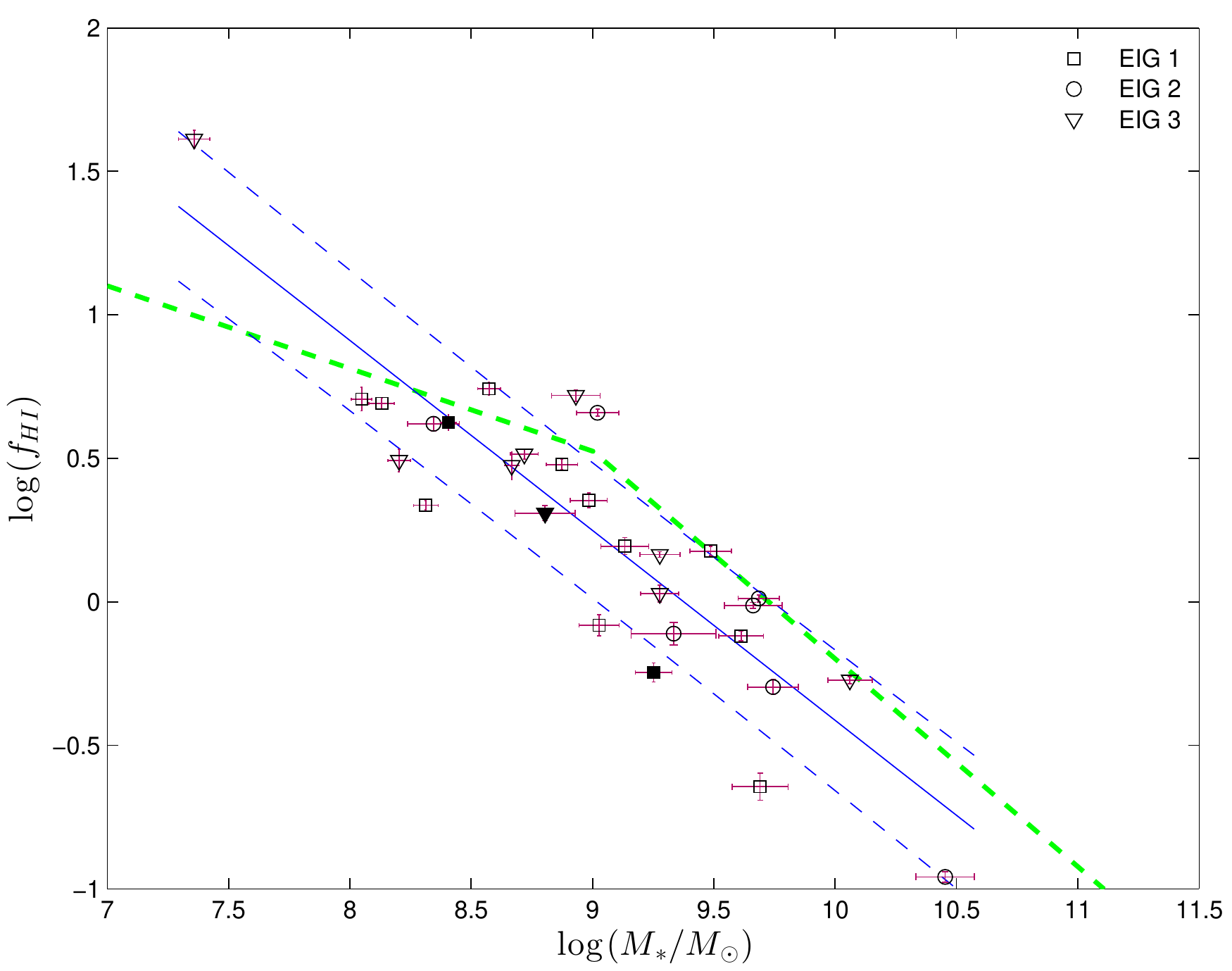}
\caption [HI content vs.~stellar mass]
{
  HI content ({\MHI} in the upper chart, and $f_{HI}$ in the lower chart) vs.~stellar mass, {\Mstar}, of the EIGs. 
  The blue solid lines \markChange{show} a linear fit to the EIGs' {\MHI} to {\Mstar} data. The dashed blue lines show the $\pm 1\sigma$ deviation from this fit.
  The green thick dashed lines show the fit found by \cite[eq.~1]{2012ApJ...756..113H} for {\MHI} to {\Mstar} of star-forming ALFALFA galaxies.
  Filled symbols indicate EIGs classified as early-types. 
  \label{f:Res_HI_Mstar}
}
\end{centering}
\end{figure*}

The HI gas content can be combined with the {\SFR}-to-{\Mstar} relation of Figure \ref{f:Res_MainSeq} in an attempt to investigate its connection to the main sequence of star-forming galaxies. This is done by breaking the {\SFR}-to-{\Mstar} relation into a relation between the star formation and the HI gas content (Figure \ref{f:Res_SF_MHI}) and a relation between the HI content and {\Mstar} (Figure \ref{f:Res_HI_Mstar}).
Figure \ref{f:Res_SF_MHI} shows {\SFR} and \markChange{star formation efficiency ($\SFE \equiv \SFR / \MHI$)} vs.~the HI mass, {\MHI}, of the EIGs.
Figure \ref{f:Res_HI_Mstar} shows {\MHI} and $f_{HI} \equiv \MHI /  \Mstar$ vs.~the stellar mass, {\Mstar}, of the EIGs.

%
%

The average deviation of EIGs from the {\SFR}-to-{\MHI} fit of \cite{2012ApJ...756..113H} is $-0.04 \pm 0.09\,\dex$ (with standard deviation of {0.5\,\dex}). Despite the EIG's small range of $\log \left( \MHI / \Msun \right)$ this indicates that the {\SFR}-to-{\MHI} relation of EIGs is similar to that of the general population of star-forming galaxies.

The near-unity slope in the {\SFR}-to-{\MHI} fit translates to a near-zero slope in {\SFE} to {\MHI} (lower chart of Figure \ref{f:Res_SF_MHI}), implying that the {\SFE} may be (statistically) independent of {\MHI}. To test this hypothesis, the Pearson product-moment correlation coefficient between {\SFE} and {\MHI} of the EIGs was calculated. The resultant correlation coefficient, $-0.14$, is insignificant. If {\SFE} and {\MHI} are not correlated, there is 0.52 chance of finding a correlation coefficient measurement at least this high. Therefore, the EIGs' measured data, supports independence of {\SFE} and {\MHI}.

The following linear relation was fitted to the measured {\MHI} vs.~{\Mstar}  EIG points (marked by blue solid lines in Figure \ref{f:Res_HI_Mstar}): 

\begin{equation}
\log \left( \MHI / \Msun \right) \cong 0.34 \cdot \log \left( \Mstar / \Msun \right) + 6.20 
\label{e:Res_MHI_Mstar_fit} 
\end{equation}

The expected standard deviation in $\log \left( \MHI / \Msun \right)$ around this fit is marked in the figure by dashed blue lines ({0.25} on average).
The average $\log \left( \MHI / \Msun \right)$ deviation of EIGs from the {\MHI}-to-{\Mstar} fit of \cite{2012ApJ...756..113H} is $-0.16 \pm 0.05$, implying that for a given stellar mass, {\Mstar}, the HI mass, {\MHI}, of EIGs is slightly lower, on average, than that of the general population of star-forming galaxies. However, some of this deviation may be a result of the difference between the {\Mstar} estimation method used by \cite{2012ApJ...756..113H} and the one used here.

\cite{2011Ap.....54..445K} analysed the {\MHI}-to-{\Mstar} relation for the 2MIG catalogue of isolated galaxies. The galaxies of the 2MIG catalogue have a range of {\Mstar} higher than that of the EIG sample (with the bulk in the range $9.5 < \log \left( \Mstar / \Msun \right) < 11.5$). Most of these higher mass and less isolated galaxies have $\MHI < \Mstar$ \citep[Fig. 10]{2011Ap.....54..445K}. This is in agreement with the results shown in Figure \ref{f:Res_HI_Mstar}, where for $\log \left( \Mstar / \Msun \right) > 9.5$ most galaxies have $\log \left( f_{HI} \right) < 0$. For their higher {\Mstar} sample \cite{2011Ap.....54..445K} found a slope ($1.00 \pm 0.04$) in the linear fit of $\log \left( \MHI / \Msun \right)$ as function of $\log \left( \Mstar / \Msun \right)$ which is larger than the slope found here.

\vspace{12pt}

The following log-log predictor for {\SFR}, using {\Mstar} and {\MHI}, was fitted to the EIGs data by a partial least-squares regression:

\begin{equation}
\begin{IEEEeqnarraybox*}{lCl}
\log \left( \SFR / \MsunPerYr \right) 
  & \cong &
    0.580 \cdot \log \left( \Mstar / \Msun \right) + \\
  &       &
    0.209 \cdot \log \left( \MHI / \Msun \right) - 7.95
\end{IEEEeqnarraybox*}
\label{e:Res_SFR_predictor} 
\end{equation}

\begin{figure*}
\begin{centering}
\includegraphics[width=14cm,trim=0mm 0mm 0mm 0, clip]{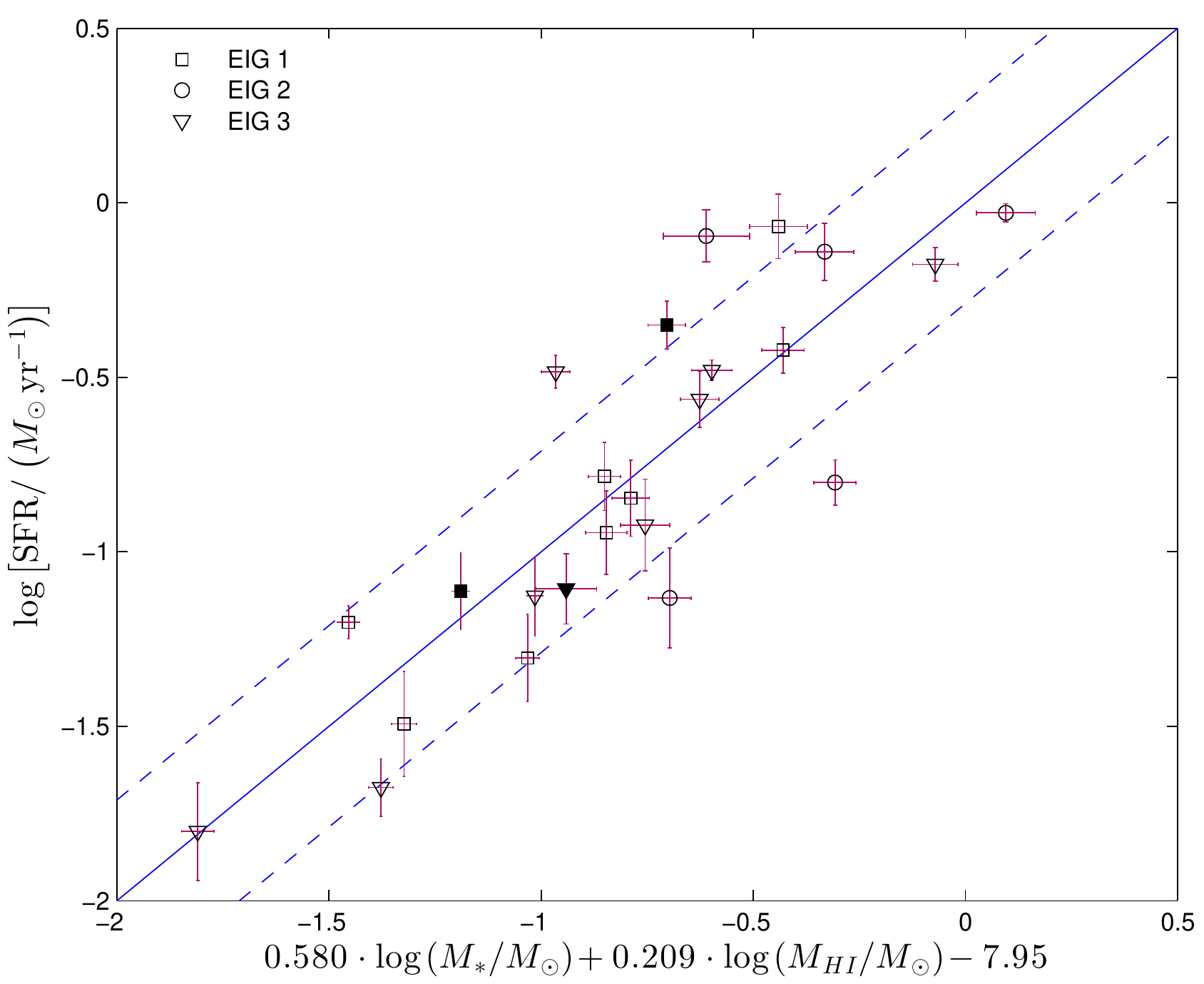}
\caption [{\SFR} predictor]
{
  Star formation rate, {\SFR}, vs.~the {\Mstar} and {\MHI} partial least-squares regression predictor of \eqref{e:Res_SFR_predictor}.
  The blue solid line shows the one-to-one line. The dashed blue lines show the $\pm 1\sigma$ deviation from the one-to-one line.
  Filled symbols indicate EIGs classified as early-types.
  \label{f:Res_SFR_predictor}
}
\end{centering}
\end{figure*}

The EIGs' {\SFR}s are shown vs.~this predictor in Figure \ref{f:Res_SFR_predictor}. The expected standard deviation around this predictor is {0.29\,\dex} (shown in the figure as dashed blue lines). It is somewhat lower than the {0.5\,\dex} {\SFR}-to-{\MHI} standard deviation around the \cite{2012ApJ...756..113H} relation, and the {0.4\,\dex} standard deviation in the {\SFR} to main sequence difference (Figure \ref{f:Res_MainSeq}). Therefore, \eqref{e:Res_SFR_predictor} that considers both {\Mstar} and {\MHI} is a more accurate estimate of {\SFR} compared to predictors based on {\Mstar} or on {\MHI} alone. It applies to EIGs, but is possibly also a good estimate for galaxies in denser regions that have not interacted with neighbours in the last few {\Gyr}.

\subsection{Morphology}
\label{s:Res_Morphology}

It is evident from Table \ref{T:Res_Morphology} that EIGs are typically late types (20 were classified as late-types, compared to six unknown and five early-types).
Based on the 20 EIGs (65 per cent) that were classified as late-types, we conclude with 0.95 confidence that the probability for an EIG to be a late-type is $\geq$0.47 (using Wilson score interval for binomial distribution).
Similarly, based on the five EIGs (16 per cent) classified as early-type we conclude with 0.95 confidence that the probability for an EIG to be an early-type is $\geq$0.07.
For comparison, \cite{2012AJ....144...16K}\rem{ and \cite{2012PhDT.........5K}} identified three early-type galaxies (5 per cent) of the 60 isolated galaxies of the VGS sample.
\cite{2006A&A...449..937S} found in the most isolated sub-sample of AMIGA that\rem{ 82 per cent of the galaxies are late-types (Sa--Sd) while} 14 per cent are early-types (E--S0).
\cite{2013AstBu..68..243K} found that 5 per cent of the LOG sample galaxies are early-types (E--S0/a).

Of the five early-types, four are part of the EIG-1 subsample (galaxies that passed the isolation criterion using both NED and ALFALFA HI data as described in section \ref{sec:Introduction}). These make an early-type fraction significantly larger in the EIG-1 subsample (27 per cent) than in the entire EIG sample (16 per cent).
The only early-type galaxy that is not part of the EIG-1 subsample, EIG 3s-01, passed the isolation criterion in the ALFALFA $\alpha$.40 dataset (but was slightly short of passing it in the NED dataset).

This means that none of the five early-type EIGs have ALFALFA (high HI content) neighbours within {3\,\Mpch}.
We can, therefore, conclude (with 0.94 confidence\rem{1-0.5^4~=0.94}) that EIGs lacking high HI content neighbours within {3\,\Mpch} have a higher tendency to be early-types, compared to EIGs that have such neighbours.
As discussed in section \ref{s:Res_MassHisgrms}, such EIGs with no high HI content neighbours tend to have a lower HI content compared to ones with high HI content neighbours.
This lower HI content may be linked to the fact that EIG-1 galaxies tend more to be early-types.
From the EIG-1 subsample classification it can be concluded that for a galaxy that passes the strict isolation criterion of the EIG-1 subsample, the probability to be early-type is $\geq$0.11 (using Wilson score interval with 0.95 confidence).

\vspace{12pt}

Only one of the early-types, EIG 1a-04, is not classified as blue in Figure \ref{f:Res_ColorMass} but is rather a `green valley' galaxy. Three others (1s-02, 1s-12 and 3s-01) are blue, and the colour classification of the last, EIG 3s-01, is unknown. This means that an extremely isolated early-type galaxy has a probability $\geq$0.30 of being blue (with 0.95 confidence).
This may be compared to the $0.057 \pm 0.004$ fraction of blue galaxies found by \cite{2009MNRAS.396..818S} in the low-redshift early-type galaxy population\markChange{, and to the $\sim$0.20 fraction of blue galaxies found by \cite{2016A&A...588A..79L} in their sample of isolated early-types}.

All five early-type EIGs are within {0.5\,\dex} of the main sequence in Figure \ref{f:Res_MainSeq}. From this we conclude that an extremely isolated early-type galaxy has a probability $\geq$0.57 (with 0.95 confidence) of fitting the main sequence of star-forming galaxies to within {0.5\,\dex}.

\vspace{12pt}

A large fraction of the EIGs show asymmetric star formation, and many show strong compact star-forming regions (see Figures \ref{f:RHacolorImgEIG-1}, \ref{f:RHacolorImgEIG-2} and \ref{f:RHacolorImgEIG-3} and Tables \ref{T:EIG_PlgnHaFlux} and \ref{T:EIG_PlgnEW}). This indicates that star formation is a stochastic process that may occur unevenly across a galaxy in a given time, even in the most isolated galaxies.
Sources of the randomness of star formation may include uneven `fuelling' of gas, and collisions with very small satellites (e.g., with $\Mhalo < 10^{9}\,\Msunh$) that could not be detected around the EIGs studied here.

\section{Discussion and conclusions}
\label{s:DisConc}

We have found surprising environmental dependencies of the HI content, \MHI, and of the morphological type of EIGs (sections \ref{s:Res_MassHisgrms} and \ref{s:Res_Morphology} respectively).
It is generally accepted that galaxies in cluster environments typically have atomic gas
deficiencies \citep{1998ARA&A..36..189K}, while void galaxies are typically gas-rich \markChange{\citep{2011MNRAS.415.1797C, 2012AJ....144...16K, 2014ApJ...788...29L\rem{section 3.2.1}}}. It is also generally accepted that early-type galaxies are more abundant in clusters than in isolated environments \citep{2009MNRAS.393.1324B}.
It was, therefore, expected that a sample of the most isolated galaxies (subsample EIG-1) would be the most gas-rich and would contain the lowest fraction of early-types.

However, contrary to these expectations, we have found that EIG-1 galaxies, which lack neighbours with significant {\MHI} content at distances {$<$3\,\Mpch}, tend to have lower {\MHI} compared to EIG-2 galaxies that have such neighbours (the average {\MHI} of EIG-1 galaxies is lower than that of EIG-2 galaxies with $2.5\,\sigma$ confidence).
\cite{2014MNRAS.444.3559M} have found a similar {\MHI} environmental dependence, in which their sample of void galaxies\rem{ (less isolated than the EIGs)} showed a tendency for lower {\MHI} compared to their sample of wall galaxies.

Similarly unexpected, we have found that the most isolated galaxies (subsample EIG-1) have a higher tendency to be early-types compared to EIG-2 galaxies (with 0.94 confidence).
To the best of our knowledge this is the first time where an isolated galaxies' sample shows a higher fraction of early-types compared to a less isolated sample.

\markChange{These findings do not contradict the results of \cite{2011MNRAS.415.1797C, 2012AJ....144...16K, 2014ApJ...788...29L} and \cite{2009MNRAS.393.1324B} which compared isolated galaxies with galaxies in clusters or with the general population of galaxies. Here we compared between two galaxy populations of extreme isolation levels, and showed that the trends of increased {\MHI} and decreased early-type fraction with the increase of the isolation level reverse at extreme isolation (or when the isolation is tested also with respect to the {\MHI} of possible neighbours).}

There is considerable evidence from cosmological simulations that the spins and major axes of haloes are correlated with the direction of the walls or filaments in which they reside \citep[e.g.,][]{2007ApJ...655L...5A, 2009ApJ...706..747Z, 2012MNRAS.427.3320C, 2014MNRAS.443.1274L}. For low-mass haloes ($\log \left[ \Mhalo / \left( \Msunh \right) \right] < 13$ according to \citealt{2009ApJ...706..747Z}, or $\log \left[ \Mhalo / \left( \Msunh \right) \right] < 12.6$ according to \citealt{2012MNRAS.427.3320C}), the halo spin is more likely to be aligned with the closest filament. This preferred spin direction is probably a result of the direction from which material is accreted to the halo. As shown in \markChange{Figure \ref{f:MhaloFromEIG_I}}, simulation analysis indicates that almost all EIGs reside in haloes that would be considered low-mass in this respect, and are, therefore, expected to have spins that correlate fairly well with the direction of the filaments and walls closest to them.

We speculate that this effect may be connected to the low abundance of early-types in the EIG-2 subsample and to its higher average {\MHI} compared to the EIG-1 subsample. Underdense filaments and walls may be the hosts of EIG-2 galaxies. The halo spins induced by their filament or wall environment may significantly reduce their early-type fraction and possibly also increase their {\MHI} (because the gas may spend more time before reaching their centres). If they indeed reside in filaments or walls, they are also expected to have some neighbours with similar tendency for being late-type and containing significant amounts of {\MHI}. 

We further speculate that a significant fraction of EIG-1 galaxies are not parts of filaments or walls, but rather reside in environments with no preferred direction for accreting material (e.g., at the junction points between filaments so extremely underdense that no galaxies were detected in them). This may increase their probability of being early-types and may also affect them in such a way that they would contain less HI on average (either because there is not much gas in their environment or because the available HI gas is accreted faster to the halo's centre and forms stars more quickly).
Further study of the early-type EIGs found in this work may be of interest, since if they indeed reside at junction points between filaments, they may resemble cluster early-types at early stages of their development.

\vspace{12pt}

Both the early-type and late-type EIGs follow the same colour-to-{\Mstar} relation (Fig.~\ref{f:Res_ColorMass}), SFR-to-{\Mstar} `main sequence' relation (Fig.~\ref{f:Res_MainSeq}) and {\MHI}-to-{\Mstar} relation (Fig.~\ref{f:Res_HI_Mstar}), and fit the SFR predictor of eq.~\eqref{e:Res_SFR_predictor} (Fig.~\ref{f:Res_SFR_predictor}). 
This indicates that the mechanisms and factors governing star formation, colour and the {\MHI}-to-{\Mstar} relation are similar in early-type and late-type EIGs.
It further indicates that the morphological type of EIGs is not governed by their {\MHI} content, {\SFR} or colour. EIGs with high {\MHI} content, high SFR or blue colour are not necessarily late-types.

\vspace{12pt}

Our observations indicate that EIGs typically fit the `main sequence of star forming galaxies' found by \cite{2012ApJ...756..113H}\rem{ for a general sample detected by ALFLFA, SDSS and GALEX}.
This indicates that the extreme isolation of the EIGs does not affect their {\SFR}\rem{ or their {\MHI}} considerably compared to field galaxies.
This is supported by \cite{2016arXiv160108228B} who found no significant difference in SF between void galaxies of the VGS sample and field galaxies.

We have found that EIGs follow a colour-to-{\Mstar} relation, in which EIGs with {\Mstar} smaller than $10^{(10.6 \pm 0.9)}\,\Msun$ are typically `blue cloud' galaxies irrespective of their morphological type (Figure \ref{f:Res_ColorMass}). Since {\Mstar} of most EIGs is below this threshold, most of the EIGs are blue.
A similar result was found by \cite{2009MNRAS.393.1324B} who found that in low density environments low {\Mstar} galaxies are mostly blue, while galaxies with high {\Mstar} are mostly red (irrespective of morphology). This is contrary to what is found in high density environments, where galaxies are mostly red irrespective of their {\Mstar} and morphology.

\vspace{24pt}

With respect to the `Nature vs.~Nurture' question, which was the primary driver of this work, we conclude the following: It is well known that cluster environments have a strong effect on star formation, colours and morphologies of galaxies.
With the exception of these high density environments the {\SFR} is not significantly affected by the environment, i.e.~the `main sequence of star-forming galaxies' holds in a range of environments from walls to the most extremely isolated regions measureable. Outside high density regions, the colours of galaxies are mostly related to their stellar mass, {\Mstar}, and are less affected (if at all) by the environment.

We have found that the HI content, {\MHI}, and the morphological type of galaxies do depend on their environment. In the most isolated environments, where no neighbours with significant {\MHI} are present (to a distance of {3\,\Mpch}), galaxies tend more to be early-types and have lower {\MHI}, on average, compared to less isolated environments. We speculate that this might reflect the large scale structure of these extremely isolated regions. Late-type and high-{\MHI} galaxies may be more abundant in underdense filaments and walls, while early-type and lower {\MHI} galaxies may be more abundant at the junctions of filaments so extremely underdense that no galaxies were detected in them.

\section*{Acknowledgements}

{\rem{ALFALFA}}
We are grateful to Martha Haynes, Riccardo Giovanelli and the entire ALFALFA team for providing an unequalled HI data set.

{\rem{NED}}
This research has made use of the NASA/IPAC Extragalactic Database (NED) which is operated by the Jet Propulsion Laboratory, California Institute of Technology, under contract with the National Aeronautics and Space Administration.
{\rem{SDSS}}
Funding for SDSS-III has been provided by the Alfred P. Sloan Foundation, the Participating Institutions, the National Science Foundation, and the U.S. Department of Energy Office of Science. The SDSS-III web site is http://www.sdss3.org/.
\rem{this is part of the Official SDSS-III Acknowledgement, which can be found on: http://www.sdss3.org/collaboration/boiler-plate.php}

{\rem{GALEX}}
This research has made use of observations made with the NASA Galaxy Evolution Explorer.
GALEX is operated for NASA by the California Institute of Technology under NASA contract NAS5-98034.
\rem{IRAS - used for downloading 2MASS, WISE and Spitzer data}
This research has also made use of the NASA/IPAC Infrared Science Archive (IRSA), which is operated by the Jet Propulsion Laboratory, California Institute of Technology, under contract with the National Aeronautics and Space Administration.
{\rem{2MASS}}
This publication makes use of data products from the Two Micron All Sky Survey (2MASS), which is a joint project of the University of Massachusetts and the Infrared Processing and Analysis Center/California Institute of Technology, funded by the National Aeronautics and Space Administration and the National Science Foundation.
{\rem{WISE}}
This publication makes use of data products from the Wide-field Infrared Survey Explorer (WISE), which is a joint project of the University of California, Los Angeles, and the Jet Propulsion Laboratory/California Institute of Technology, funded by the National Aeronautics and Space Administration.

\addcontentsline{toc}{chapter}{Bibliography}
\bibliographystyle{mn2e}
\bibliography{references}

\appendix
\section{\\EIG Specific Data}
\label{App:EIGdata}

This appendix contains general notes for some of the EIGs.

\subsection*{EIG 1s-05}

No optical counterpart could be identified for EIG 1s-05 (an ALFALFA object).
In the Wise Observatory images, no {\Halpha} emission was identified around the ALFALFA coordinates.
Within one arcminute from the ALFALFA coordinates of EIG 1s-05, all galaxies detected by SDSS have ${\SDSSg} > 21.6$, and none have spectroscopic redshifts. All GALEX detected objects in the same region have ${\magFUV} > 24$ and ${\magNUV} > 21$.
EIG 1s-05 may, therefore, be a `dark galaxy' with an extremely high HI to stellar mass ratio and a very low SFR. It may also be an \markChange{ALFALFA} false detection, even though its SNR is 8.1 and it is considered a `code 1' object, i.e. a source of SNR and general qualities that make it a reliable detection \citep{2011AJ....142..170H}.

\subsection*{EIG 1s-09}

SDSS DR10 shows an edge-on galaxy, SDSS J112157.63+102959.6, {$\sim$13\,\arcsec} east of the centre of EIG 1s-09. The angular size of SDSS J112157.63+102959.6 is similar to that of EIG 1s-09. Its magnitude is ${\SDSSg} = 18.6$, compared to ${\SDSSg} = 16.9$ of EIG 1s-09. The redshift of SDSS J112157.63+102959.6 is unknown. Although there is a possibility that SDSS J112157.63+102959.6 is a close neighbour of EIG 1s-09, this seems unreasonable, since tidal tails are neither visible in the SDSS images nor in the images shown in figure \ref{f:RHacolorImgEIG-1} \markChange{(which combine 40 minute exposure in the {\R} band and 120 minute exposure in an \Halpha band, both using the WO {1\,meter} telescope)}.

\subsection*{EIG 1s-10}

SDSS DR10 shows two objects at an angular distance of {$\sim$6\,\arcsec} from the centre of EIG 1s-10. One is north of EIG 1s-10, and is classified as a star by SDSS DR10. The second, classified as a galaxy, is south-west of EIG 1s-10. Both objects do not have measured redshifts. Although there is a possibility that one or both of these are galaxies merging with EIG 1s-10, this seems unreasonable, since tidal tails are neither visible in the SDSS images nor in the images shown in figure \ref{f:RHacolorImgEIG-1} \markChange{(which combine 100 minute exposure in the {\R} band and 260 minute exposure in an \Halpha band, both using the WO {1\,meter} telescope)}.

\subsection*{EIG 1s-11}

The only redshift measurement found for EIG 1s-11 is from \cite{1993A&AS...98..275B} that quotes \cite{1987ApJS...63..247H}. This is a HI measurement made at the Arecibo observatory. The HI-profile for the galaxy was not published by \cite{1987ApJS...63..247H}. It is possible that the measurement ($4725 \pm 10$\,\kms) is a result of HI-confusion, and that EIG 1s-11 is actually a part of the Virgo cluster.

\subsection*{EIG 1s-14}

EIG 1s-14 is projected close to a bright foreground star, which prevented SDSS from measuring its spectrum.
This also affected the accuracy of measurement of its magnitudes and {\Halpha} flux here. Its relative {\Halpha} flux uncertainty was 0.3. Its estimated uncertainty in {\SDSSu}{\SDSSg}{\SDSSr}{\SDSSi}{\SDSSz} was {0.1\,\magnitude}.

\subsection*{EIG 1a-02}

SDSS DR10 shows a galaxy, SDSS J005629.17+241913.3, {$\sim$2\,\arcmin} west of EIG 1a-02 with unknown redshift. The angular size of SDSS J005629.17+241913.3 is not very different from that of EIG 1a-02. Its magnitude is ${\SDSSg} = 16.6$, compared to ${\SDSSg} = 17.0$ of EIG 1a-02.
Although there is a possibility that SDSS J005629.17+241913.3 is a close neighbour of EIG 1a-02, this \markChange{is probably not the case}, since no tidal tails or other signs of interaction are visible in the SDSS images. \markChange{However, since EIG 1a-02 was not imaged using the WO {1\,meter} telescope we cannot be certain of this, as tidal tails and other fine structure features are hard to see or detect in shallow images like those of SDSS.}

\subsection*{EIG 1a-04}

{\Halpha} images of EIG 1a-04 showed strong star formation in LEDA 213033, a galaxy separated by {107\arcsec} from EIG 1a-04. Since LEDA 213033 has no measured redshift, its distance from EIG 1a-04 is unknown. The fact that it shows emission in the two narrow {\Halpha} filters used for the measurement, indicates that its redshift is $\cz \cong 6000 \pm 1500\,\kms$. Therefore, the probability that it is less than {300\,\kms} away from EIG 1a-04 is estimated to be $\sim$0.1.
No sign of interaction between EIG 1a-04 and LEDA 213033 was detected.

\subsection*{EIG 2s-04}

The {\Halpha} flux of EIG 2s-04 could not be measured due to a foreground star ($\SDSSr = 14.30$) at a projected distance of {12\arcsec}.
Although EIG 2s-04 has a GALEX measurement (in NUV only) it was not used, since it is contaminated with flux from this foreground star.

\subsection*{EIG 2s-06}

A foreground star of magnitude $\SDSSr = 15.6$, which is comparable to that of EIG 2s-06, is projected close to the EIG's centre. Its presence interfered with the photometric measurements, somewhat reducing the measured flux. The SDSS automatic photometry of EIG 2s-06 did not produce reliable results; the EIG was identified as two galaxies separated by the foreground star. 
GALEX measurements for this EIG were not used, since they also are contaminated by the foreground star. \markChange{The morphological type of EIG 2s-06 was not classified because of the foreground star.}

\subsection*{EIG 3s-06}

This is the only EIG that passes the isolation criterion using the ALFALFA dataset, but had neighbours closer than {3\,\Mpch} in the NED dataset. It was classified as part of subsample EIG-3s, because all of its NED neighbours are more than {2\,\Mpch} away from it.

\section{\\Modelled EIG properties}
\label{App:ModelledEIG_Properties}

The SFH, dust attenuation and stellar mass of the EIGs were estimated by fitting a model to their UV-to-near-IR SEDs as described section \ref{sec:ResultsModel}.
Figure \ref{f:ResGalMC_1D} shows, for each of the modelled EIGs, the marginalized posterior distributions of $Age_{1}$, $\tau^{-1}$, \EBV, $\Mass_{2}$ and {\Mstar}.
The extreme $\tau^{-1}$ values where $\left| \tau \right| \ll Age_{1}$ represent scenarios in which the first stellar population was created in a short burst. The low $\tau^{-1}$ values (where $\left| \tau \right| \gg Age_{1}$) represent an almost constant star formation for the first (main) stellar population.

\begin{figure*}
\begin{centering}
  \includegraphics[width=15.8cm,trim=0mm 0mm 0mm 0, clip]{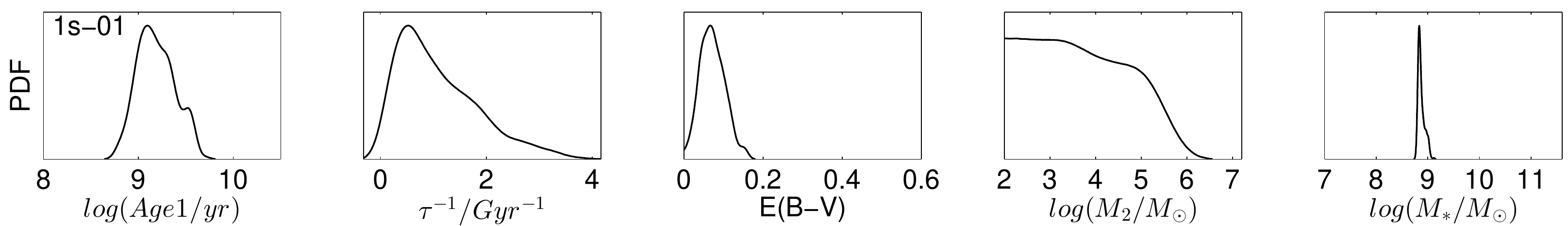}
  \includegraphics[width=15.8cm,trim=0mm 0mm 0mm 0, clip]{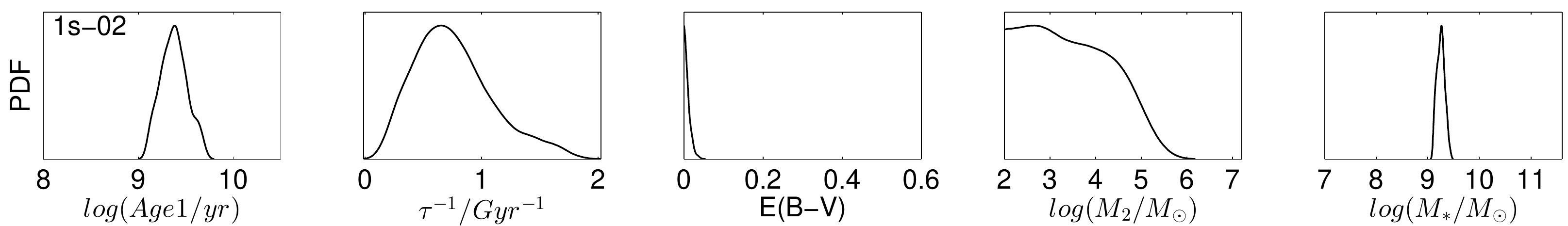}
  \includegraphics[width=15.8cm,trim=0mm 0mm 0mm 0, clip]{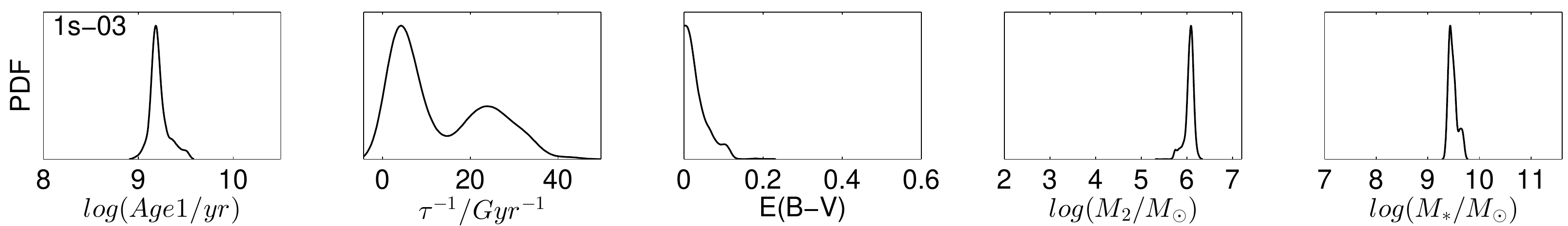}
  \includegraphics[width=15.8cm,trim=0mm 0mm 0mm 0, clip]{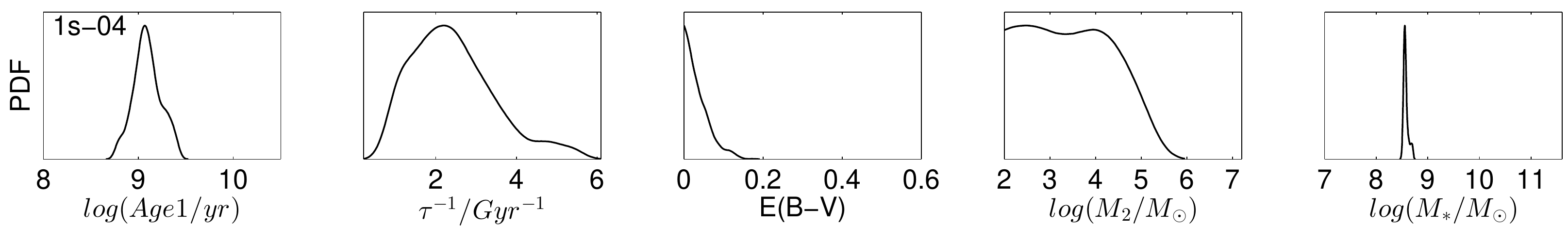}
  \includegraphics[width=15.8cm,trim=0mm 0mm 0mm 0, clip]{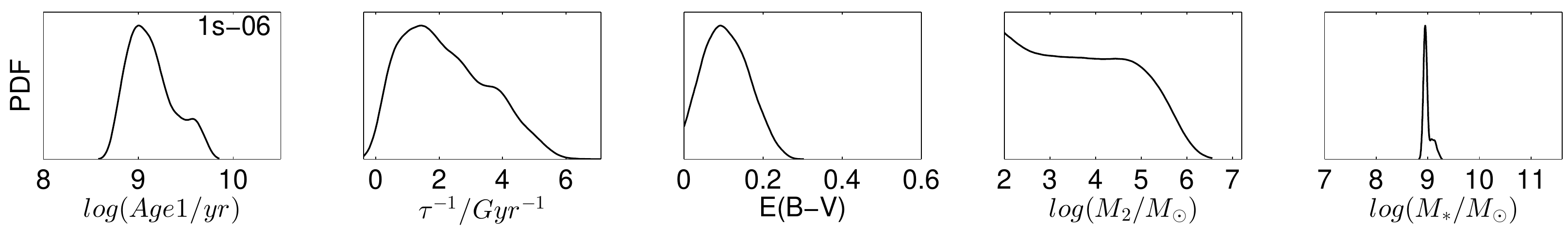}
  \includegraphics[width=15.8cm,trim=0mm 0mm 0mm 0, clip]{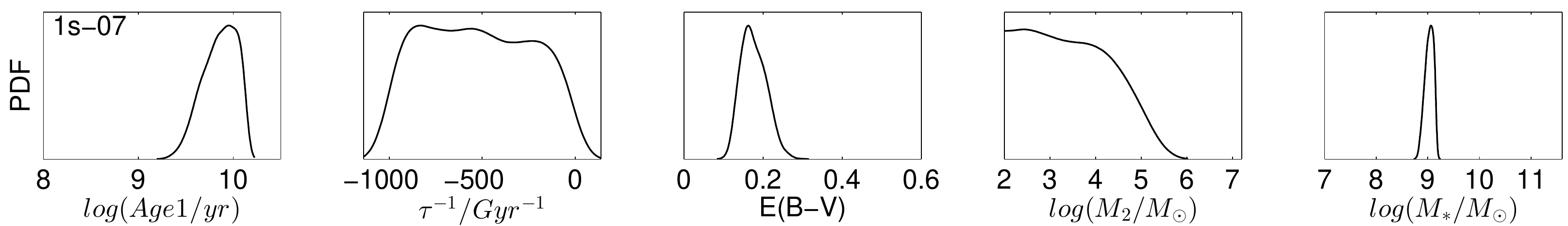}
  \includegraphics[width=15.8cm,trim=0mm 0mm 0mm 0, clip]{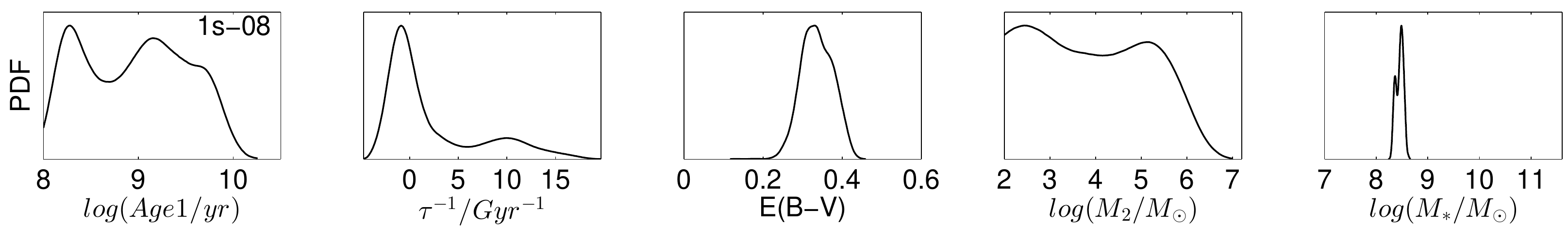}
  \includegraphics[width=15.8cm,trim=0mm 0mm 0mm 0, clip]{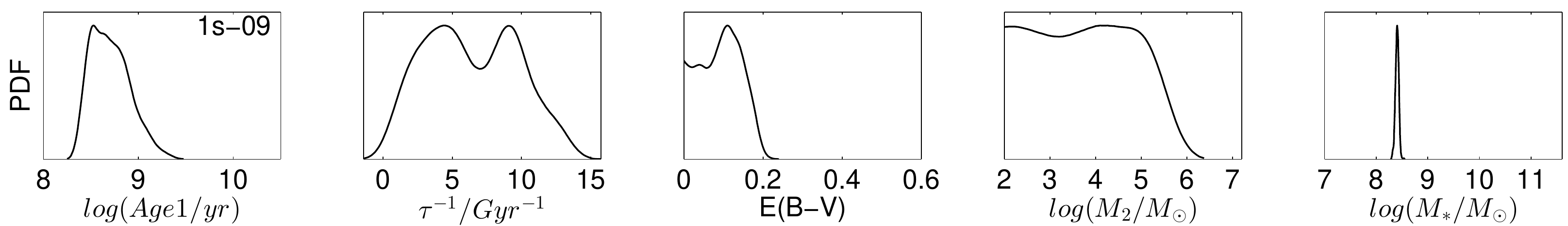}
  \includegraphics[width=15.8cm,trim=0mm 0mm 0mm 0, clip]{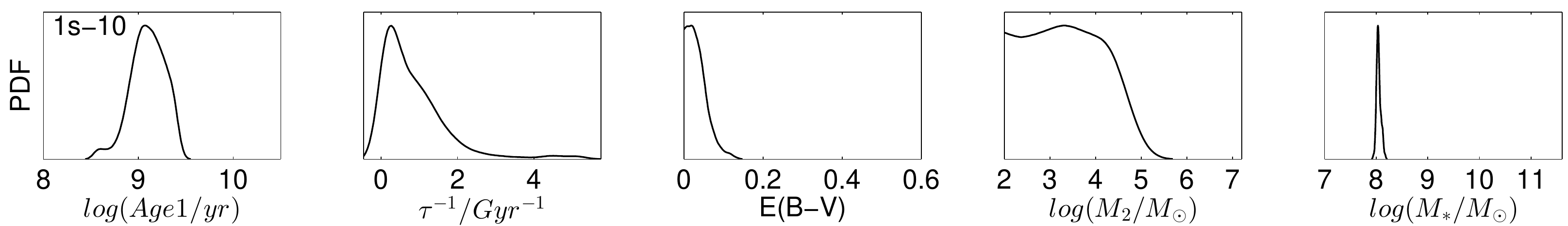}
  \caption [Modelled one-dimensional marginalized posterior distributions]
  {
    Modelled one-dimensional marginalized posterior distributions of $Age_{1}$, $\tau^{-1}$, \EBV, $\Mass_{2}$ and $M_{*}$. Data for each EIG is plotted in a separate row.\label{f:ResGalMC_1D}
  }
\end{centering}
\end{figure*}

\begin{figure*}
\begin{centering}

  \includegraphics[width=15.8cm,trim=0mm 0mm 0mm 0, clip]{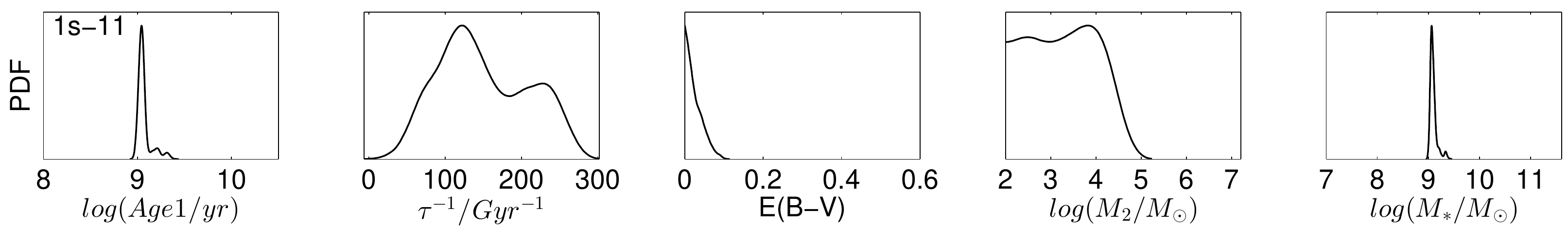}
  \includegraphics[width=15.8cm,trim=0mm 0mm 0mm 0, clip]{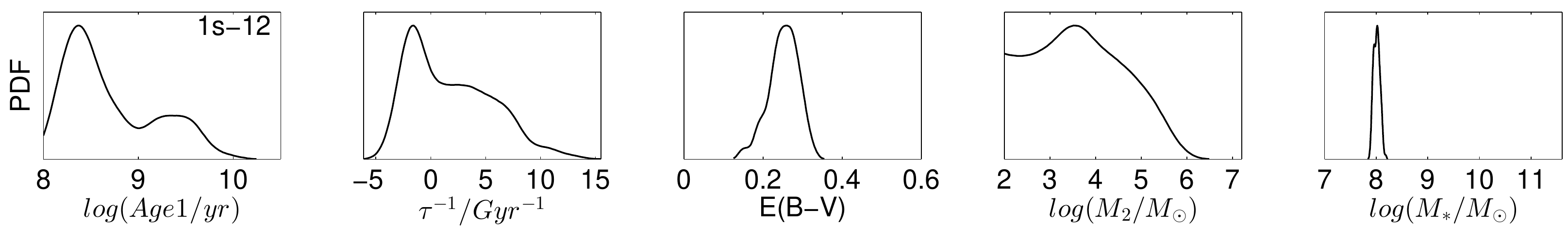}
  \includegraphics[width=15.8cm,trim=0mm 0mm 0mm 0, clip]{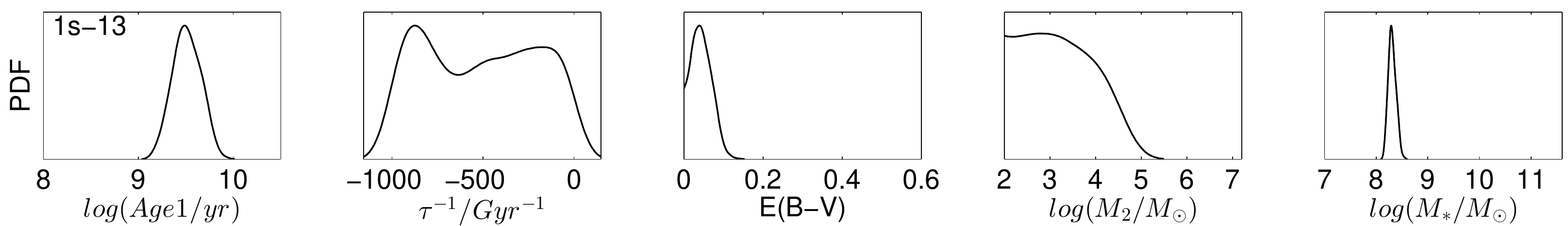}
  \includegraphics[width=15.8cm,trim=0mm 0mm 0mm 0, clip]{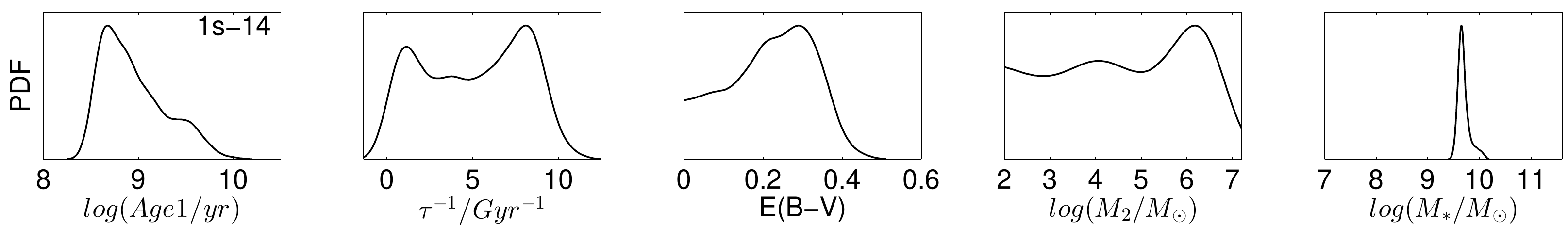}
  
  \includegraphics[width=15.8cm,trim=0mm 0mm 0mm 0, clip]{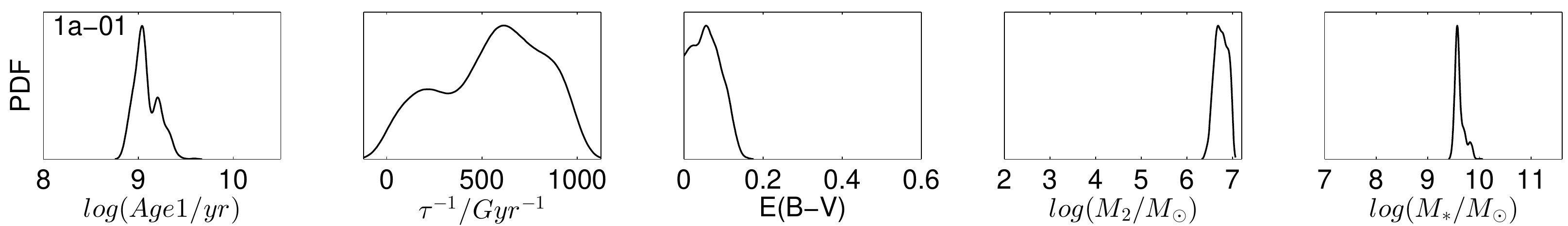}
  \includegraphics[width=15.8cm,trim=0mm 0mm 0mm 0, clip]{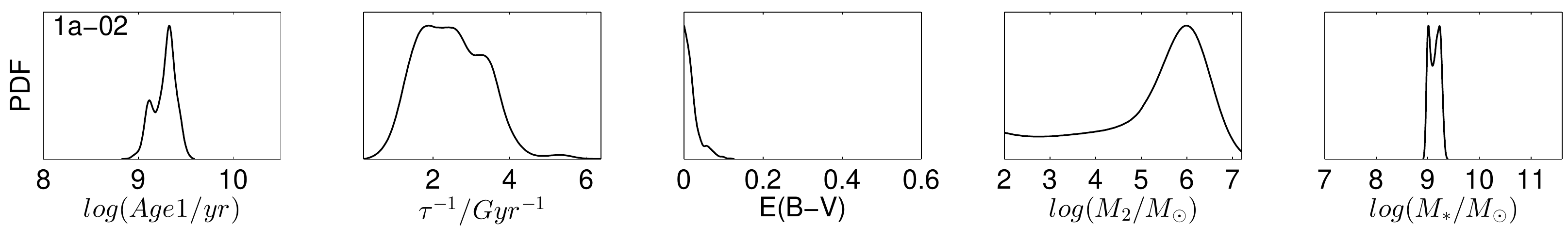}
  \includegraphics[width=15.8cm,trim=0mm 0mm 0mm 0, clip]{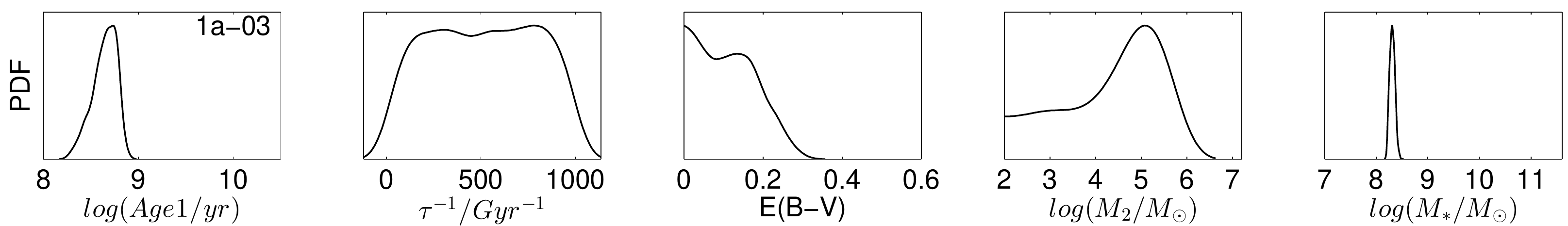}
  \includegraphics[width=15.8cm,trim=0mm 0mm 0mm 0, clip]{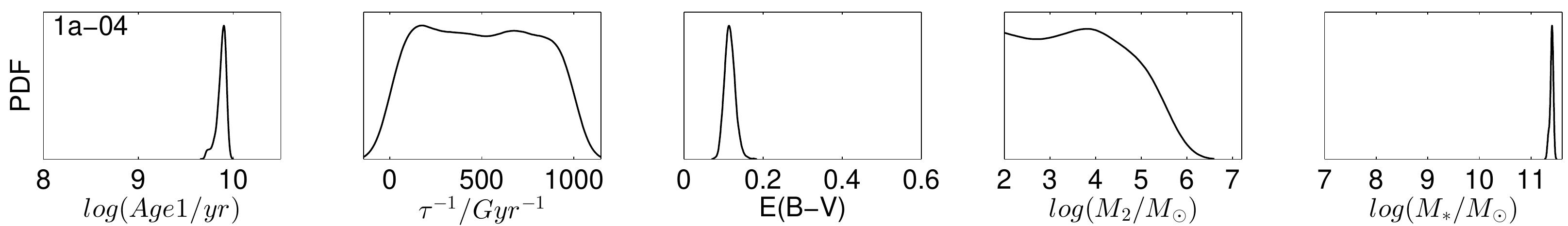}
  \includegraphics[width=15.8cm,trim=0mm 0mm 0mm 0, clip]{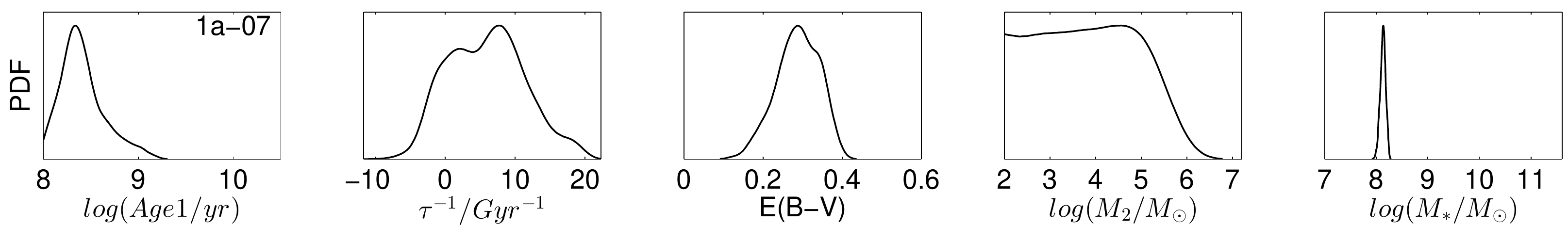}

  \contcaption
  {
    \label{f:ResGalMC_1D_2}
  }
\end{centering}
\end{figure*}

\begin{figure*}
\begin{centering}

  %
  \includegraphics[width=15.8cm,trim=0mm 0mm 0mm 0, clip]{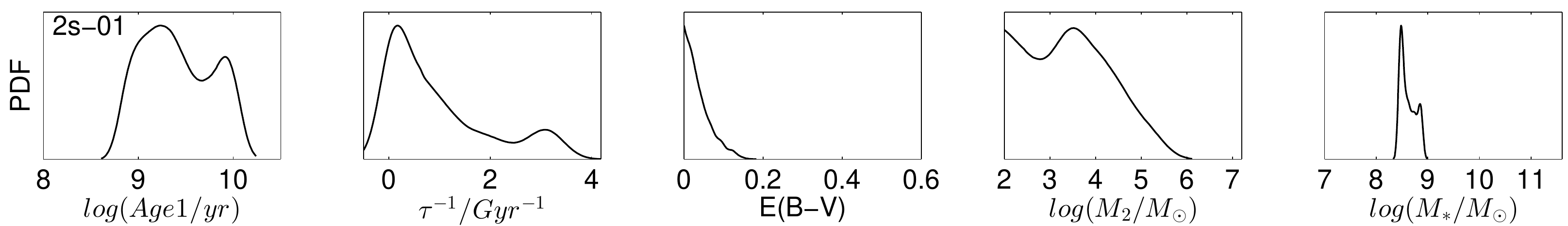}
  \includegraphics[width=15.8cm,trim=0mm 0mm 0mm 0, clip]{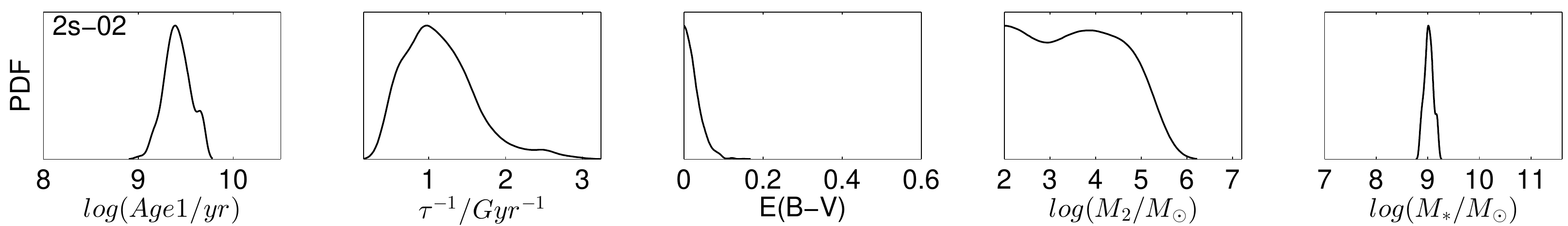}
  \includegraphics[width=15.8cm,trim=0mm 0mm 0mm 0, clip]{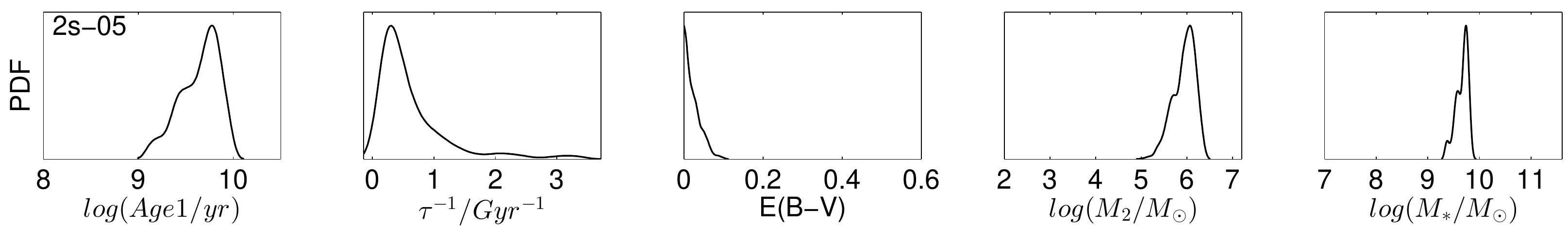}
  \includegraphics[width=15.8cm,trim=0mm 0mm 0mm 0, clip]{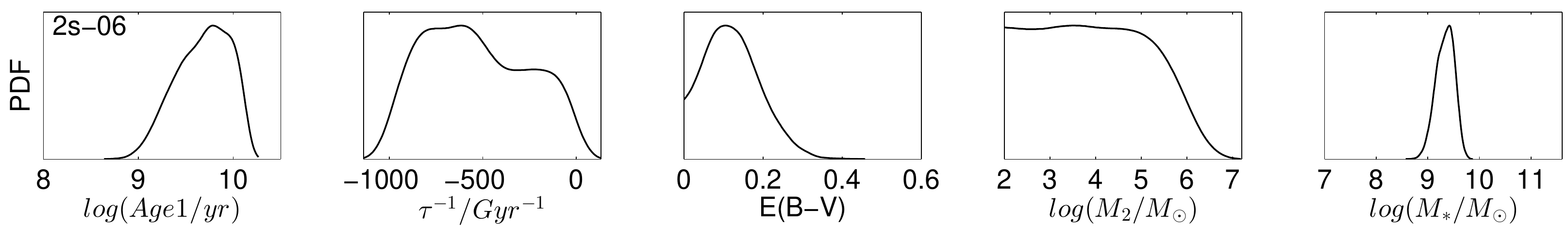}
  \includegraphics[width=15.8cm,trim=0mm 0mm 0mm 0, clip]{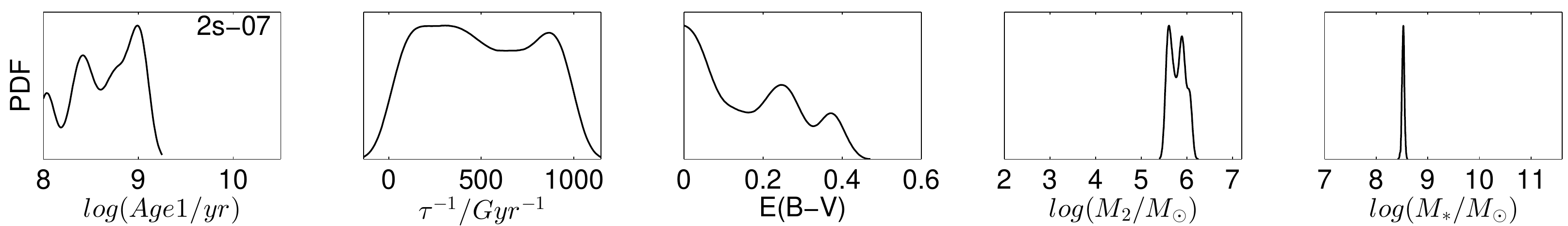}
  \includegraphics[width=15.8cm,trim=0mm 0mm 0mm 0, clip]{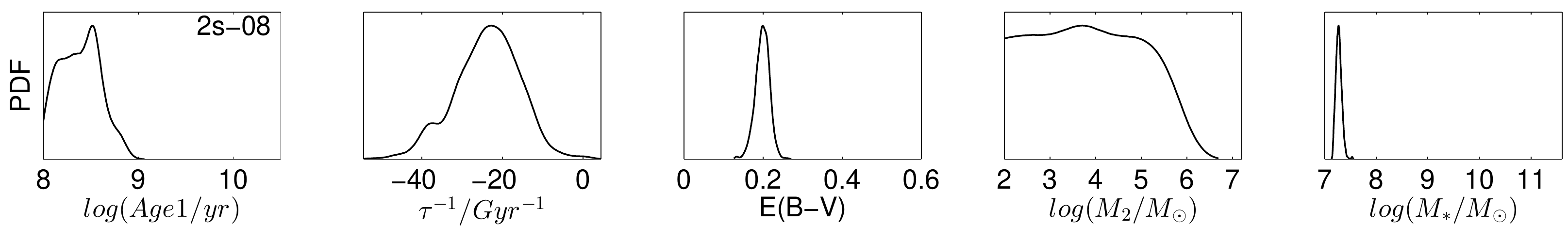}
  \includegraphics[width=15.8cm,trim=0mm 0mm 0mm 0, clip]{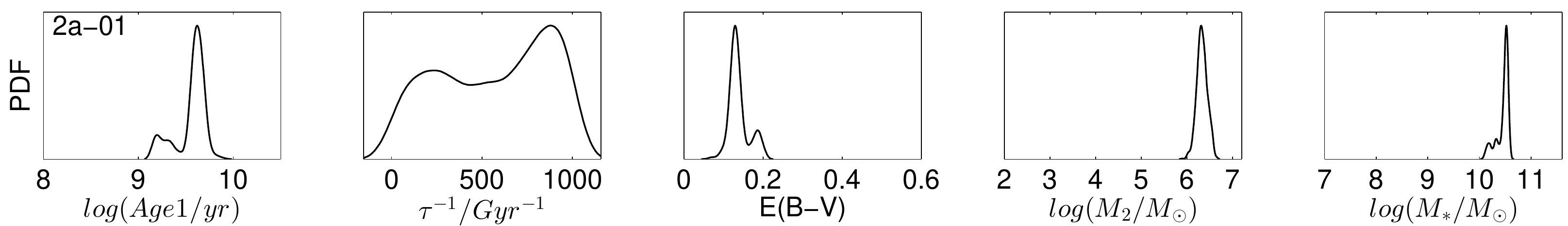}
  \includegraphics[width=15.8cm,trim=0mm 0mm 0mm 0, clip]{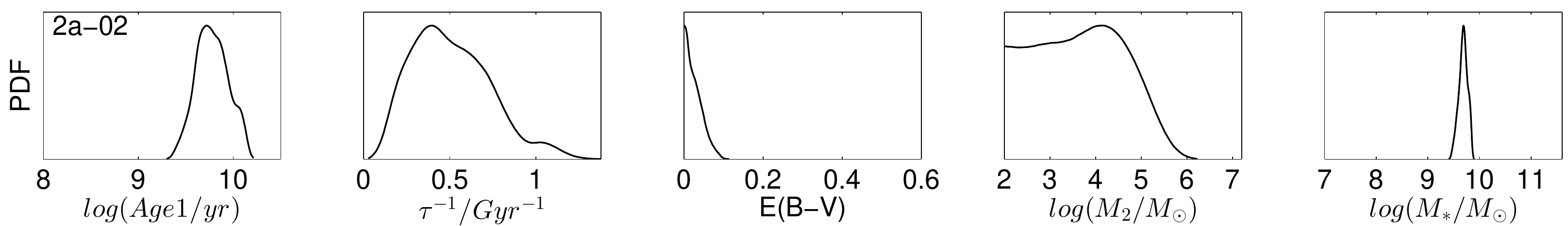}
  \includegraphics[width=15.8cm,trim=0mm 0mm 0mm 0, clip]{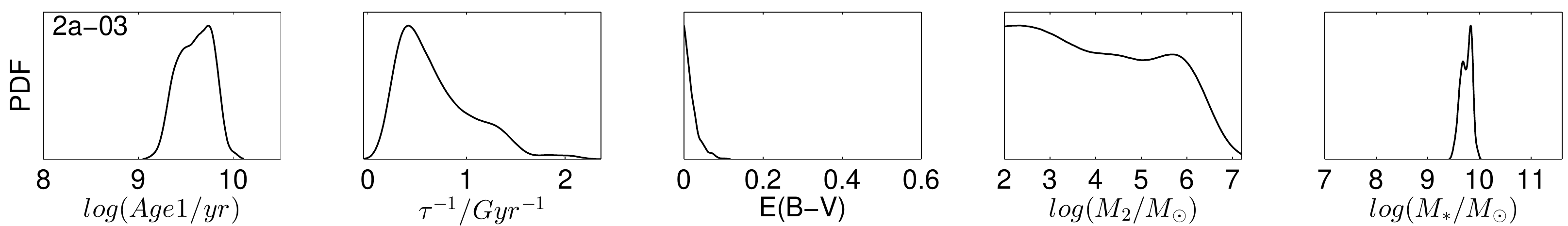}
  \includegraphics[width=15.8cm,trim=0mm 0mm 0mm 0, clip]{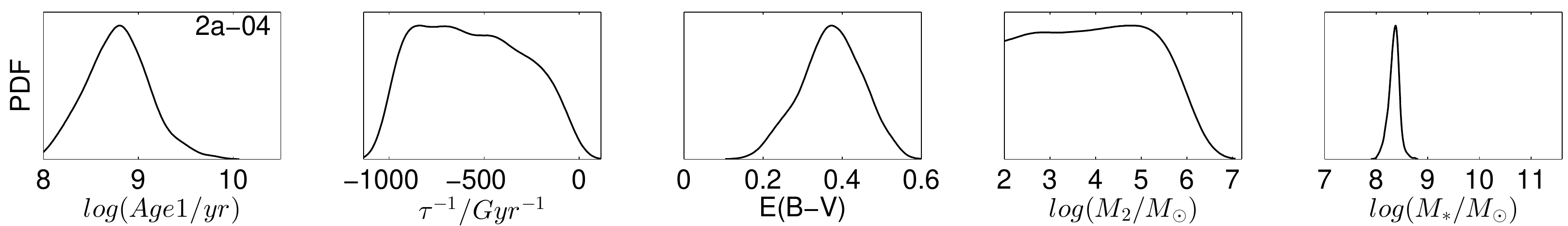}

  \contcaption
  {
  }
\end{centering}
\end{figure*}

\begin{figure*}
\begin{centering}

  \includegraphics[width=15.8cm,trim=0mm 0mm 0mm 0, clip]{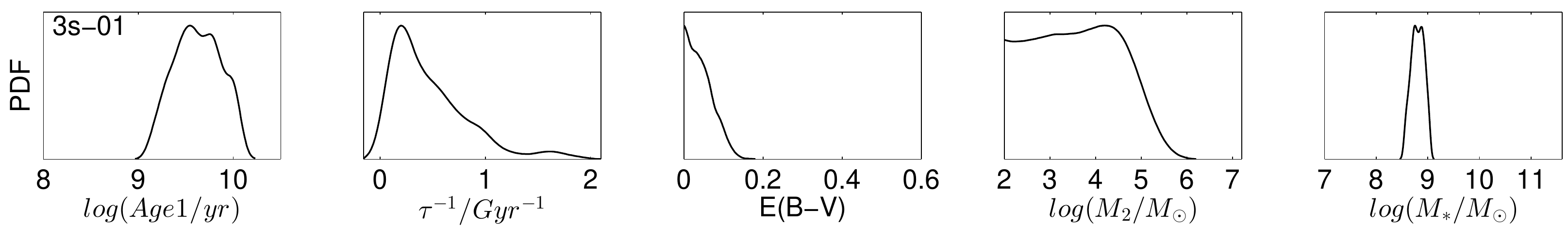}
  \includegraphics[width=15.8cm,trim=0mm 0mm 0mm 0, clip]{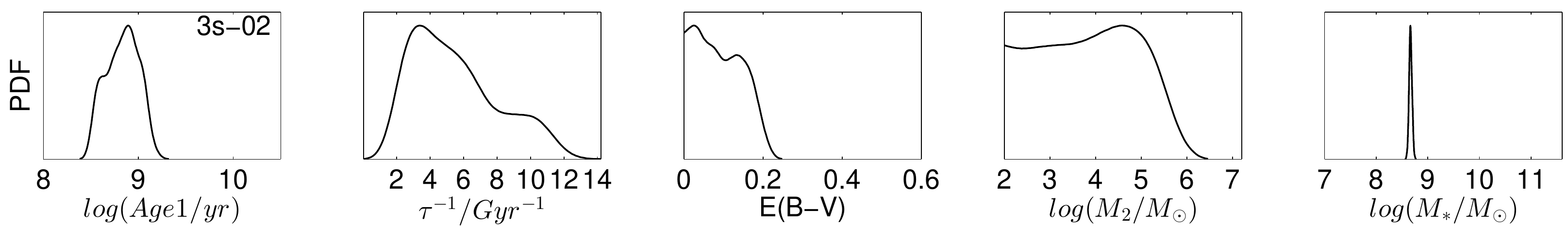}
  \includegraphics[width=15.8cm,trim=0mm 0mm 0mm 0, clip]{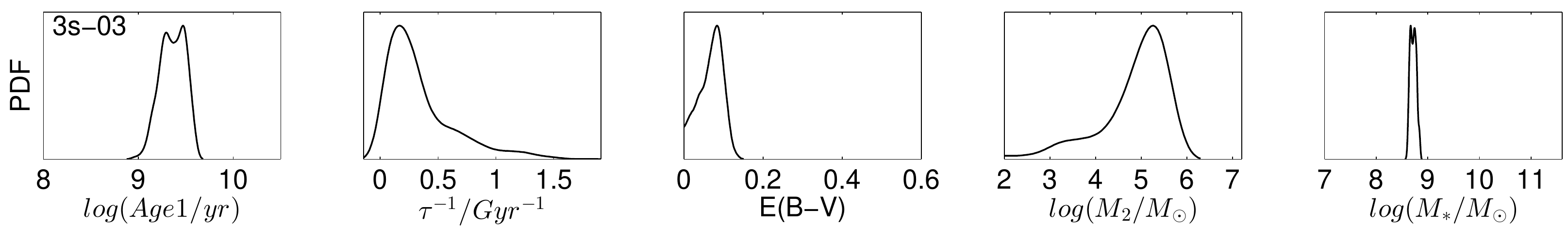}
  \includegraphics[width=15.8cm,trim=0mm 0mm 0mm 0, clip]{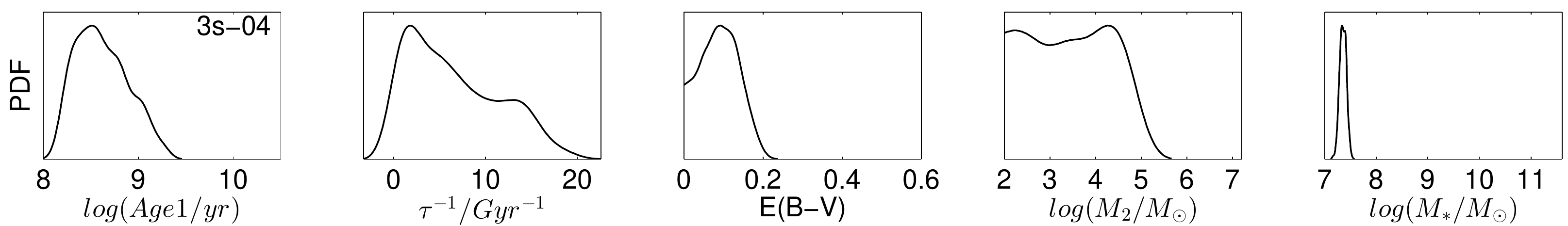}
  \includegraphics[width=15.8cm,trim=0mm 0mm 0mm 0, clip]{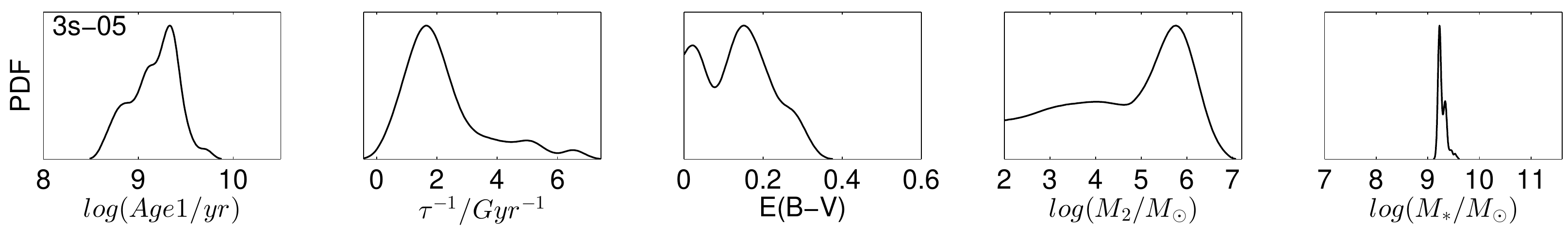}
  \includegraphics[width=15.8cm,trim=0mm 0mm 0mm 0, clip]{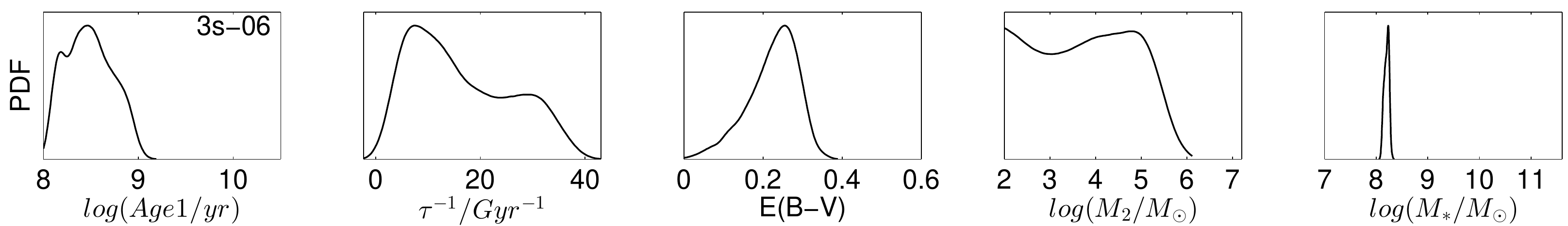}
  \includegraphics[width=15.8cm,trim=0mm 0mm 0mm 0, clip]{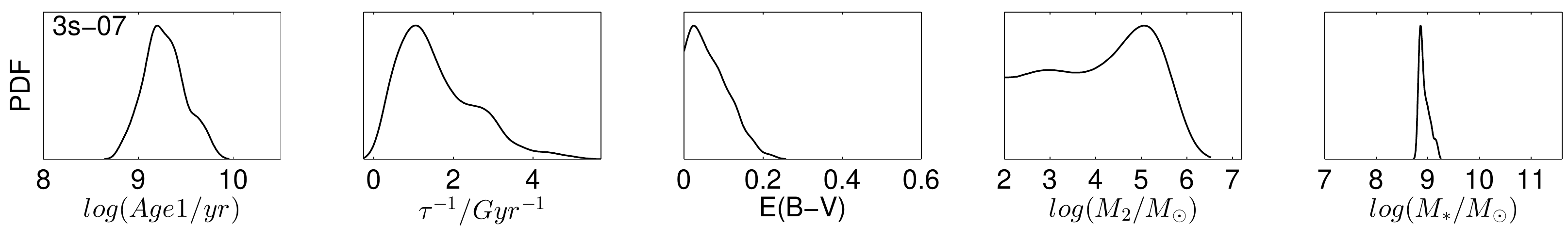}
  \includegraphics[width=15.8cm,trim=0mm 0mm 0mm 0, clip]{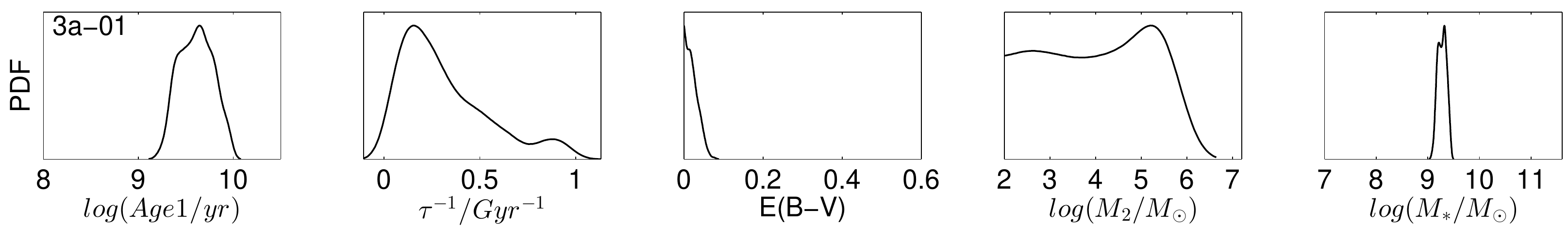}
  \includegraphics[width=15.8cm,trim=0mm 0mm 0mm 0, clip]{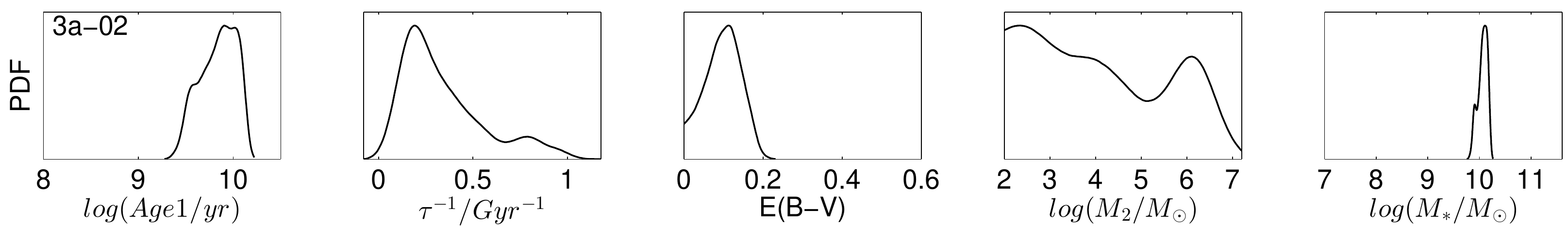}

  \contcaption
  {
  }
\end{centering}
\end{figure*}

\label{lastpage}

\end{document}